%% file: VortexUntwisted_paper.tex
\newif \iffinalversion

\finalversiontrue

\iffinalversion
\documentclass[acmtog,table]{acmart}
\else
\documentclass[acmtog,anonymous,review]{acmart}
\acmSubmissionID{252}
\fi

\usepackage{booktabs} % For formal tables
\setcitestyle{nosort,square} % nosort to allow for manual chronological ordering

\setcopyright{rightsretained}
\acmJournal{TOG}
\acmYear{2022}\acmVolume{41}\acmNumber{6}\acmArticle{241}\acmMonth{12} \acmDOI{10.1145/3550454.3555459}
\input macro.tex

\begin{document}
% Title portion
%\title{Untwisted Clebsch Variables for Vortex Filaments}
\title{Hidden Degrees of Freedom in Implicit Vortex Filaments}

\begin{teaserfigure}
    \centering
    \includegraphics[width=\textwidth]{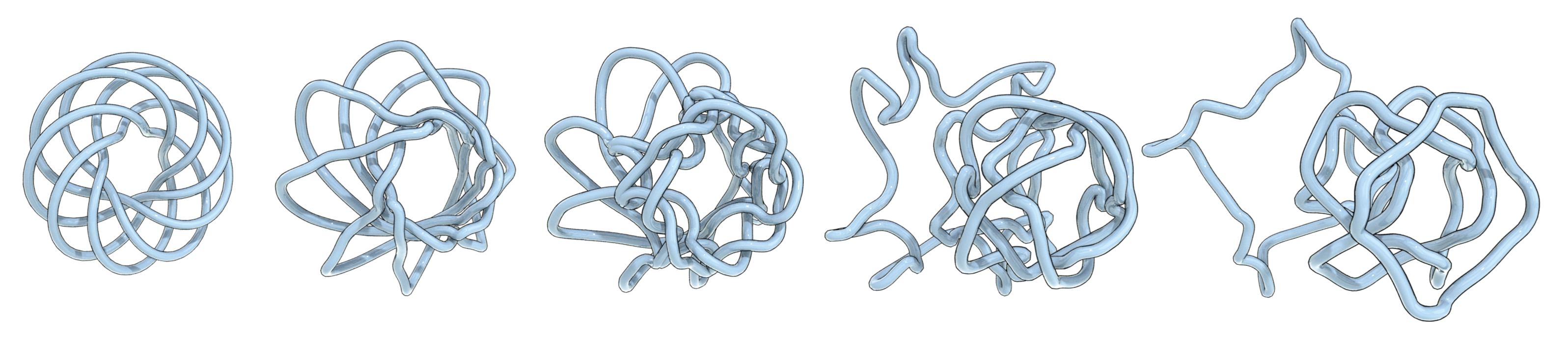}
    \caption{\stt{Our method evolves vortex filaments while automatically and robustly accounting for topological changes between curves. This image illustrates the}\rev{The time-}evolution of a highly knotted vortex filament according to the rules of inviscid fluid dynamics. \rev{Our implicit description represents filaments as the zero levelsets of space-time complex-valued functions, which automatically handles topological changes of curves without yielding singularities.}} 
    \label{fig:torus}
\end{teaserfigure}

% DO NOT ENTER AUTHOR INFORMATION FOR ANONYMOUS TECHNICAL PAPER SUBMISSIONS TO SIGGRAPH 2019!
\author{Sadashige Ishida}
\affiliation{%
  \institution{Institute of Science and Technology Austria}
  \country{Austria}
}
\email{sadashige.ishida@ista.ac.at}

\author{Chris Wojtan}
\affiliation{%
  \institution{Institute of Science and Technology Austria}
  \country{Austria}
}
\email{wojtan@ista.ac.at}

\author{Albert Chern}
\affiliation{%
  \institution{University of California, San Diego}
  \country{USA}
}
\email{alchern@eng.ucsd.edu}

\begin{abstract}
% This paper presents a novel method for simulating vortex filament dynamics for fluid simulation.
%This paper presents a description of curve dynamics for applications to vortex filaments in fluid simulation.
This paper presents a new representation of curve dynamics, with applications to vortex filaments in fluid dynamics. 
% Instead of explicitly representing these filaments with explicit curve geometry, we represent curves {\em implicitly} with a novel co-dimensional 2 level set method.
%Instead of explicitly representing these filaments with explicit geometry and equation of motion for filament curves, we represent curves {\em implicitly} with a new co-dimensional 2 level set description.
Instead of representing these filaments with explicit curve geometry and Lagrangian equations of motion, we represent curves {\em implicitly} with a new co-dimensional 2 level set description.
% We describe how naive approaches for implicit curve dynamics suffer from overwhelming numerical stability problems, and we identify several redundant mathematical degrees of freedom in both the velocity field and implicit geometry representation that can be tailored specifically to avoid these pitfalls. 
%We explain how naive approaches for implicit curve dynamics suffer from overwhelming numerical stability problems. 
%Our implicit representation admits several redundant mathematical degrees of freedom in both the configuration and the dynamics of the curves that can be tailored specifically to avoid these pitfalls.
% In contrast to naive approaches for implicit curve dynamics, which suffer from overwhelming numerical stability problems,
% our implicit representation admits several redundant mathematical degrees of freedom in both the configuration and the dynamics of the curves, which can be tailored specifically to improve numerical robustness.
Our implicit representation admits several redundant mathematical degrees of freedom in both the configuration and the dynamics of the curves, which can be tailored specifically to improve numerical robustness, in contrast to naive approaches for implicit curve dynamics that suffer from overwhelming numerical stability problems.
%\cSI{Thanks a lot for the edit, Chris. Now the flow is very natural. I made a tiny change here as the previous text could imply that naive approches do not have degrees of freedom while they do not exploit degrees of freedom that are hidden.}
Furthermore, we note how these hidden degrees of freedom perfectly map to a {\em Clebsch representation} in fluid dynamics. 
% Motivated by these observations, we introduce novel strategies for generalizing velocity extrapolation and level-set redistancing which successfully regularize sources of numerical instability, particularly in the twisting modes around curve filaments.
Motivated by these observations, we  introduce {\em untwisted} level set functions and {\em non-swirling} dynamics which successfully regularize sources of numerical instability, particularly in the twisting modes around curve filaments.
\stt{The result}\rev{A consequence} is a novel simulation method which produces stable dynamics for large numbers of interacting vortex filaments and effortlessly handles topological changes and re-connection events. 
%\cSI{I changed it so becasue the simulation method is not the only result of this paper. For example, the definitions of twist for our Clebsch variables and the fact that twist is invariant in the equivalence class are important results too.}
% \cSI{I made minor changes to emphasize that we introduce an implicit representation for both the configuration and the dynamics of filaments from the begnining. Chris's original version is commented out, so feel free to rollback. } \cCW{Slight modifications to Sadashige's draft to improve the flow.}
\end{abstract}

%
% The code below should be generated by the tool at
% http://dl.acm.org/ccs.cfm
% Please copy and paste the code instead of the example below.
%


%
% End generated code
%

\maketitle

\input introduction

\input relatedwork

%\input overview
\input ClebschVariables

\input algorithms

\input applications
\input conclusion

% Bibliography
\bibliographystyle{ACM-Reference-Format}
\bibliography{VortexUntwisted_bibliography}

\input appendix

\end{document}

%% file: macro.tex
\usepackage{picinpar,moresize,xfrac,graphpap,dcolumn,wrapfig,graphicx}
\microtypesetup{stretch=30,shrink=30}
\DeclareGraphicsExtensions{.png,.pdf,.jpg}
\usepackage{subcaption}
\usepackage{stackengine}

\usepackage{tikz,pgfplots}
\usepgfplotslibrary{colormaps}
\usetikzlibrary{pgfplots.colormaps}
\usepackage{rotating}
\pgfplotsset{compat=newest}
\pgfplotsset{plot coordinates/math parser=false}
\newlength\figureheight
\newlength\figurewidth
\usepackage{soul,array,calc,url,ragged2e,graphpap}
\usepackage{booktabs} % better tabular
\usepackage{tabularx}
\usepackage{colortbl}
\usepackage{multirow}
\usepackage{hyphenat}
\usepackage{transparent}
\usepackage{enumerate}
\usepackage{mathrsfs}

\usepackage[figure]{hypcap}
\usepackage{mathtools}
\mathtoolsset{centercolon}
\usepackage{nicefrac}
\usepackage{units}
\usepackage[normalem]{ulem}
\usepackage{cancel}
\usepackage{pbox}

\usepackage[Option]{overpic}

\makeatletter
\DeclareFontFamily{OMX}{MnSymbolE}{}
\DeclareSymbolFont{MnLargeSymbols}{OMX}{MnSymbolE}{m}{n}
\SetSymbolFont{MnLargeSymbols}{bold}{OMX}{MnSymbolE}{b}{n}
\DeclareFontShape{OMX}{MnSymbolE}{m}{n}{
    <-6>  MnSymbolE5
   <6-7>  MnSymbolE6
   <7-8>  MnSymbolE7
   <8-9>  MnSymbolE8
   <9-10> MnSymbolE9
  <10-12> MnSymbolE10
  <12->   MnSymbolE12
}{}
\DeclareFontShape{OMX}{MnSymbolE}{b}{n}{
    <-6>  MnSymbolE-Bold5
   <6-7>  MnSymbolE-Bold6
   <7-8>  MnSymbolE-Bold7
   <8-9>  MnSymbolE-Bold8
   <9-10> MnSymbolE-Bold9
  <10-12> MnSymbolE-Bold10
  <12->   MnSymbolE-Bold12
}{}
\let\llangle\@undefined
\let\rrangle\@undefined
\DeclareMathDelimiter{\llangle}{\mathopen}%
                     {MnLargeSymbols}{'164}{MnLargeSymbols}{'164}
\DeclareMathDelimiter{\rrangle}{\mathclose}%
                     {MnLargeSymbols}{'171}{MnLargeSymbols}{'171}
\makeatother

\usepackage{fancyvrb,listings}
\usepackage{algorithm}
\usepackage{algorithmicx}
\usepackage{algpseudocode}

\let\ForEach\ForAll

\algrenewcommand\alglinenumber[1]{\footnotesize #1:}
\makeatletter  
 \renewcommand{\ALG@name}{\small Algorithm} 
\makeatother 

\newtheoremstyle{mine}{3pt}{3pt}{\itshape}{}{\bfseries}{.}{.5em}{}
\theoremstyle{mine}
\newtheorem{theorem}{Theorem}
\newtheorem{lemma}{Lemma}
\newtheorem{proposition}{Proposition}

\newtheorem{definition}{Definition}
\newtheorem{example}{Example}
\newcommand{\figref}[1]{Figure~\ref{#1}}
\newcommand{\tabref}[1]{Table~\ref{#1}}

\newcommand{\secref}[1]{Section~\ref{#1}}

\renewcommand{\algref}[1]{Algorithm~\ref{#1}}
\newcommand{\thmref}[1]{Theorem~\ref{#1}}

\newcommand{\exref}[1]{Example~\ref{#1}}

\def\cf{\emph{cf.}}
\def\ie{\emph{i.e.}}

\def\CC{\mathbb{C}}

\def\RR{\mathbb{R}}
\def\SS{\mathbb{S}}

\def\bU{\mathbf{U}}
\def\bV{\mathbf{V}}

\def\cF{\mathcal{F}}

\def\cH{\mathcal{H}}

\def\cN{\mathcal{N}}

\def\sF{\mathscr{F}}
\def\sG{\mathscr{G}}

\def\bm{\mathbf{m}}
\def\bn{\mathbf{n}}

\def\bp{\mathbf{p}}

\def\bu{\mathbf{u}}
\def\bv{\mathbf{v}}
\def\bw{\mathbf{w}}
\def\bx{\mathbf{x}}

\usepackage{bbm}
\DeclareSymbolFont{bbold}{U}{bbold}{m}{n}
\DeclareSymbolFontAlphabet{\mathbbold}{bbold}
\renewcommand{\Im}{\operatorname{Im}}
\renewcommand{\Re}{\operatorname{Re}}

\def\det{\operatorname{det}}

\usepackage[draft]{pdfcomment}

\iffinalversion
\newcommand{\COMMENT}[3]{}  
\else
\newcommand{\COMMENT}[3]{\textcolor{#2}{{[\textbf{#1:} \textsl{#3}}]}}  
\fi

\defineavatar{TODO}{color=red,open=true}
\newcommand{\TODO}[1]{\COMMENT{TODO}{red}{#1}}
\defineavatar{cSI}{color=blue,open=true}
\newcommand{\cSI}[1]{\COMMENT{SI}{blue}{#1}}
\defineavatar{cAC}{color=olive,open=true}
\newcommand{\cAC}[1]{\COMMENT{AC}{olive}{#1}}
\defineavatar{cCW}{color=cyan,open=true}
\newcommand{\cCW}[1]{\COMMENT{CW}{cyan}{#1}}

\setstcolor{red}%Specify strikethrough'scolor 

\iffinalversion
\newcommand{\stt}[1]{}%for final version.
\newcommand{\sttEqn}[1]{}%for final version for equations.

\newcommand{\rev}[1]{{#1}}

\else
\newcommand{\stt}[1]{\st{#1}}%for revision.
\newcommand\sttEqn[1]{\setbox0=\hbox{$#1$}%
\rlap{\raisebox{.45\ht0}{\textcolor{red}{\rule{\wd0}{1pt}}}}#1} 

\newcommand{\rev}[1]{\textcolor{red}{#1}}

\fi

%% file: introduction.tex
%!TEX root = VortexUntwisted_paper.tex

\iffinalversion
\section{Introduction}
\else
\section{ \sout{NEW} Introduction}
\fi

The deformation of space curves is an interesting topic in many subjects such as differential geometry, low-dimensional topology, classical and quantum fluid mechanics, and electromagnetism. One example from fluid mechanics is the dynamics of vortex filaments.
In a nearly inviscid fluid, vorticity originates from {\em codimension-1} interfaces or obstacle surfaces. The vortex sheets subsequently roll up into {\em codimension-2} vortex filaments, due to the Kelvin--Helmholtz instability.  Hence, most physical inviscid fluids have their vorticity concentrated into a sparse set of space curves, rather than distributed evenly throughout space.
Based on this observation, certain physical equations model fluids only with dynamically deforming space curves. Many fluid simulation methods take advantage of this sparsity structure.

%\cCW{Intro could benefit from a simple didactic figure illustrating a codim2 surface rolling up into a codim1 filament.}

% \cSI{Maybe make the introduction of vortex filaments a bit succinct as it is just an example of deformaiton of curves though it is an important example that this paper is motivates by. }
% Due to certain physical nature of inviscid fluids, there is a tendency of fluid configurations that vorticity is concentrated into one-dimensional geometries such as vortex shedding or the Kelvin Helmholtz instability. \cSI{what is the {\it certain physical nature} here?} To capture this, many take the description that the fluid state is governed solely by such a codimension-2 concentration of vorticity called vortex filaments. c

One major challenge for an explicit (Lagrangian) filament-based fluid solver is to handle reconnection events when filaments collide.
Without any reconnection, the total length of filaments can grow exponentially, exploding the computational cost and halting the solver. Hence, existing explicit filament simulators include a tedious process of collision detection followed by non-differentiable heuristic geometry surgeries.

To that end, implicit (Eulerian) curve representations are more appealing.
The recently emerging \emph{Clebsch representation} expresses vortex lines as level sets of a 2-dimensional-valued function called \emph{Clebsch variables} 
\cite{Clebsch:1859:IHG}.
Like any level set method, topological changes of level set geometries occur gracefully.  
The difficulty, however, in a Clebsch-based fluid solver is in the dynamics of the Clebsch variables.  The Clebsch variables satisfy the transport equation advected with the fluid velocity, which unfortunately behaves in a swirling motion with a high-spatial frequency and singularities near the vortex filaments.
Such a rough transporting vector field is hard to resolve accurately in a computational grid.  Even if the transport equation is computed accurately, the level set function will quickly evolve into a twisted and distorted function that is difficult to deal with.

This paper develops a new 
% and more computationally stable 
approach for describing the geometry and dynamics of filaments with implicit curve functions. Our main insight is that the problem has a huge number of redundant degrees of freedom:  both the velocity field and the level set function (i.e., the Clebsch representation) can be varied in ways that do not change the solution. We exploit these additional degrees of freedom to ensure stable numerical simulation and automatic handling of topological changes, without sacrificing 
%speed or 
accuracy. 
% \cCW{I rewrote the paragraph more succinctly, while trying to keep Sadashige's main points.}
% \cSI{After saying we exploit the degree of freedom for both representation and dynamics, I think we should say what kind of specific choices we make for both of them: {\it untwisted} Clebsch variable and smooth non-swirling dynamics. }
% \cAC{In particular, what we can achieve by exploiting these degrees of freedom is to obtain a Clebsch representation or a codimensional-2 level set method that is untwisted and non-swirling both in the function and the dynamics.}
In particular, we choose an {\em untwisted} Clebsch representation for the level set geometry, and {\em non-swirling} dynamics for advecting  vortex filaments. We regularize these functions by identifying and constraining hidden degrees of freedom in their representations, allowing us to greatly improve numerical robustness compared to naive implementations. 

Our algorithm is the first method for animating implicit vortex filament geometry with automatic topological changes. Our mathematical formulation offers new tools for future research on fluid simulation and curve geometry processing, and our results show greatly improved stability compared to a standard level set implementation, with fewer user parameters than for explicit Lagrangian filament techniques.

%% file: relatedwork.tex
%!TEX root = VortexUntwisted_paper.tex
\section{Related Work}

% \cCW{I think related work goes great right after the intro, where we can fill in the details that we hinted at in the intro. More info on past methods for filament simulation, curve representations, etc.}
% \cSI{I talked with Chris about this. At this point, our related work seciton is Section 5 as comparing with other systems like the (nonlinear) Schr\"odinger equation, quantum Euler equation, or the Marsden-Weinstein form requires some preliminaries, which we will give secitons following the introduction. So we have options like
% \begin{itemize}
%   \item Keep Related Work in Section 2 but limit ourselves very high-level, and give some math to Marseden-Weinstein in a later section. So, this splits the discussion into very brief ones and involved ones. 
%   \item Have all the related work at a later section as we are doing now.
% \end{itemize}
% }

A majority of numerical schemes for the evolution of high-codimension geometries are developed in vortex methods in fluid animation and computational fluid dynamics.
These vortex-capturing schemes seek a representation for the vorticity and solve their governing equations of motion.

\paragraph{Explicit vortex methods}
In previous vortex methods, vortices are represented either as
particles \cite{gamito1995two,park2005vortex,Selle:2005:VPM,zhang2014pppm,Angelidis:2017:MVF},
filaments \cite{Cottet:2000:VM,Angelidis:2005:SSB,Weissmann:2009:FBS,Weissmann:2010:filaments,Padilla:2019:BRIC},
segments \cite{Chorin:1990:HRV,Xiong:2021:VSC}, 
sheets \cite{brochu2012linear,Pfaff:2012:LVS,da2015double} or 
volumes \cite{Elcott:2007:SCP,Zhang:2015:RMV}.
Vortex particle methods
represent vortices as a disconnected point cloud. However, the strength of vortex per particle or per unit volume \cite{Zhang:2015:RMV} undergoes a numerically unstable \emph{vortex stretching}, requiring an artificial clamping or diffusion that sacrifices accuracy.
The stretching problem is avoided by representing vorticity per filament, segment, or sheet, or per unit area using differential 2-forms \cite{Elcott:2007:SCP}.  However, describing vortex explicitly (Lagrangian method) with filaments, segments and sheets comes with a cost of sophisticated and heuristic treatment for changes of vortex topology.  Volumetric (Eulerian) methods \cite{Elcott:2007:SCP} do not require managing topological changes, but they do not have a handle on codimensional structures. 
Our codimension-2 level set method is an Eulerian method that can represent sharp filament structures without any additional difficulty from vortex reconnection.

% \cAC{Maybe mention Keenan's repulsive curve paper, discrete elastic rod paper somewhere.  It's also fine not to mention them as they are not really relevant.} \cCW{I think the current writing is excellent and gets right to the point. I think adding other papers would be a distraction.}

\paragraph{Clebsch representations}
Another Eulerian representation of vorticity is to describe vortex lines as the level sets of a 2D-valued function known as the Clebsch variable \cite{Clebsch:1859:IHG,Lamb:1895:HD}.  The representation was not widely adopted since all \(\RR^2\)-valued Clebsch variables can only describe fluids with zero \emph{helicity} \cite{Chern:2017:IF}.  The helicity problem is solved by using a sphere-valued Clebsch variable \cite{Kuznetsov:1980:TMC} which can represent nonzero (though quantized) values of helicity. Since the recognition of these ``spherical'' Clebsch maps, they have become an established method for vortex representation in computer graphics \cite{Chern:2016:SS,Chern:2017:IF,Yang:2021:CGF}.  However, Clebsch variables represent a {\em smooth vorticity field} by the continuum of level sets of all values. It remains a challenge to represent a {\em sharp codimension-2 filament}, especially with a limited grid resolution. By contrast, our method represents a sharp filament just as the \emph{zero set} of a %\(\CC\)-valued 
complex-valued
function.   Moreover, we show in \secref{sec:UntwistedClebschVariables} 
%\cAC{A corollary of Section 4.1 is that the helicity is the writhe, which can take any value.} 
that our representation can actually resolve continuous values of helicity (as opposed to the quantized values mentioned earlier).

\paragraph{Dynamics of Clebsch variables}
In addition to the implicit representation of vorticity, Clebsch variables also play significant roles in a variational and Hamiltonian formulation for the incompressible Euler equation \cite{Clebsch:1859:IHG,Lamb:1895:HD,Morrison:1998:HDI,Chern:2017:FD}.
In short, one (of many possible) governing equation(s) for the Clebsch variable is exceedingly simple: the Clebsch variable is advected by the fluid velocity.  This equation of motion is recently adopted by Yang et al.~\shortcite{Yang:2021:CGF}.  However, referred to as the \emph{Lagrangian chaos}, a direct transportation by the fluid velocity quickly stirs and twists any variable to a distorted one unresolved by the finite computational grid \cite{Qu:2019:ECF}. The Clebsch variable is no exception under such dynamics.
The method of \emph{Schr\"odinger's Smoke} \cite{Chern:2016:SS} bypassed the Lagrangian chaos: its total energy (Hamiltonian) includes the Dirichlet energy of the Clebsch variable, which is therefore bounded for all time.  
However, while the dynamics of \emph{Schr\"odinger's Smoke} appear to be similar to that of Euler's equation, it is only an approximation to the Euler fluid. 
There is still a large degree of freedom in the Clebsch representation and its dynamical system.  Finding an equation of motion for the Clebsch variable that both describe\rev{s} the correct Euler fluid\stt{and} without Lagrangian chaos is an unexplored research topic.  Our paper describes an instance of a Lagrangian-chaos-free dynamical system for an implicit representation of vortex filaments.

\paragraph{Implicit filament representations}
We represent filaments as the zero set of a complex-valued function. These zero sets of a complex phase field are widely studied in condensed matter physics as \emph{topological defects} appearing in superfluids and superconductors \cite{Bethuel:1994:GLV,Pismen:1999:VNF}. These topological defect models also facilitate singularity placements in flow analysis and geometry processing \cite{weissmann2014smokerings,Solomon:2017:BEO,Palmer:2020:ARV}. 
Complex phase field models are taken more generally as high-codimensional level set representations by \cite{Ambrosio:1996:LSA,Ruuth:2001:DGM,Burchard:2001:MOC,Min:2004:LLS}.
%However, in the physics or geometry processing literature, the level set functions have other physical meaning, leaving little rooms for a smoother representative.
However, in the physics and geometry processing literature, the level set functions have specific physical meanings such as the phase of a wave, leaving little room for a smoother representative. %\cCW{$\leftarrow$ I didn't understand this sentence, so I tried to make it clearer. Please tell me (or fix it) if I got the meaning wrong.} 
In most cases, the phase field has norm 1 except for a sudden dip to 0 near the filaments, creating a configuration that is difficult to resolve efficiently on a computational grid. 
In the level set method literature,
the norm-1 condition is often adopted for (re)initializing the level set functions despite the discontinuity \cite{Ruuth:2001:DGM}. The complex phase is constructed locally with little discussion about global topological obstruction. Burchard et al.~\shortcite{Burchard:2001:MOC} addressed the challenges in reinitializing the multi-component level set functions; by mimicking the codimension-1 signed distance functions, they propose a sophisticated reinitialization by solving a ``manifold eikonal equation'' along the isosurface of each function component. Unfortunately the process will not resolve the twists of the framed curve.
To our knowledge, there has not been a thorough discussion about the degrees of freedom in the implicit filament representations or in their dynamics until now.

In our work, we explore the degrees of freedom of both the codimension-2 level set functions and their equations of motion.  We further provide a simple reinitialization method comparable to the codimension-1 signed distance function.  Comparisons show that exploiting these degrees of freedom are essential to a robust simulation.

%% file: ClebschVariables.tex
%!TEX root = VortexUntwisted_paper.tex
% \section{Idea of our algorithm}
% \cSI{This is just a rough sketch of ideas.}

% Our setting in a broad sense is a level-set method with codimension 2.  This means that we express the time evolution of space curves by representing the curves as level-sets of some two functions and evolve these functions in time.

% In order to model physical phenomena with this setting, we need to specify the functions and their time evolution. For example, quantum vortices appearing in the Bose-Einstein condenstates of superfluids are modeled as the singularities of the Gross-Pitaevski equation.

% In the sequel, we will make a particular choice of the functions and their evolution law to relate our codimensional expression to the dynamics of (quantum) ideal fluids.

\section{Representations for evolving curves}
\label{sec:representation}
%A union of closed space curves is the central object in our simulation.  In this section, we describe an implicit representation for the object and its dynamics, and discuss the degrees of freedom in the representation. 
% In this section, we see representations for space curves and their dynamics. Then we discuss their degrees of freedom, which we will exploit.
% 
We now begin our description of \stt{how we simulate}implicit filament dynamics.
The main mathematical object \stt{in our simulation}is a union of closed space curves.  This section describes an implicit representation for these curves and their dynamics, as well as the degrees of freedom in the representation. 

% \cCW{I tried to work on making this more accessible. Some might not be familiar with the notation, others will benefit from some text interpretation. Although you've done a great job of expressing yourself concisely, it will likely fly over the head of a large fraction of readers. I am trying to draw out some important points and sometimes restate them in ways that will be more familiar to SIGGRAPH readers. Please check if I interpreted and explained these accurately.} 

\subsection{Representations for curve configuration}
Let the physical domain be an open region \(M\subset\RR^3\).  
We use $\gamma$ to represent a collection of $m$ closed curves
\begin{align}
    \label{eq:gamma}
     \gamma\colon\textstyle\bigsqcup_{i=1}^m\SS^1\rightarrow M,
\end{align}
where $\bigsqcup$ denotes a disjoint union, \(\SS^1\) is the topological circle, $m$ is the number of filaments, and $\gamma$ is the mapping from the 1D curves into 3D.
The configuration space \(\sF\) of these filaments is the space of all possible placements of these curves:
\begin{align}
    \label{eq:StateSpace}
    \sF = 
    \left.
    \bigsqcup_{m=1}^\infty\left\{
    \gamma\colon\textstyle\bigsqcup_{i=1}^m\SS^1\rightarrow M
    \right\}
    \right/\text{reparametrizations}.
\end{align}
Although it is straightforward to represent curves {\em explicitly} via objects \(\gamma\) in \eqref{eq:StateSpace} as parameterized curves 
or their discrete counterparts, topological changes of curves such as splitting or merging are difficult to describe mathematically and algorithmically.  For instance, the number \(m\) of components can change 
%under reconnection events.
when curves reconnect or split apart.
% \cAC{I replaced the original \(\bigoplus_{m=1}^\infty\) by a \(\bigsqcup_{m=1}^\infty\).  There is no linear structure for direct sum.}

% An explicit representation of a filament is a space curve i.e. a continuous map from \(\SS^1=\left\lbrace[0,2\pi] / 0 \sim2\pi\right\rbrace\) to a three dimensional configuration space \(M\) up to reparametrization. Hence the state space of the collections of filaments is given as
% \begin{align}
%      \sF= \bigoplus_{m=1} ^\infty \left\lbrace \gamma \colon\textstyle\bigsqcup_{i=1}^m\SS^1 \rightarrow \text{interior}(M) \cup \{\infty\} \right\rbrace \Big/\text{reparametrizations}.
% \end{align}
% \cSI{Maybe insteady of defining \(\sF\), we could just say "the collection of finite number of closed and open space curves (that connect to the boundary of \(M\) )." to avoid equations. But we want to use this state space \(\sF\) later in this section and in a later section to relate other spaces, so repeating "the collection of ..." everytime is tedious.
% }
% This representation is straightforward, but the topological changes of curves such as splitting or merging must be heuristically taken care of. In addition, as many types of filament dynamics requires spatial derivatives, curves often need manual cleaning via resampling with conditions such as maximum/minimum edge length or angles.

\begin{figure}[tbp]
    \begin{minipage}[b]{1.0\linewidth}
         \centering
         \includegraphics[width=0.95\linewidth, tics=0,
% ,trim={4cm 2cm 6cm 4cm}
,trim={0cm 0cm 0cm 0cm}
, clip]
{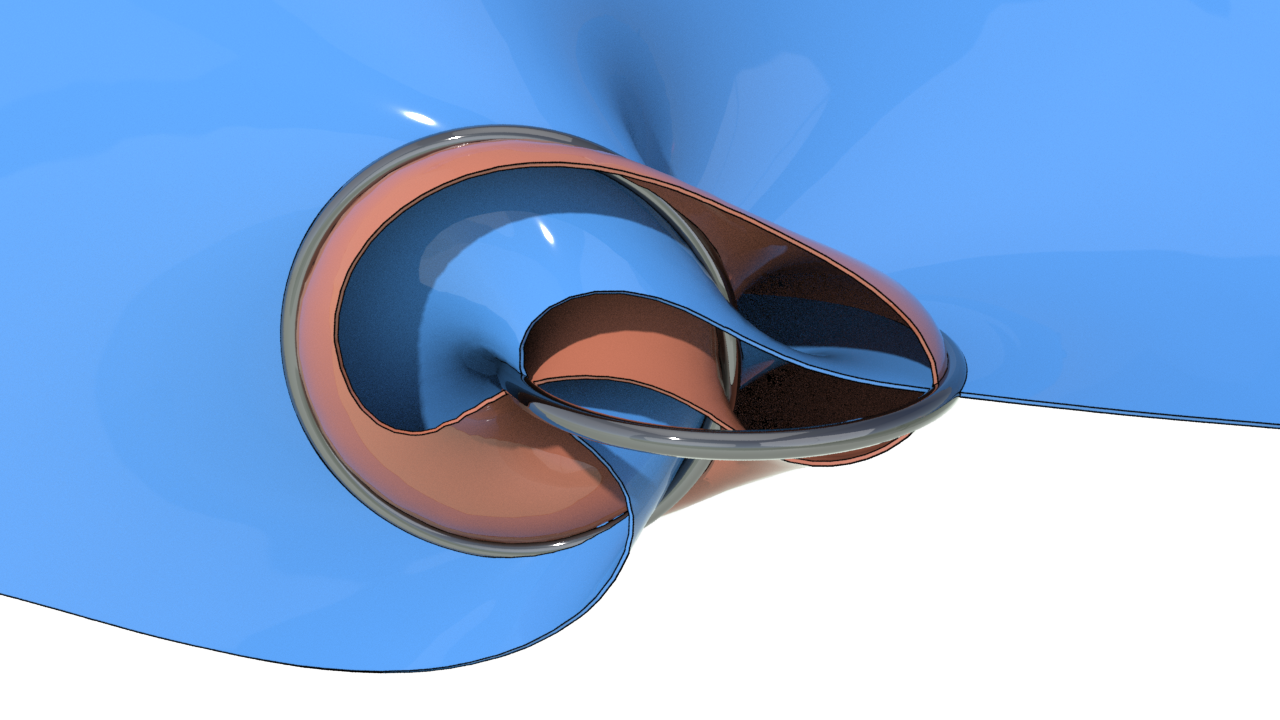}
% {{images/illustrations/level_surfaces_and_ribbons/level_surface_re_im}.jpg}
     %   \caption{obstacle curves}
     \end{minipage}
%       \begin{minipage}[b]{1.0\linewidth}
%           \centering
%           \includegraphics[width=0.95\linewidth, tics=0,
% ,trim={4cm 2cm 6cm 4cm}
% , clip]
% {{images/illustrations/level_surfaces_and_ribbons/level_surface_re}.png}
%       %   \caption{obstacle curves}
%       \end{minipage}
%       \begin{minipage}[b]{1.0\linewidth}
%           \centering
%           \includegraphics[width=0.95\linewidth, tics=0,
% ,trim={4cm 2cm 6cm 4cm}
% , clip]
% {{images/illustrations/level_surfaces_and_ribbons/level_surface_im}.png}
%       %   \caption{obstacle curves}
%       \end{minipage}

    \caption{An example of level surfaces of two linked curves. The blue and red surfaces are \rev{cross sections of} \(\{\textup{Im} \psi =0 \}\) and  \(\{\textup{Re} \psi =0 \}\) respectively. 
    % \cSI{If you think we don't need all of them. Feel free to use only the one with both real and imaginary level surfaces.}
    }\label{fig:level_surfaces}
\end{figure}

Instead of relying on an explicit curve representation, our work adopts an {\em  implicit} representation for the elements in \(\sF\).
We model every collection of closed curves in \(M\) as the zero set of a complex-valued level set function \(\psi\colon M\rightarrow\CC\):
\begin{align}
    \gamma = \{p\in M\,\vert\,\sttEqn{\psi_p}\rev{\psi(p)} = 0\} = \{\Re\psi = 0\}\cap\{\Im \psi=0\},
\end{align}
In other words, an alternative definition for $\gamma$ is the set of all points $p$ where {\em both} the real and imaginary components of a level set function $\psi$ evaluate to zero as in \figref{fig:level_surfaces}. We find it useful to draw an analogy to the scalar-valued level sets commonly used in computer graphics applications: the zero level-set of a scalar-valued function represents shapes of codimension 1 ({\em a.k.a.} surfaces), while our complex-valued level set has twice as many variables and thus represents shapes of codimension 2 (curves).
% In this work, we give an implicit representation to a collection of curves \(\gamma\). We represent \(\gamma\) as a codimension-2 level set, meaning that we employ a pair of real-valued functions where the intersection of their zero level surfaces is \(\gamma\). Complex-valued functions \(\{\psi:M\rightarrow \CC^1\}\) are convenient as we can simply set
% \begin{align}
%      \gamma=\{\psi=0\} = \{\textup{Re} \psi=0\} \cap \{\textup{Im} \psi=0\}.
% \end{align}
 
We next note that different functions $\psi$ can represent the same collection of curves if they share the same zeros. 
To formulate this redundancy precisely, we define the following equivalence relation \(\sim\) on the function space \((M\rightarrow\CC)\).
We say \(\psi_1\sim\psi_2\) if and only if \(\psi_1 = \varphi\psi_2\) for some nowhere-vanishing function \(\varphi\colon M\to\CC\setminus\{0\}\).
%To formulate this, we say, two  functions are equivalent if 
%\(\psi_1\sim\psi_2\colon\Leftrightarrow\psi_1 = \varphi\psi_2\) for some \(\varphi\colon M\to\CC \setminus \{0\}\).  
\stt{In particular}\rev{This means} \(\psi_1\sim\psi_2\) if and only if they share the same zero set.

\iffalse
After disregarding this pointwise rescaling, we have a one-to-one correspondence
\begin{align}
     \sF\cong\sG \coloneqq \{\psi\colon M\rightarrow\CC\}/\sim.
 \end{align}
 The degrees of freedom in expressing an element in \(\sF\cong\sG\) by a representative \(\psi\) is the choice of a non-vanishing complex-valued rescaling function \(\varphi\). 
\fi

% \cCW{I removed the mathematical definition of congruence between configuration spaces since it is fairly obvious from the writing and only adds potential confusion if you don't take the effor to define and explain your symbols. I also removed the sentence following it about the free degree of freedom, since it is redundant with the text above.}
 
% \cCW{Update after meeting Sadashige: The equation said more than I realized, so need to find a way to restore the lost information while giving the reader complete confidence that he/she knows what is the important information they should take away from it.}

% \cAC{I moved the regularity assumption up here}
 
%\subsubsection{Regularity}
%\label{sec:Regularity}
 For the implicit representation to work properly, we make a regularity assumption that the level set functions \(\psi\) are sufficiently ``nice'' around the zero set.  Precisely, we assume \(\psi\) is smooth, and
 the differential \(d\psi\vert_p\colon T_pM\rightarrow\CC\cong\RR^2\) is surjective, i.e. \stt{it's}\rev{its} matrix form has rank 2 at any point \(p\)  on the zero-set.
%  it's first spatial derivative has rank 2 at all points on the zero-set.
%\cSI{The differential being a linear map between vector spaces and its surjectivity are important, so I retrieved this.}
% 
\footnote{The full-rank requirement does not modify the degrees of freedom in the rescaling \(\varphi\) since \(d(\varphi\psi)\vert_p = d\varphi\vert_p\underbrace{\psi_p}_{=0} + {\varphi_p} d\psi\vert_p = \varphi_p d\psi\vert_p\) and a non-zero complex multiplication by \(\varphi_p\) always preserves the rank of \(d\psi\vert_p\).}
%\cCW{simplified the explanation in the text and demoted the subsubsection to a paragraph. The footnote and remaining text still preserve the mathematical info needed for later proofs.}

%  \cAC{How about the following:}
To summarize, our configuration space of filaments \(\sF\) from \eqref{eq:StateSpace} is replaced by the space of complex-valued functions \(\psi\) modulo a multiple of a non-vanishing function:
 \begin{align}
     \label{eq:GDefinition}
      \sG \coloneqq \{\psi\colon M\rightarrow\CC\}/\sim.
  \end{align}
While \(\sF\) and \(\sG\) describe the same configuration space of filaments (\(\sF\cong\sG\)), the objects in \(\sG\) are much more continuous compared to the disjoint spaces of \(\sF\).  

\paragraph{Relation to Clebsch representations}
The representation of codimension-2 curve geometries in 3D is known in fluid dynamics as \emph{Clebsch representations} \cite{Clebsch:1859:IHG,Lamb:1895:HD,Chern:2017:IF,Yang:2021:CGF}. 
For a fluid flow with a smooth vorticity field, the vortices are geometrically depicted as fibrous vortex lines diffusely distributed over the fluid domain.  A Clebsch representation aims at an implicit representation for such fibrous structure.  The representation uses a map \(s\colon M\rightarrow\Sigma\) from the 3D fluid domain \(M\) to a 2-dimensional manifold \(\Sigma\) with a measure \(\sigma\) 
%\cAC{just \(\sigma\) and remove \(\Omega^2(\Sigma)\)}
 to describe the \emph{vortex lines} as preimages \(s^{-1}\{p\}\) of points \(p\in\Sigma\), and the density of the vortex lines 
%  \sout{(\ie\@ the vorticity 2-form)} \cAC{\(\leftarrow\) remove} 
 as the pullback \(s^*(\sigma)\) of the measure \(\sigma\).
A smooth Clebsch map \(s\) and a smooth measure \(\sigma\) yields a smooth distribution of vortex lines.  To achieve a more singular and concentrated vorticity field such as vortex filaments, one would consider \(s\) with larger derivatives \cite{Marsden:1983:COV} (\(s\) sweeps out more measure \(\sigma\) over a small area in \(M\)). 

In our setup, we want to represent singular curves with a Dirac-\(\delta\) density, instead of diffused distribution of vortex lines.  Previous considerations in Clebsch representations would set the Clebsch map \(s\) with enormous derivative.  By contrast, we obtain such concentrated filaments by setting \(\sigma\) singular while keeping the Clebsch map smooth.   Our complex level function \(\psi\colon M\rightarrow\CC\) is a Clebsch map with the target space \(\Sigma = \CC\) equipped with a \(\delta\)-measure \(\sigma = \delta_0\) at the origin.

An important discussion about Clebsch representations \cite{Chern:2017:IF} is whether or not a fluid configuration can be represented with the choice of \(\Sigma\) and \(\sigma\).
Previous Clebsch representations \cite{Chern:2017:IF,Yang:2021:CGF} adopt \(\Sigma=\SS^2\), since the more straightforward choice of \(\Sigma = \RR^2\cong\CC\) with the standard area measure \(\sigma\) can only represent fluid flows with zero \emph{helicity}.  The helicity obstruction is reduced for \(\Sigma=\SS^2\) as it can admit a discrete set of nonzero helicity \cite{Chern:2017:IF}.  Our Clebsch representation can represent any space curve without any obstruction.  In particular, the helicity of a vortex filament is proportional to its \emph{writhe} \cite{Arnold:1998:TMH} which can take any real value.

\subsection{Representations for curve dynamics}
\label{sec:RepresentationsForFilamentDynamics}
%\cSI{I believe this section is now easy to understand to most readers who know elementary differential geometry, assumed in many graphics papers. }

%Next, we discuss the representation for the dynamics of the filaments.
In the explicit representation, a first-order time evolution of curves  \(\gamma\colon(\bigsqcup^{m}\SS^1)\times\RR\rightarrow M\) can be described by an  equation of the form:
% In the explicit representation, the motion of time-dependent curves
%  \(\gamma\colon(\bigsqcup^{m}\SS^1)\times\RR\rightarrow M\) can be described by a dynamical system of the form:
%In the explicit representation of time-dependent curves \(\gamma\colon(\bigsqcup^{m}\SS^1)\times\RR\rightarrow M\), a dynamical system of first-order in time is described by an equation of the form
\begin{align}
    \label{eq:ExplicitDynamics}
    {\partial\gamma\over\partial t}(s,t) = \bV_{\gamma_t}(s),\quad s\in\textstyle\bigsqcup\SS^1, t\in\RR.
\end{align}
%\cAC{
%\begin{align}
%    \label{eq:ExplicitDynamics}
%    {d\gamma\over d t}(s,t) = \bV_{\gamma_t}(s),\quad s\in\textstyle\bigsqcup\SS^1, t\in\RR.
%\end{align}}
Here, $s$ is the parameterization of the curve, $t$ is time, and the \emph{velocity} \(\bV_{(\cdot)}\colon\sF\rightarrow (\bigsqcup\SS^1\rightarrow\RR^3)\) is a dynamical model that tells the filament how to move based on the current filament shape and position.
%\cCW{$\gamma$ is a Lagrangian variable, right? Should this be the total derivative wrt time, instead of the partial?}\cAC{In both Lagrangian and Eulerian the time derivatives are all partial derivatives, because there are more than one coordinate.  The difference is what the rest of the variables that is fixed.  Here the Lagrangian parameter \(s\) is fixed during the partial.  Similarly, in continuum mechanics, the material derivative is a time derivative where the Lagrangian coordinates are fixed.  So in a way it doesn't matter.  If \({d\gamma\over dt}(s,t) = \bV_{\gamma_t}(s)\) causes less confusion, I'm fine with it.}

%For a time evolutionary parametrized curve \(\gamma\colon \SS^1 \times \RR \rightarrow M \), a dynamical system of first-order in time is described simply by specifying the velocity \(\partial_t \gamma (s,t)\) for each point \((s,t)\). 

\begin{example}
    \label{ex:BiotSavart}
% \cAC{I encapsulated specific fluid model as an example. I also directly introduce Biot Savart and Rosenhead Moore as models to be evaluated on curves.}
%Vortex filaments are an example in fluid mechanics. A set of filaments \(\gamma=\bigcup_j \gamma_j\) define a velocity field in \(\RR^3\)
In the context of fluid dynamics, important examples for the velocity model \(\bV\) are the ones that govern the motion of vortex filaments.  When \(M=\RR^3\), \ie\@ there are no obstacles or boundaries, the velocity models are the Biot--Savart model
% \begin{align}\label{eq:biot_savart}
%      \bv_{\gamma}^{\rm BS}(x) \coloneqq \frac{\Gamma}{4 \pi} \sum_{j} \oint \gamma_{j}^{\prime}(s) \times \frac{x-\gamma_{j}(s)}{\left|x-\gamma_{j}(s)\right|^{3}} d s.
% \end{align}
\begin{align}\label{eq:biot_savart}
     \bV_{\gamma}^{\rm BS}(s) \coloneqq \frac{\Gamma}{4 \pi}  \oint \gamma^{\prime}(\tilde s) \times \frac{\gamma(s)-\gamma(\tilde s)}{\left|\gamma(s)-\gamma(\tilde s)\right|^{3}} d\tilde s
\end{align}
and the more regular Rosenhead--Moore model \cite[pp.~213]{Saffman:1995:VD}
%or an approximated field via the Rosenhead-Moore kernel,
% \begin{align}\label{eq:Rosenhead-Moor}
%      \bv_\gamma^{\rm RM}(x)\colon\coloneqq \frac{\Gamma}{4 \pi}  \sum_j\oint \gamma_{j}^{\prime}(s) \times \frac{x-\gamma_{j}(s)}{\sqrt{a^{2}+\left| x-\gamma_{j}( s)\right|^{2}}^{3}} d s
%  \end{align}
\begin{align}\label{eq:Rosenhead-Moor}
     \bV_\gamma^{\rm RM}(s)\coloneqq \frac{\Gamma}{4 \pi}  \oint \gamma^{\prime}(\tilde s) \times \frac{\gamma(s)-\gamma(\tilde s)}{\sqrt{e^{-\nicefrac{3}{2}}a^{2}+\left| \gamma(s)-\gamma(\tilde s)\right|^{2}}^{3}} d\tilde s
 \end{align}
where the constants $\Gamma$ and $a$ are the vortex strength and vortex thickness respectively, and the integrating measure \(d\tilde s\) is the arclength element (set \(a=0\) for the Biot--Savart model). 
%\cCW{This equation assumes $\Gamma$ and $a$ are global constants, not dependet on $\tilde s$. I guess spatial variation gets captured by $\gamma'$, so $\Gamma$ is just a scale factor. But $a$ should conceivably vary (at least per connected component). I think I'd prefer to leave it as is to avoid over-explaining, even if it's not the most general form of the equations. Opinions?}
%\cAC{I'm taking both \(\Gamma\) and \(a\) as constants.  I don't think we need to dive into details here.}
Note that  \eqref{eq:biot_savart} and \eqref{eq:Rosenhead-Moor} are the restrictions at the curve of the entire fluid velocity field over the 3D domain
\begin{align}
    \label{eq:RMAll}
    \bU^{\rm RM}_{\gamma}(\bx)
    \coloneqq \frac{\Gamma}{4 \pi}  \oint  \frac{\gamma^{\prime}(\tilde s) \times(\bx-\gamma(\tilde s))}{\sqrt{e^{-\nicefrac{3}{2}}a^{2}+\left| \bx-\gamma(\tilde s)\right|^{2}}^{3}} d\tilde s,\quad\bx\in\RR^3
\end{align}
That is, \(\bV^{\rm RM}_\gamma(s) = \bU^{\rm RM}_\gamma(\gamma(s)).\)

% with a real constant \(\Gamma\) called {\it intensity} and a non-negative constant \(a\) called the {\it thickness} of the filaments. 
%  In this paper, we call both \autoref{eq:biot_savart} and \autoref{eq:Rosenhead-Moor} the Biot-Savart velocity field. 
% Now with the velocity field \(\bv_\gamma(x)\) either in \autoref{eq:biot_savart} or \autoref{eq:Rosenhead-Moor}, we can describe a time-evolution of the filaments by
%  \begin{align}
%      \partial_t{\gamma_i}(s,t)=\bv_\gamma(\gamma_i(s,t)).
% \end{align}
% Works such as Pinkall \& Weissman simulate filaments by discretizing this equation.
\end{example}

Now, we translate the dynamical system \eqref{eq:ExplicitDynamics} into an evolution equation for a time-dependent complex level function \(\psi\).
First, we note that \stt{the general form of the evolution for } \rev{the evolution of} \(\psi\) \stt{is an \emph{advection equation} due to} \rev{around the zeros is given as the \emph{transport equation} along some vector field, which is formally}
% \rev{the evolution of \(\psi\) around the zeros is given as the \emph{transport equation} along some vector field, which is formally}
 the following lemma.
\stt{As a remark the representation of the dynamics for }\(\sttEqn{\psi}\) \stt{can be encapsulated into a \emph{vector field}}
\begin{lemma}\label{lem:transport_equation}
    For any time-dependent complex level set function \(\psi\) with 
    %assumptions in \secref{sec:Regularity}, 
    the regularity assumptions in \secref{sec:RepresentationsForFilamentDynamics},
    there exists a neighborhood \(U\subset M\) of the zero set of \(\psi\) and a vector field \(\bv\colon U\rightarrow\RR^3\) such that
    \begin{align}
        \label{eq:AdvectionInU}
        {\partial\psi\over\partial t} + \bv\cdot\nabla\psi = 0\quad\text{in \(U\)}.
    \end{align}
\end{lemma}
\begin{proof}
    %Using \secref{sec:Regularity} 
    By the regularity assumptions in \secref{sec:RepresentationsForFilamentDynamics},
    there exists a neighborhood \(U\) of the zero set of \(\psi\) where the 3D-to-2D linear map \(d\psi\) is full-rank and thus surjective.  
    %Hence at every point \(\bx\in U\) there exists \(\bv_\bx\in\RR^3\) so that \(d\psi\vert_\bx(\bv_\bx)\) equals the prescribed value of \(-\nicefrac{\partial\psi}{\partial t}\vert_{\bx}\). 
    Hence at every point \(\bx\in U\) and for any value of \(\nicefrac{\partial\psi}{\partial t}\rev{(\bx)}\sttEqn{\vert_{\bx}}\), there exists \rev{a vector} \(\sttEqn{\bv\vert_{\bx}}\rev{\bv_{\bx}}\in\RR^3\) such that \(d\psi\vert_\bx\rev{(\bv_{\bx})}\sttEqn{\bv\vert_\bx} = -\nicefrac{\partial\psi}{\partial t}\rev{(\bx)}\sttEqn{\vert_\bx}\).
    With this construction, we obtain a vector field \(\bv\colon U\rightarrow\RR^3\) satisfying \(\nicefrac{\partial\psi}{\partial t} + \bv\cdot\nabla\psi=0\) in \(U\).
\end{proof}
Hence the representation of the dynamics for \(\psi\) can be encapsulated into a \emph{vector field}.
Observe that \stt{under the  advection equation}\sttEqn{\eqref{eq:AdvectionInU},}the zero level curve \(\gamma\) of \(\psi\) evolves by \(\nicefrac{\partial\gamma}{\partial t} = \bv\vert_\gamma\) \rev{under the transport equation \eqref{eq:AdvectionInU}, and that \(\operatorname{ker} (d\psi)\) at each point on \(\gamma\) is spanned by the tangent \(\gamma'\) as \(\gamma'\) lies on the tangent spaces of both \(\{\textup{Im} \psi =0 \}\) and  \(\{\textup{Re} \psi =0 \}\)}.  By matching these induced curve dynamics \rev{in Lemma \ref{lem:transport_equation}} with \eqref{eq:ExplicitDynamics} we conclude:
\begin{theorem}
    \label{thm:VelocityCondition}
    The zero level curve \(\gamma\) of \(\psi\) evolves according to \eqref{eq:ExplicitDynamics} if and only if \(\psi\) satisfies \eqref{eq:AdvectionInU} for some vector field \(\bv\) that agrees with the curve velocity at the curve:
    \begin{align}
        \label{eq:VelocityCondition}
        \bv(\gamma(s)) = \bV_\gamma(s) + f(s)\gamma'(s)
    \end{align}
    where \(f\gamma'\) is the tangent vector \(\gamma'\) multiplied by an arbitrary scalar function \(f\).
    The degrees of freedom of the dynamics for \(\psi\) are the degrees of freedom for choosing \(\bv\) with the condition \eqref{eq:VelocityCondition}.
\end{theorem}

Essentially, the only velocities that really matter for the evolution of the curves are the velocities located {\em on the zero level set}. Furthermore, the locations of the curves will not change if we slide them around their tangent direction (like spinning a circle around its axis of symmetry), so we only need to pin down their {\em normal and bi-normal components}. So we have a huge number of velocity variables (3 for each point in the 3D domain) with very few constraints (2 for each point on the 1D curves). This under-determined system gives us a {\em redundancy} in possible velocity fields for curve dynamics, which is largely unexplored by previous work. %\cCW{trying to make reader pay attention to why this matters, but don't want to bash previous work}. \cAC{Perfect}

%\cCW{Great structure here. I reworded to make it more concise.}\cAC{Perfect}
\iffalse
Let us apply \thmref{thm:VelocityCondition} to vortex filament dynamics (\cf\@ \exref{ex:BiotSavart}) in a straightforward and, as we will soon argue, naive way.
Since we already know that the velocity model \eqref{eq:Rosenhead-Moor} is the restriction of vector field \eqref{eq:RMAll} in the ambient 3D space, we can simply assign the dynamics for \(\psi\) as
\begin{align}
    \label{eq:NaiveAdvection}
    {\partial\over\partial t}\psi +\nabla_\bv\psi = 0,\quad\bv(\bx) = \bU^{\rm RM}_\gamma(\bx).
\end{align}
That is, the level set function, or more well-known in this context the Clebsch variable, is advected by the fluid velocity.
However, solving \eqref{eq:NaiveAdvection} is numerically unstable as the \(\bv=\bU^{\rm RM}_\gamma\) field is highly swirling around \(\gamma\) and an advection by \(\bv\) twists up \(\psi\) rapidly. 

Fortunately, according to \thmref{thm:VelocityCondition} there is a huge degrees of freedom for switching the dynamics of \(\psi\) without changing the effective dynamics on \(\gamma\). We will redesign the transport velocity \(\bv\) of \eqref{eq:NaiveAdvection} in the next section.
\fi
Let us apply \thmref{thm:VelocityCondition} to vortex filament dynamics (\cf\@ \exref{ex:BiotSavart}): Plugging in \eqref{eq:RMAll} for $\bv$ gives us the following dynamics for $\psi$:
\begin{align}
    \label{eq:NaiveAdvection}
    {\partial\over\partial t}\psi +\bv\cdot\nabla\psi = 0,\quad\bv(\bx) = \bU^{\rm RM}_\gamma(\bx).
\end{align}
This is the most straightforward way to do it: simply advect the level set $\psi$ in the exact same way as the rest of the fluid. However, we know that \(\bv=\bU^{\rm RM}_\gamma\) is an extremely sensitive function to deal with numerically --- it tends to infinity near $\gamma$, has unbounded derivatives, and rapidly changes direction in very tight swirls. Small errors in $\gamma$ inevitably create huge errors in velocity, making simulations unstable, as demonstrated in our accompanying video. Fortunately, according to \thmref{thm:VelocityCondition} we now know that there \rev{are} infinitely many velocity fields that will all theoretically give us the same filament motions; our mission in the next section is to swap out this unstable Biot-Savart velocity field for one that is much more numerically robust.

\section{Untwisted Clebsch variables and non-swirling dynamics}
\label{sec:untwisted}

%\cCW{The introduction paragraph here was redundant with the closing paragraph of the previous section, so I commented it out.}\cAC{Ok}
%In the previous section, we observe the degrees of freedom in the complex level set functions and their dynamical systems.  In this section, we describe a method to select a representative among the freedom. The choice aims at reducing the potential \emph{twisting} and \emph{swirling} in the variables that are difficult to capture under discretization.  

%In this section, we define our \textit{untwisted} level functions (Clebsch variables) and non-swirling dynamics for simulating deforming curves.
\subsection{Untwisted Clebsch variables}
\label{sec:UntwistedClebschVariables}
% \cAC{Begin new writing}

% Suppose we have a union of space curves \(\gamma\).  We construct a complex level set function \(\psi\colon M\rightarrow\CC\) that
% \cAC{I think we need to establish that the situation is ``if \(\gamma\) is given find a good \(\psi\).''}

Like the common codimension-1 real-valued level set methods, the implicit representation benefits from the regularity of the level set function.  There, a level set function is well-conditioned if the magnitude of the gradient is close to one.  For that reason, the level set function is typically initialized as the \emph{signed distance function}, and this property is typically maintained as the level set is evolved (called \emph{re-distancing}). 

For our codimension-2 complex level set representation, we shall also characterize a set of desirable qualities of the complex level set function \(\psi\colon M\rightarrow\CC\). Due to the higher codimension, the discussion involves the notion of \emph{twist} from the mathematical  \emph{ribbon theory}. Finally, we describe a concrete construction of \(\psi\) that will be used for initialization and re-distancing. 

\subsubsection{Conditioning of a complex level set function}
We want \(\psi\) to be continuous everywhere and non-zero outside the curves \(\gamma=\{\psi=0\}\). 
% \cCW{, to avoid accidentally creating new filaments from spurious zero-crossings}. \cSI{I think accidental craetion does not happen just by \(\psi \approx 0\)  if \(\psi\) is both continous and non-zero because real and imaginary level surfaces are created only when the imaginary and real parts of \(\psi\) jump over zero within a cell. Accidental creation happens only if \(\psi\) goes to nearly-zero below numerical precision. I guess this is why we had "bounded away from zero" and you added some explanation in the comment. But I'm not sure if we need this level of precise discussion about numerical issues here. Also I find talking about craetion of a curve from \(\psi\) here a bit tautological as we are talking about creation of \(\psi\) from curves \(\gamma\) here.
% So I just changed the original text from "bounded away from zero" to "non-zero".}
% \cCW{Sounds good! I just wanted to give the reader some explanation for {\em why} we want what this opening sentence says we want. I suppose it is implied strongly enough now.}
Near the 
%zero set
curves \(\gamma\), we want the differential \(d\psi\) to be well-conditioned:
if we only consider the function restricted to the plane spanned by the curve normal and bi-normal, \(d\psi\vert_{\gamma\bot}\) is close to an isometry; that is \(d\psi\vert_{\gamma^\bot}\colon\gamma^\bot\to\CC\cong\RR^2\) has singular values \(\approx(1,1)\).
Notice, however, that even when \(d\psi\vert_{\gamma^\bot}\) is well-behaved on each normal plane, the level set function \(\psi\) can still exhibit significant {\em shearing} if the complex phase varies significantly along the tangent direction.
%due to a large variation of its complex phase along the tangent.

This variation of complex phase along the curve's tangent can be seen more intuitively via the geometry of the surface \(S_{\psi}\) formed by the level sets of the positive real part of \(\psi\) (\(S_{\psi} = \{\Re(\psi)> 0, \Im(\psi)=0\} = \{\arg(\psi)= 0\}\)).  Note that the boundary of the surface \(S_{\psi}\) is \(\gamma\), as illustrated in \figref{fig:ribbon}. The tangential variation of the phase \(\arg(\psi)\) is embodied by the twist of \(S_{\psi}\) at its boundary \(\gamma\).

For the specific method presented in this paper, we take a simple construction of \(\psi\) that is known to have a smooth \(S_\psi\) with little twisting at \(\gamma\).  Readers who are only concerned about the method can skip ahead to \secref{sec:SolidAngleDistanceFunction}; for
researchers interested in extending this work, we will now lay out a few geometric and topological properties about the twist.

\subsubsection{The twist of a complex level set function}
To discuss the twist precisely, we consider the normal vector \(U_\psi\) of the curve \(\gamma\) that is tangent to the surface \(S_{\psi}\).
\begin{definition}[\(\psi\)-induced framing]
    Each complex level set function \(\psi\colon M\rightarrow\CC\) for \(\gamma\) gives rise to a normal vector field \(U_\psi\colon\gamma\to\RR^3\), \(|U_\psi|=1\), \(\langle \gamma',U_\psi\rangle = 0\) that points to the direction where \(\psi\) is real and positive.  Explicitly, at each point on \(\gamma\),
    \begin{align}
        U_\psi \coloneqq \frac{(d\psi\vert_{\gamma^\bot})^{-1}(1)}
        {|(d\psi\vert_{\gamma^\bot})^{-1}(1)|}.
    \end{align}
\end{definition} 
This normal vector field \(U_\psi\) makes \(\gamma\) a \emph{framed curve}.
\begin{definition}[Twist]
    The \emph{twist} of \(\psi\) at each point on \(\gamma\) is given by
    \begin{align}
        \omega \coloneqq U_\psi'\cdot(\gamma'\times U_\psi).
    \end{align}
    Here, \((\cdot)'\) is the derivative along curve's tangent relative to arclength.
\end{definition} 

When \(d\psi\vert_{\gamma^\bot}\colon \gamma^\bot\to\CC\) is conformal (which is true in our case \rev{introduced in Section \ref{sec:SolidAngleDistanceFunction}} since it is isometric), the twist directly relates to the derivative of \(\psi\) along the curve:
 \begin{align}
     \label{eq:TwistDPsi}
     \omega = -(d\theta)(\gamma')= -\Re\left({(d\psi)(\gamma')\over i\psi}\right),\quad\theta = \arg\psi,
 \end{align}
in a small neighborhood of \(\gamma\).

\begin{figure}[htbp]
    % \begin{minipage}[b]{1.0\linewidth}
    %      \centering
    %      \includegraphics
    %      [width=0.95\linewidth,tics=0,trim={0cm 4cm 0cm 4cm},clip]
    %      {{images/illustrations/level_surfaces_and_ribbons/Active_Render.0002.46}.png}
    %  \end{minipage}
    %  \begin{minipage}[b]{1.0\linewidth}
    %      \centering
    %      \includegraphics
    %      [width=0.95\linewidth,tics=0,trim={0cm 4cm 0cm 4cm},clip]
    %      {{images/illustrations/level_surfaces_and_ribbons/Active_Render.0002.38}.png}
    %  \end{minipage}
     \begin{minipage}[b]{1.0\linewidth}
         \centering
         \includegraphics
         [width=0.98\linewidth,tics=0,trim={6cm 7cm 6cm 6cm},clip]
         % {{images/illustrations/level_surfaces_and_ribbons/Active_Render.0002.41}.png}
         {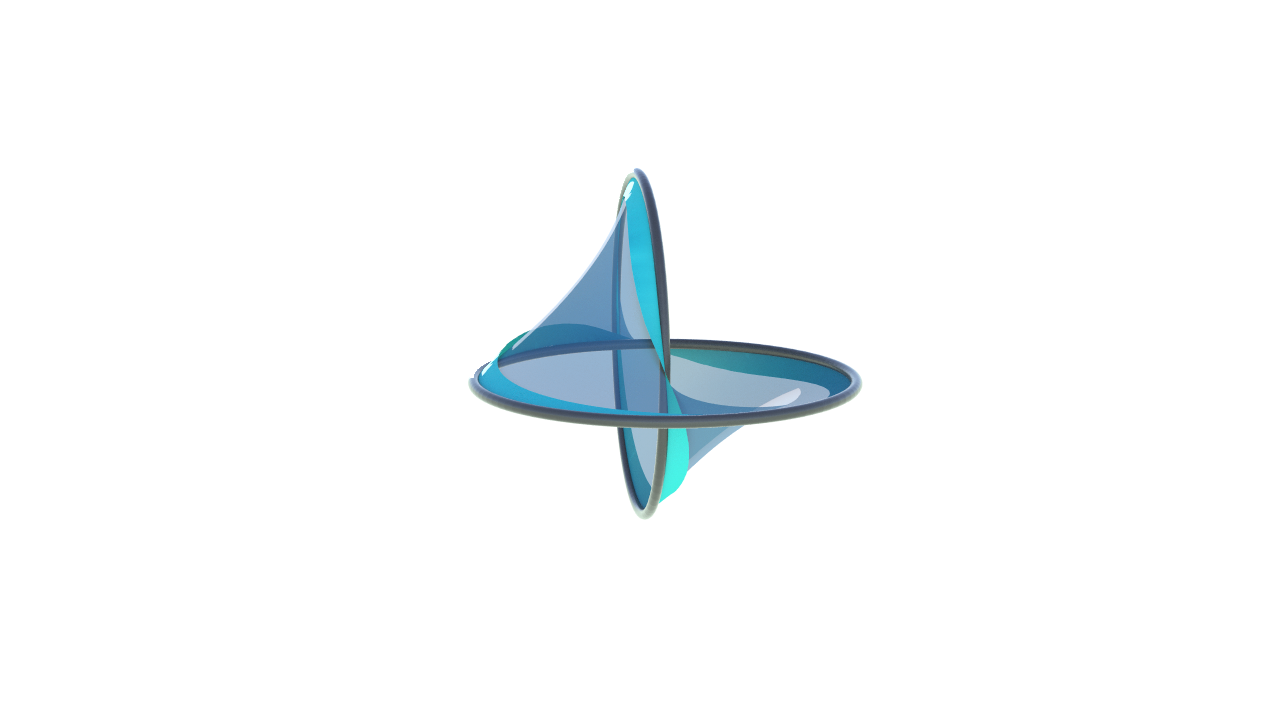}
        %  {{images/illustrations/level_surfaces_and_ribbons/Active_Render.0002.48_color_adjusted}.jpg}
     \end{minipage}
    \caption{The surface \(S_\psi\) (translucent blue) and the 
    %\cCW{twisted} 
    ribbon \(U_\psi\) (opaque cyan).
    %\cSI{I suggest not having "twisted" as this figure is actually our untwisted Clebsch variable.}
    \label{fig:ribbon}
    }
\end{figure}

Now, a natural consideration for designing \(\psi\) is to minimize the twist.  For example, one may try to construct \emph{Bishop's parallel frame} \cite{bergou2008elastic}, which has no twist.  However, one would find the construction \emph{impossible} for a closed curve and a union of closed curves in general.  This is due to the following theorem that the \emph{total twist}
\begin{align}
    \operatorname{Tw}(\psi)\coloneqq {1\over 2\pi}\oint_{\gamma}\omega
\end{align}
is fixed by the curve geometry \(\gamma\in\sF\) or the equivalent class of level functions \([\psi]\in\sG\) and is generally non-zero.

\begin{theorem}[Invariance of total twist]
    Suppose \(M\) is simply connected.  Let \([\psi]\in\sG\). Then any representative \(\psi\in[\psi]\) has the same total twist \(\operatorname{Tw}(\psi)\).
\end{theorem}
\begin{proof}
    In the special case that \(\gamma\) is a single connected curve in \(\RR^3\), the statement is the result of the C\u{a}lug\u{a}reanu Theorem: The total twist of the frame \(\operatorname{Tw}(\psi)\) and the total writhe of the curve \(\operatorname{Wr}(\gamma)\) must add up to the linking number between the two boundaries of the ribbon swept out by the frame \(U_\psi\).  Here, we note that any closed curve admits a Seifert surface i.e. a compact, connected, and oriented surface spanning the curve 
    \cite{
        Seifert1935seifertsurface,
        musasugi2008knottheory}.
    \rev{Without loss of generality, we can assume a constant phase-shift to \(\psi\) so that \(S_{\psi}\) is a Seifert surface of \(\gamma\).} 
     Since the ribbon \(U_\psi\) lies in a Seifert surface\(\sttEqn{S_\psi}\)\stt{ with }\(\sttEqn{\partial S_\psi=\gamma}\), the linking must be zero.  Therefore, \(\operatorname{Tw}(\psi) = -\operatorname{Wr}(\gamma)\), which only depends on \(\gamma\). See \figref{fig:ribbon} for a visualization of \(S_\psi\) and \(U_\psi\). For the definitions of quantities for curves such as writhe, twist, or linking numbers, we refer to 
    \cite{Arnold:1998:TMH}.
    
    In the general case, observe that under any change of representative \(\tilde\psi = \varphi\psi\) in \([\psi]\) with a nowhere-vanishing \(\varphi\colon M\rightarrow\CC\setminus \{0\}\), the twist transforms according to \(\tilde\omega = \omega - d_{\gamma'}(\arg(\varphi))\).  Since \(\varphi\) has no zero in a simply connected \(M\), the angle \(\arg(\varphi)\) can be defined as a real-valued function globally over \(M\) rather than a \(2\pi\)-ambiguous angle-valued function. In particular, Stokes' theorem applies and the integral along the closed curve vanishes: \(\oint_{\gamma}d(\arg(\varphi)) = 0\). So \(\oint_\gamma\tilde\omega = \oint_\gamma\omega\).
\end{proof}
% \cAC{I'm still struggling to write this part... What's the significance of the theorem?}

Note that even though the total twist is fixed, strong {\em local} twist can still be present.  
 
\subsubsection{Solid-angle distance function}
\label{sec:SolidAngleDistanceFunction}

\begin{figure}
    \centering
    \hspace{20pt}
    \setlength\unitlength{1pt}
    \begin{picture}(110,110)
        \put(0,0){\includegraphics[width=0.6\columnwidth]{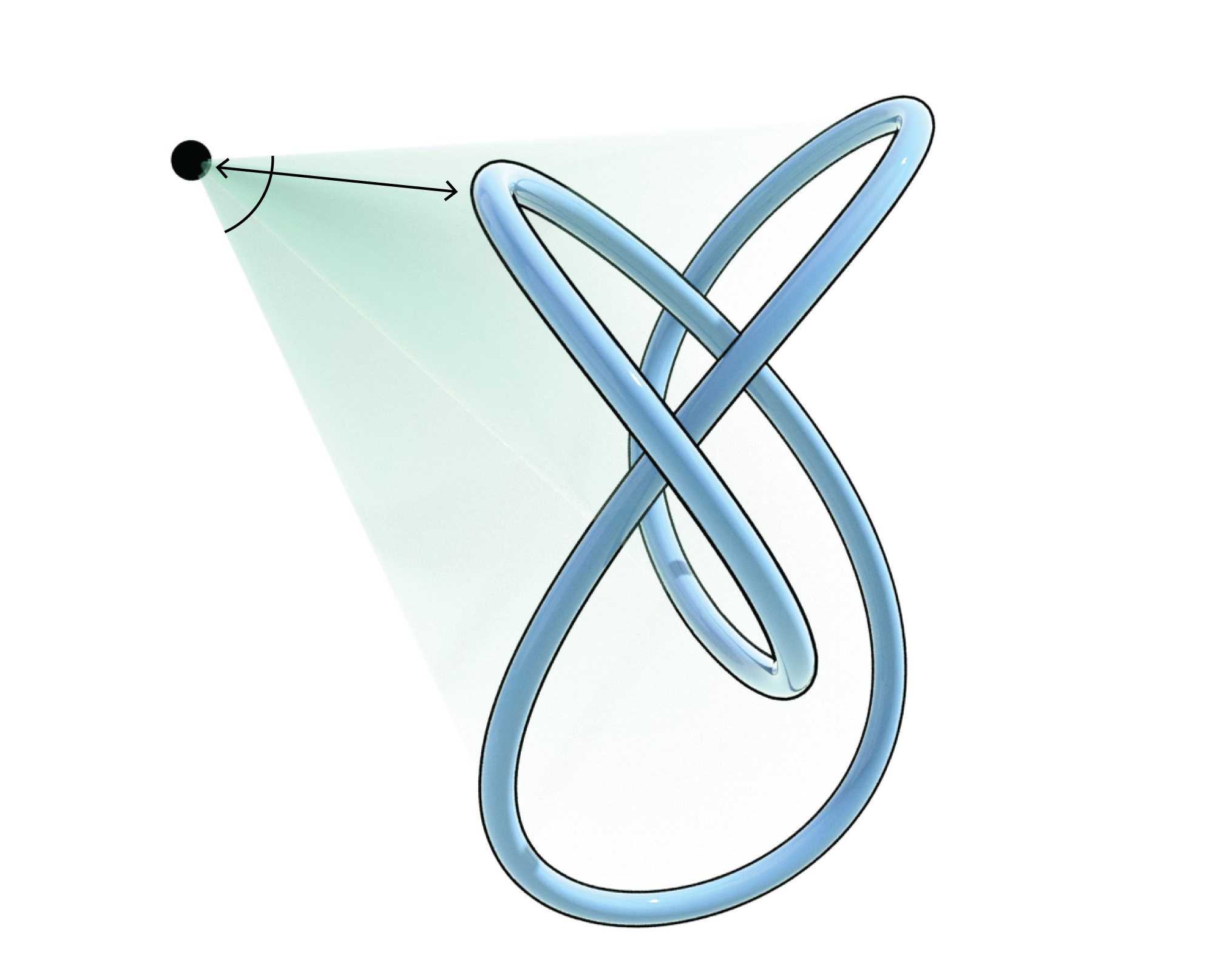}}
        % \put(0,0){\includegraphics[width=0.6\columnwidth]{images/illustrations/projected_curve/SolidAngle.jpg}}
        \put(105,65){\small\(\gamma\)}
        \put(15,100){\small\(\bx\)}
        \put(-10,80){\small\(\operatorname{SolidAngle}(\gamma;\bx)\)}
        \put(25,105){\small\(\operatorname{dist(\gamma,\bx)}\)}
        \put(-50,40){\small\(\psi(\bx) = \operatorname{dist}(\gamma,\bx)e^{{i\over 2}\operatorname{SolidAngle}(\gamma;\bx)}\)}
        %\graphpaper(0,0)(110,110)
    \end{picture}
    \caption{The solid-angle distance function \(\psi\) for a space curve \(\gamma\) is constructed by the distance and the angle subtended by the curve.}
    \label{fig:projection_onto_sphere}
\end{figure}
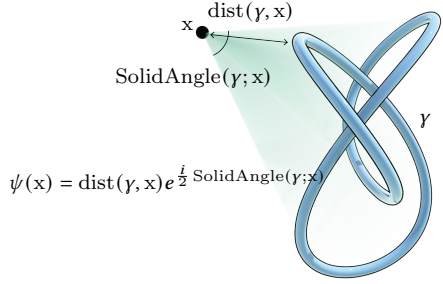

As explained above, we desire a complex level set function $\psi$ that is numerically well-conditioned: it should be nearly isometric near the curve $\gamma$ and non-zero outside $\gamma$. In order to isolate these properties and enforce them explicitly, we model $\psi$ with a complex wave function:
\begin{align}
    \psi(\bx) = r(\bx) e^{i\theta(\bx)}.
\end{align}
Similar to \emph{signed distance functions} for the codimension-1 level set functions, we set
\begin{align}
    |\psi(\bx)| = r(\bx) \coloneqq \operatorname{dist}(\gamma,\bx),\quad\bx\in M\subset\RR^3.
\end{align}
This setup ensures that the value \(\psi(\bx)\) is non-zero for \(\bx\notin\gamma\).
What remains is a choice for the complex phase \(\theta\colon M\setminus\gamma\to\SS^1 = \RR_{\text{mod }2\pi}\),
%
%\begin{align}
%    \psi(\bx) = r(\bx) e^{i\theta(\bx)}.
%\end{align}
%We set the complex phase as 
which we set to
the \emph{half solid-angle subtended by \(\gamma\)}:
\begin{align}
    \theta(\bx) \coloneqq {1\over 2}\operatorname{SolidAngle}(\gamma;\bx)\mod 2\pi
\end{align}
where \(\operatorname{SolidAngle}(\gamma;\bx)\in\RR_{\text{mod $4\pi$}}\) is \rev{a dimensionless quantity given by} the signed spherical area \footnote{\rev{Our definition of the solid angle is modulated by \(4\pi\) as the spherical polygon of projected curves may cover the sphere multiple times.}} enclosed by the projection of \(\gamma\) on the unit sphere centered at \(\bx\) \rev{i.e. \(\operatorname{Proj}\gamma_{\bx}(s):=(\gamma(s)-\bx)/|\gamma(s)-\bx| \)} \sttEqn{(\figref{fig:projection_onto_sphere})}.
We call this construct of \(\psi\) the \emph{solid-angle distance function} \rev{(\figref{fig:projection_onto_sphere})}.

\begin{figure}[htbp]
    \centering

    \includegraphics[width=0.95\columnwidth]{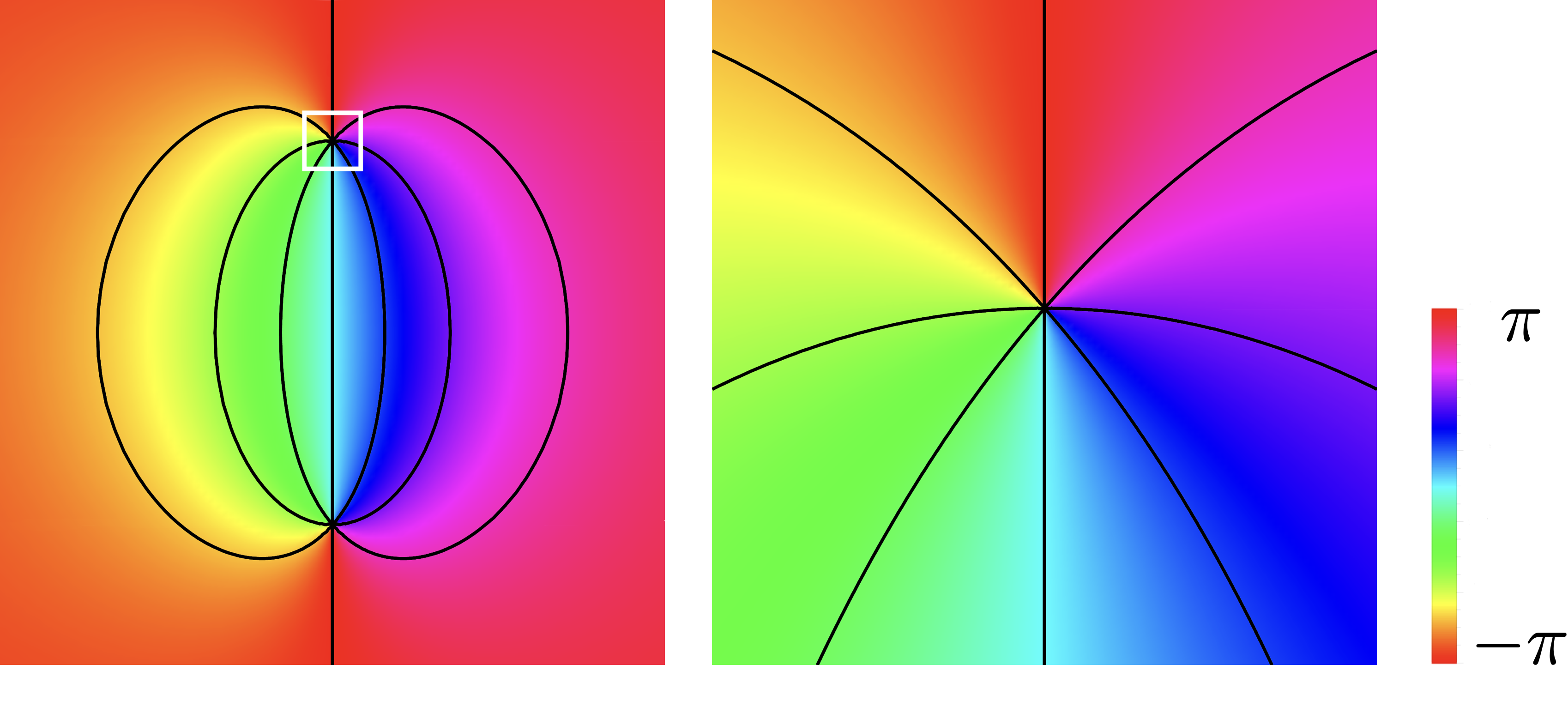}
    \caption{Plotting $\theta$ on a 2D plane which intersects a circular vortex ring at two points (left). The color indicates the value of the $\theta$, and the black lines are its level curves. The curves meet where the filament intersects the plane. Zooming into the white box (right) shows evenly-spaced curves closer to the filament, where $e^{i\theta}$ resembles the complex plane $\CC$.}
    \label{fig:solid_angle}
\end{figure}

The solid-angle distance function meets our desired conditions for the codimension-2 level set representation.  On each normal plane \(\theta\vert_{\gamma^\bot}\) is asymptotically the 2D angle function about the zero \(\gamma\), so that \(d\psi\vert_{\gamma^\bot}\colon \gamma^\bot\to\CC\) is close to an isometry.  
%\cCW{This point would be clearer with a better illustration (like a 2D visualization of $\theta$ for a simple ring, with samples at $\theta=0, \theta=\pi/2, \theta=\pi$, and $\theta=3\pi/2$ with visualizations of the filament projected onto the unit sphere.). Otherwise there is a lot for the reader to figure out for themselves, or just blindly trust. The visualizations we do have are beautiful but too complicated to ease the reader into understanding.} 
\stt{Moveover}\rev{Moreover}, as studied by \cite{Binysh_2018solidangle}, 
% \cAC{cite \url{https://iopscience.iop.org/article/10.1088/1751-8121/aad8c6/pdf}}
 the surface \(S_\psi=\{\theta=0\}\) features little twist at \(\gamma\). \figref{fig:solid_angle} illustrates this concept.
 
\rev{We note that our choice of the Clebsch variables \(\psi\) is an instance of many reasonable ones that exploit degrees of freedom in curve representations rather than the optimal one for a speficic dynamics such as vortex filaments.
Nevertheless, our \(\psi\) has 
%good 
a number of desirable
properties, which we list in Appendix ~\ref{appendix:property_untwisted_Clebsch}.
}

\subsection{Non-swirling dynamics}
\label{sec:NonSwirlingDynamics}
In \secref{sec:UntwistedClebschVariables}, we leveraged the degrees of freedom in the complex level set function \(\psi\) to design a sufficiently regular implicit representation.  
%Here, we exploit the similar freedom to construct evolution equations that show more regular dynamics.
Here, we exploit similar degrees of freedom to construct evolution equations that produce theoretically equivalent dynamics but are more numerically robust. 
%\cCW{The equations are only more {\em numerically} regular, right? Otherwise they should be exactly the same if solved exactly. I changed the previous sentence to point this out more clearly.}\cSI{Yes, this it correct.}

\begin{figure}[htbp]
     \begin{minipage}[b]{1.0\linewidth}
          \centering
          \includegraphics
          [width=0.95\linewidth,tics=0,trim={0cm 4cm 0cm 4cm},clip]
          {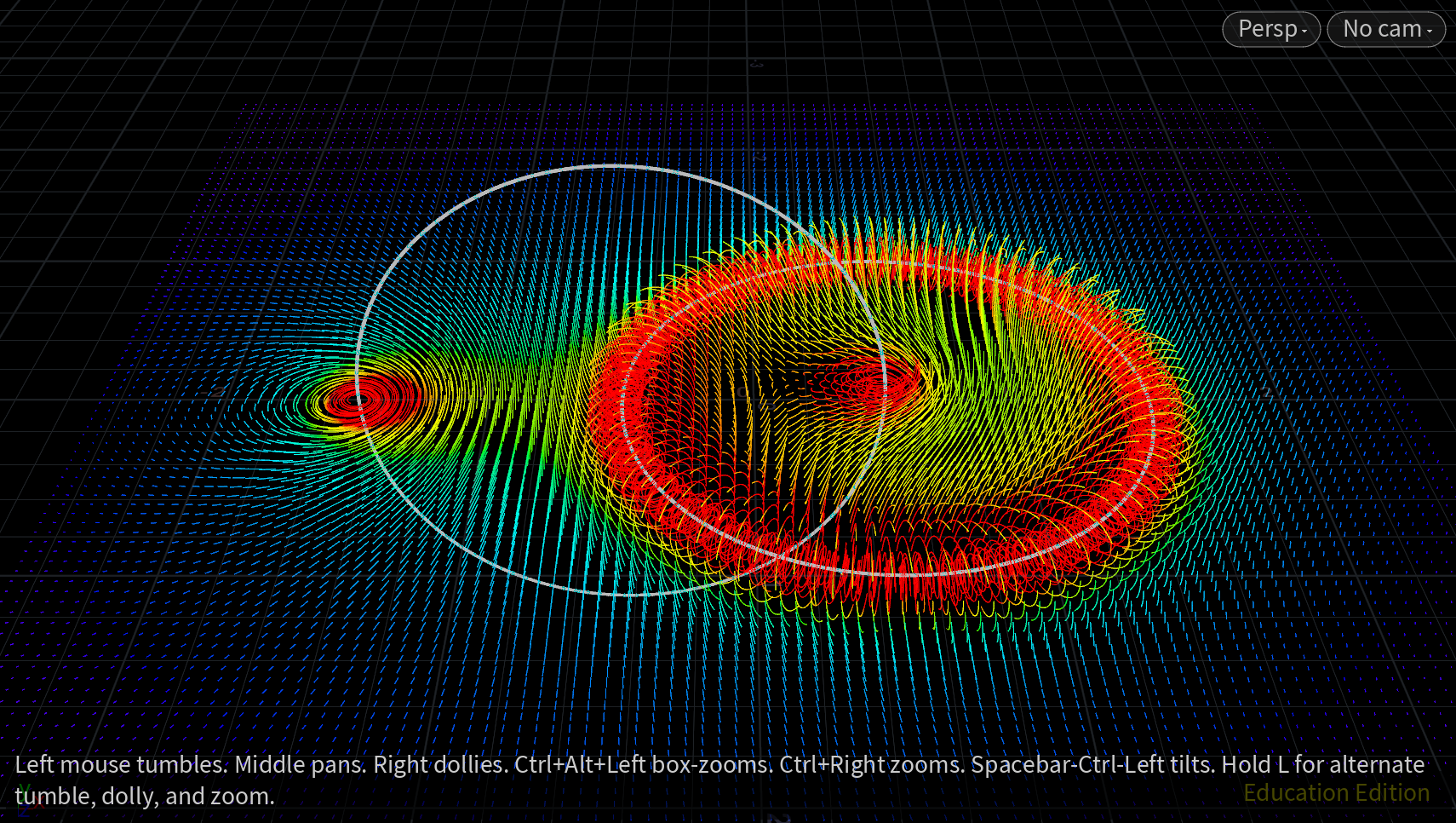}
        %   {{images/illustrations/vector_fields/bs_t2}.jpg}
      \end{minipage}
      \begin{minipage}[b]{1.0\linewidth}
          \centering
          \includegraphics
          [width=0.95\linewidth,tics=0,trim={0cm 4cm 0cm 4cm},clip]
          {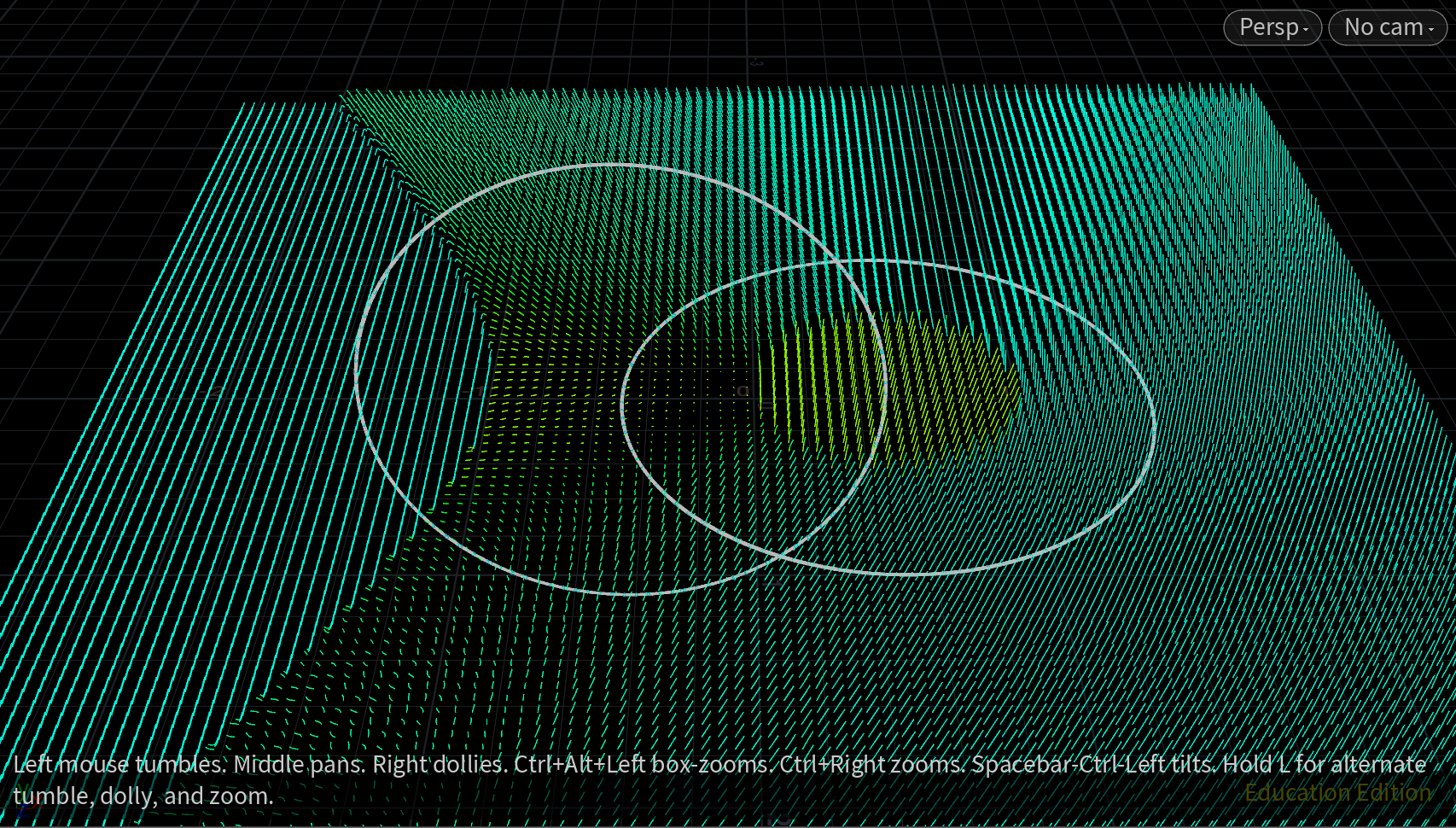}
        %   {{images/illustrations/vector_fields/ne_t2}.jpg}
      \end{minipage}
      \begin{minipage}[b]{1.0\linewidth}
          \centering
          \includegraphics
          [width=0.95\linewidth,tics=0,trim={0cm 4cm 0cm 4cm},clip]
          {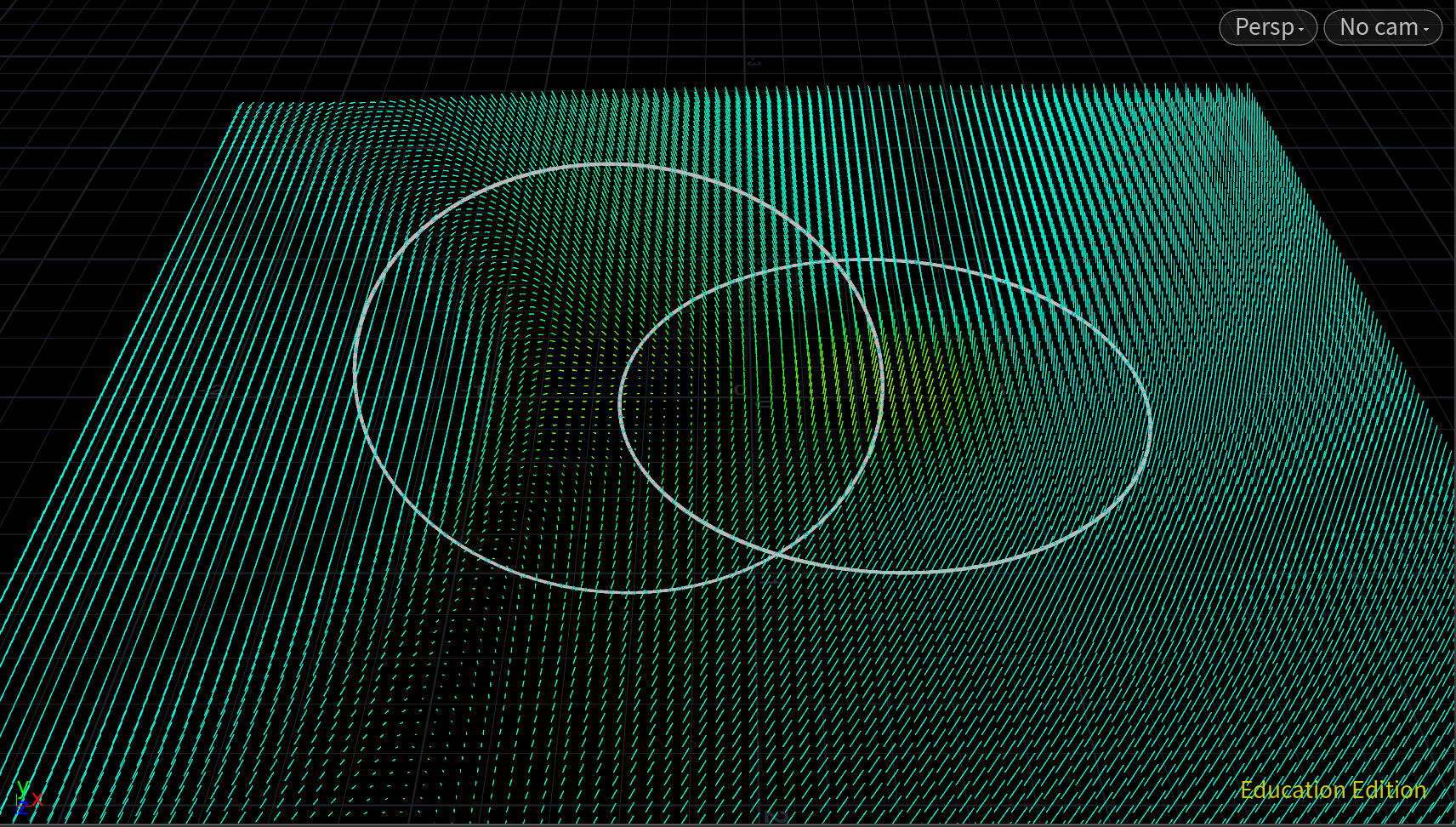}
        %   {{images/illustrations/vector_fields/good_t2}.jpg}
      \end{minipage}
     \caption{Different vector fields for two linked rings. The original Rosenhead--Moore model (top), the nearest point velocity field (middle), and a smooth  weighted average field (bottom). While these three velocity fields coincide on the filaments, they differ significantly outside the filaments.
     %\cCW{moved this figure closer to where it appears in text.}
     }
     \label{fig:different_velocity_fields}
 \end{figure}

Many dynamical systems for curves already come with a known physical evolution equation.  For example, the vortex filament dynamics can be simulated with \eqref{eq:NaiveAdvection}, \ie\@ by advecting \(\psi\) using the Biot--Savart or the Rosenhead--Moore flow (\exref{ex:BiotSavart}).   Hence, redesigning the equation may seem unnecessary.  However, when it comes to numerically advancing the variables, the highly oscillatory or discontinuous nature of the Biot--Savart and the Rosenhead--Moore flows near the vortex core (as illustrated in \figref{fig:different_velocity_fields}) can cause significant interpolation error.  We point out that these errors are avoidable by redesigning the flow of the advection.

%As we have learned in the previous section, we are allowed to modify the dynamics of Clebsch variable \(\psi\). This can be done by choosing a time-dependent artificial velocity field for the transport equation (\autoref{eq:AdvectionInU}) within the degree of freedom given by \autoref{eq:VelocityCondition}.  In numerically advecting \(\psi\), it is important to choose a regular enough velocity field so it does not stir up \(\psi\). For example, when dealing with the Biot-Savart field \(\bv_\gamma\), we can exploit the fact that \(\bv_\gamma\) limited on the \(\gamma\) is continuous and nearly parallel to the binormals of \(\gamma\) although it is strongly swirling near  \(\gamma\). In fact, we can cancel the swirling motion by extrapolating the filament velocity \( \bv_\gamma | _\gamma\) to the domain outside \(\gamma\).

% A straightforward extrapolation is,
% \begin{align}
%      \bv(\bx)=  \bv_\gamma(\gamma_\bx)
% \end{align}
% where \(\gamma_\bx\) is one of the nearest points of \(\text{Image}(\gamma)\) from \(\bx\). This simple velocity field can, however, be discontinuous and irregular when the filament configuration is complex.

We consider dynamical systems as discussed in \secref{sec:RepresentationsForFilamentDynamics}.  
Suppose the evolution of the curve is given by \({\partial\gamma\over\partial t} = \bV_{\gamma}\) where \(\bV_{\gamma}\) is the velocity field defined on the curve.  The evolution for \(\psi\) must be an advection by an extension \(\bv\) of the velocity field \(\bV_\gamma\) in a neighborhood of the curve  (\cf\@ \eqref{eq:AdvectionInU}).   A straightforward construction  of \(\bv\) is a constant extrapolation.  That is, \(\bv(\bx)\) is set to \(\bV_\gamma\) at the closest point on \(\gamma\) from \(\bx\). This extrapolation is, however, singular where closest points are not unique as in the middle row of \autoref{fig:different_velocity_fields}.

% In practice, the curve is polygonal and the closest point extrapolation can be discontinuous.  
% \cSI{This happens in the continuous setting too. For example, the center of a circle is closest to any point, so the velocity field is singular.}
To gain continuity without changing the velocity on the filaments, we 
smooth away these singularities by taking
%take 
the weighted average of the filament velocity as 
\begin{align}\label{eq:non-swirling_velocity}
     \bv(\bx)\coloneqq \frac{1}{\cN(\bx)}\oint_\gamma \bV_\gamma(\gamma(s)) w(\bx,\gamma(s))\,  ds
\end{align}
%gives stable computation. 
Here, \(\cN(\bx)\) is the normalization factor
\begin{align} \label{eq:normalization_factor}
     \cN(\bx) = \oint_\gamma w(\bx,\gamma(s)) ds,
\end{align}
and \(w\) is some weight function 
that applies less smoothing as it gets closer to the filament, i.e. 
%such that  
{\(w(\bx, \gamma(s)) / \cN(\bx) \rightarrow \delta(\gamma^{-1}(\bx)-s)\)} as  {\( \textup{dist}(\bx,\gamma) \rightarrow 0\)}.
% \rev{where the inverse \(\gamma^{-1}\) is well-defined when \(\bx\) is on \(\gamma\)}
% \cSI{Honestly I think this explanation is too detailed and obvious, so we can remove it.}. \cCW{Exactly how much do you want to remove? This smoothin is our own heuristic, not a natural mathematical consequence of implicit curves, so nobody can re-derive it if we accidentally don't provide enough information; we should be sure to have enough info to make it reproducible.}
% \cSI{By "this explanation", I meant the part I added for the revision "where the inverse ..." in response to a reviewer's comment. }
% \cCW{I agree. Just stating that it is well-defined doesn't add anything; I think we should either remove it or add enough detail to justify precisely how it is well-defined. Cleaner to just remove it.}
For example, we observed that a Gaussian function with distance-dependent variance works stably:
\begin{align} \label{eq:weight_function}
    w(\bx,s)=\exp\left(-\frac{|\bx-\gamma(s)|^2}{ \sigma^2 \textup{dist}(\bx,\gamma)}\right),\quad\sigma\text{ some constant}.
\end{align}

To accelerate computation for \(\bv\), we can further multiply the integrand of \autoref{eq:non-swirling_velocity} by a smooth cutoff function which equals to \(1\) near \(\gamma\) and \(0\) far away from \(\gamma\). Then \(\bv\) is non-vanishing only near \(\gamma\).  By applying this smooth cutoff, we only need to evaluate \(\bv\) in a narrow band close to the filament.

\rev{Note that \(\bv\) is in general not divergence-free. The velocitiy near the curve is determined according to the curve velocities so that the motion of the zeros of \(\psi\) emulates the motion of the curves. Imposing an additional constraint like incompressibility to the velocitity field may trade off the fidelity to the original curve motion or the numerical smoothness of the surrounding vector field.}

%% file: algorithms.tex
%!TEX root = VortexUntwisted_paper.tex
\section{Algorithms}
\label{sec:algorithms}
In this section, we describe an algorithm for simulating filament dynamics.

Throughout the simulation, we maintain a complex level set function \(\psi\). The main algorithm computes the transport equation of \(\psi\) along a velocity field \(\bv\) in a neighborhood of the zeros of \(\psi\).  This main algorithm is accompanied by a few subroutines for evaluating the velocity and redistancing: one subroutine extracts the zero set \(\gamma\) of \(\psi\); another subroutine constructs the solid-angle distance function \(\psi\) from \(\gamma\) (\secref{sec:SolidAngleDistanceFunction}); a third subroutine evaluates the filament motion \(\bV_\gamma\) using \(\gamma\); and the last subroutine extends \(\bV_\gamma\) to a velocity field \(\bv\) in a neighborhood of \(\gamma\) (\secref{sec:NonSwirlingDynamics}).

%We store quantities such as a level function \(\psi\) and velocity \(\bv\) on 3D lattice and descritize the curves \(\gamma\) as oriented collections of line segments. 

We store the level set function \(\psi\) and the velocity \(\bv\) on a 3D lattice and discretize the curves \(\gamma\) as oriented collections of line segments. 

%The time integration is performed by computation of the transport equation, that is, advecting the level function \(\psi\) along the velocity field \(\bv\) as given in \autoref{eq:non-swirling_velocity}. As such, we evaluate \(\bv\) from \(\gamma= \{\psi=0\}\) every time \(\gamma\) is updated. We also  ensure that \(\psi\) is untwisted by setting \(\psi\) as in \autoref{eq:untwisted_Clebsch_variable}. 

\begin{algorithm}[H]
     \caption{The main time integration}
     \label{alg:time_integration}
     \begin{algorithmic}[1]
     \Require{Initial filament \(\gamma\in\sF\); %timestep \(\Deltait t>0\);
     }
     \State{\(\psi\gets\) construct \(\psi\) from \(\gamma\);}
     \Comment{\secref{sec:ComputePhase}} 
     %\For{\({\rm time} = \Deltait t, 2\Deltait t, 3\Deltait t,\ldots\) }
     \While{simulating}
     \State{\(\bv\gets\) evaluate \(\bV_\gamma\) and extend it on grids near \(\gamma\);}
     \State{\(\psi\gets\) advect \(\psi\) along \(\bv\); %for timespan \(\Deltait t\);
     } 
     \Comment{\secref{sec:Advection}}
     %\State{Reconstruct \(\gamma\) from \(\lbrace \psi=0 \rbrace\)}
     \State{\(\gamma\gets\) extract the zero set of \(\psi\);}
      \Comment{\algref{alg:construct_curve}}
     \State{\(\psi\gets\) construct \(\psi\) from \(\gamma\);}
     \Comment{redistance; \secref{sec:ComputePhase}} 
     %\State{Update \(\psi=\textup{dist}(x,\gamma) e^{i\theta (x)}\)  \par 
     %\hspace{-10pt} with new \(\textup{dist}(x,\gamma)\) and \(\theta(x)\)} \Comment{\algref{alg:compute_phase}} 
     \EndWhile
     \end{algorithmic}
\end{algorithm}

\begin{algorithm}[htb]
     \caption{Extract the zero curve \(\gamma\) from \(\psi\)}
     \label{alg:construct_curve}
     \begin{algorithmic}[1]
     \ForEach{cell \(c\)}
          \For{each face \(f\) in \(c\)}
          \State{Compute incidence \(n_f\in\{-1,0,1\}\);}  \Comment{Eq.~\eqref{eq:ArgumentPrinciple}}
               \If {\(n_f=\pm 1\)}
                    \State{Find \(\bp_f^\pm = \psi^{-1}(0)\in\RR^3\)  in \(f\);}\Comment{bilinear interp.}
               \EndIf
          
          \EndFor
          \State{Connect \(\bp_f^-\) and \( \bp_g^+ \) of some faces \(f, g\) in \(c\);}
     \EndFor     
     \end{algorithmic}
\end{algorithm}

%\cCW{By setting time = integer multiples of $\Delta t$ in Algorithm 1, we imply this method cannot be adaptively time-stepped. I think this is an unnecessary constraint and think we should be more generic. Is my proposed alteration ok?}
%\cSI{Yes, I agree. So I removed \(\Delta t\) from the algorithm table.}

\subsection{Details of the main algorithm}

\subsubsection{Advection}
\label{sec:Advection}
To advect \(\psi\) with a given flow \(\bv\), one can adopt any Eulerian advection scheme.  In our implementation, we use the modified MacCormack method \cite{Selle:2008:MCM} with 4th order Runge--Kutta back-tracing.

%To advect \(\psi\) with a given flow \(\bv\), we can basically use any advection scheme. In our implementation for the examples in this paper, we chose the MacCormack with Runge-Kutta 4 both for forward and backward advection.

\subsubsection{Construction of  \(\gamma\) from \(\psi\)}
After updating \(\psi\), we need to %construct new
update the
filaments \(\gamma\) by extracting \(\{\psi=0\}\). 
We summarize this subroutine in \algref{alg:construct_curve}, which is adopted from \cite{weissmann2014smokerings}.
% \cAC{cite smoke ring from smoke}.
% We do so as in  \algref{alg:construct_curve}. 
In our setting, 
each  vertex of \(\gamma\) lives on a face \(f\) of 
the volumetric grid.
%a volumetric cell \(c\). 
We first evaluate for each face \(f\) the \(\{-1,0,+1\}\)-valued \emph{signed intersection} \(n_f\) with the zero curve of \(\psi\) using the \emph{argument principle}: If the vertices of a face \(f\) are \(i,j,k,\ell\) in an oriented order, then
\begin{align}
    \label{eq:ArgumentPrinciple}
    n_f = {1\over 2\pi}\left(\arg(\tfrac{\psi_j}{\psi_i}) + \arg(\tfrac{\psi_k}{\psi_j}) + \arg(\tfrac{\psi_\ell}{\psi_k}) + \arg(\tfrac{\psi_i}{\psi_\ell})\right)
\end{align}
using the principal branch \(-\pi<\arg(\cdot)\leq\pi\).
% \cAC{[the original writing \(\omega_f=\sum_{i=\{0,1,2,3\}}^{\textup{mod }4} \theta_f^{i+1} - \theta_f^i\) can be problematic.  The branching is subtle.]}
Geometrically, \eqref{eq:ArgumentPrinciple} describes how many times the quadrilateral \(\psi_i,\psi_j,\psi_k,\psi_\ell\in\CC\) winds around the origin.
For each face where \(n_f\neq 0\) we evaluate the more precise location of the zero using a bilinear interpolation. That is, we regard \(f=[0,1]^2\) by scaling and \(\psi\) is bilinearly interpolated as
\begin{align}
     \psi_f(x,y)
     := & (1-x)(1-y) \psi(0,0) + x(1-y)\psi(1,0)\nonumber\\ 
     &+(1-x)y\psi(0,1)+xy \psi(1,1),
\end{align}
and  \(\psi_f:f\rightarrow \CC\) has the inverse when \(n_f=\pm 1\).
The location \(\psi_f ^{-1}(0)\)  is a vertex \(\bp_f\) of the curve \(\gamma\).
Finally, we build the edges of \(\gamma\) by running over the grid cells where we connect the pairs of zeros on the face with a consistent orientation. 
%\cCW{We are not discussing potentially ambiguous cases with 4 or 6 intersections per cell. Do we agree that this is too much low-level detail to be included here? Or is it not a problem at all?} \cSI{I think it's good to discuss it here. And actually I originally had an argument about it, so I added it back.}
%By a geometric consideration, it is easy to see that 
Each cube \(c\) may have up to two pairs of zeros with positive and negative  \(n_f\). A cube with two pairs of vertices has an ambiguity similarly to the marching cube algorithm \cite{Lorensen1987marchingcubes}. We resolve this ambiguity by connecting vertices arbitrarily in a way that preserves curve orientation.
%Within this work, w
We have not investigated higher order algorithms 
%that may better handle such an ambiguity.
to connect curves more accurately at the sub-grid scale.
% \cAC{Maybe include the bilinear interpolation formula.}
%\cCW{Please check my modifications.} \cSI{Looks great! I removed a bit of redundancy.}

%To accelerate computation, we first evaluate local vorticity \(w_f\) on each face \(f\). Note that \(w_f\) is zero or \(\pm 2\pi\). When it is \(\pm 2\pi\), we further evaluate the precise location of \(p_f^{\pm}:=\{\psi=0\}\cap f\). We finally make an edge by connecting vertices \(p_f^+\) and \(p_f^-\) within cube \(c\). 

% \begin{algorithm}[htb]
%      \caption{Construct \(\gamma\) as \(\{ \psi=0\}\)}
%      \label{alg:construct_curve}
%      \begin{algorithmic}[1]
%      \For{each cube \(c\)}
%           \For{each face \(f\) in \(c\)}
%           \State{Compute local vorticity \(\omega_f=\sum_{i=\{0,1,2,3\}}^{\textup{mod }4} \theta_f^{i+1} - \theta_f^i \)}  %\Comment{this is a comment}
%                \If {\(\omega_f=\pm 2\pi\)}
%                     \State{Find \(p_f^\pm = \psi^{-1}(0)\)  in \(f\)}
%                \EndIf
%
%           \EndFor
%           \State{Connect \(p_f^-\) and \( p_g^+ \) of some faces \(f, g\) in \(c\)}
%      \EndFor
%
%
%      \end{algorithmic}
% \end{algorithm}

\subsubsection{Construction of \(\psi\) from \(\gamma\)}
\label{sec:ComputePhase}
Given \(\gamma\), we construct \(\psi\) as described in \secref{sec:SolidAngleDistanceFunction}.  For each grid point \(\bx\) near \(\gamma\), we evaluate \(|\psi(\bx)|={\rm dist}(\gamma,\bx)\) as the distance to the closest polygon edge. 
We evaluate \(\arg(\psi(\bx)) = \theta(\bx) = {1\over 2}{\rm SolidAngle}(\gamma;\bx)\) by computing the signed area of a spherical polygon with vertices \(\{p_i\}_{i=1}^N\) that are the discrete points of \(\gamma\) projected onto the unit sphere centered at \(x\). To ensure differentiability and the local isometry properties of \(d\psi\vert_{\gamma\bot}\), the sign of $\theta$ must depend on curve orientation: we compute the {\em signed} area by introducing a pole \(Z\colon=(1,0,0)\):
\begin{align}
     {\rm SignedArea}\left( \{p_i\}_{i=1}^N\right)=\sum_{i=1}^{N} {\rm sign}\left((p_i \times p_{i+1})\cdot Z \right) {\rm Area}(p_i,p_{i+1},Z)
\end{align}
where the {\em unsigned} area of each spherical triangle \(\{q_1,q_2,q_3\}\) is computed using a standard area formula \cite{Bevis1987ComputingTA,arvo1995applications}, 
%\cCW{I prefer to cite Arvo instead of Wang \& Ramamoorthi here, since Arvo is ac classic reference, W\&R cite Arvo for the solid angle, and W\&R seemed to copy the formula incorreclty. Ok?}
% \cSI{Yes, sounds good.}
\begin{align}
     {\rm Area}(q_1,q_2,q_3)= -\pi+\sum_{i=1}^3 
     \arccos\left(
          \frac{
          \left(q_{i-1} \times q_{i}\right) \cdot\left(q_{i} \times q_{i+1}\right)} {\left\|q_{i-1} \times q_{i}\right\|\left\|q_{i} \times q_{i+1}\right\|}\right).
\end{align} 
These equations assume cyclic vertex 
indexing,
%indexing of $p$ and $q$, 
so ${p_{N+1}=p_1}$, ${q_{3+1}=q_1}$, and ${q_{1-1}=q_3}$.
% \cCW{Added this bit to make sure $q_{i-1}$ and $q{i+1}$ are defined well enough for a reader to figure it out.}\cSI{Looks good. I just removed \(p0=p_N\) as the signed area equation has only \(p_i\) and \(p_{i+1}\) so \(p_0\) does not show up. Also I fixed a typo and changed \(q_{0}\) to \(q_{1-1}\) so that it is consistent with your \(q_{3+1}\).}

\subsubsection{Extending velocity to grid}
% \cAC{Just a short paragraph reprising sec 4.2.}
In order to advect \(\psi\), we need to extend the velocity field defined on \(\gamma\) to \stt{to the entire domain} \rev{the grid points near the curves}. To produce non-swirling dynamics, we used a smooth average field (Equation \ref{eq:non-swirling_velocity}, \ref{eq:normalization_factor}, and \ref{eq:weight_function}) for the examples in this paper, unless otherwise explained. \rev{In \autoref{eq:weight_function}, we observed that \(\sigma\) ranging from \(0.1\) to \(10\) times the grid size works stably without smoothing out detailed dynamics.}

%% file: applications.tex
\section{Applications}
\label{sec:applications}

%\cCW{{\bf Simulations that would be nice to add} (in order of decreasing priority):
%\begin{itemize}
%   \item a simulation (or static image) showing what is the effect of changing the grid resolution
%   \item a result comparing to ground truth (not just other methods). Maybe a single analytical smoke ring? Or is there an analytical solution to leapfrogging rings? We can measure errors in the velocity field, ring position, kinetic energy, etc.
%\end{itemize}
%}
%\cCW{None of the above are paper killers if we don't have them, but they would make the paper stronger.}
%
%\cCW{{\bf Data that needs to be added to the paper}
%\begin{itemize}
%   \item Total simulation run time for each sim (including W\&P): runtime per frame, runtime per timestep, runtime per simulation
%   \item computational cost breakdown for the steps in Algorithm 1. Especially: what is the slowest part?
%   \item grid resolutions
%   \item memory cost (not necessary but nice to have to put grid resolution in context)
%   \item timestep size relative to W\&P if relevant (otherwise just say it's the same)
%   \item initial conditions (maybe just say they are included in a script that we will release with the code)
%   \item simulation time vs render time (approximate) (nice to have, not necessary)
%\end{itemize}
%}

This section discusses applications of our approach, specifically applied to vortex filament dynamics.
We implemented our algorithms on Houdini 18.5.759 and ran all simulations on a MacBook Pro (13-inch, 2020) with a 2.3 GHz Quad-Core Intel Core i7 processor. \stt{We plan to release our code upon publication.} \rev{For an example implementation, see}
\href{https://github.com/sdsgisd/ImplicitVortexFilaments}{https:\slash\slash{}github.com/sdsgisd\slash{}ImplicitVortexFilaments}

\begin{figure}[t]
    \centering{
       \includegraphics[width=\linewidth, trim={10cm 1cm 3cm 5cm},clip]{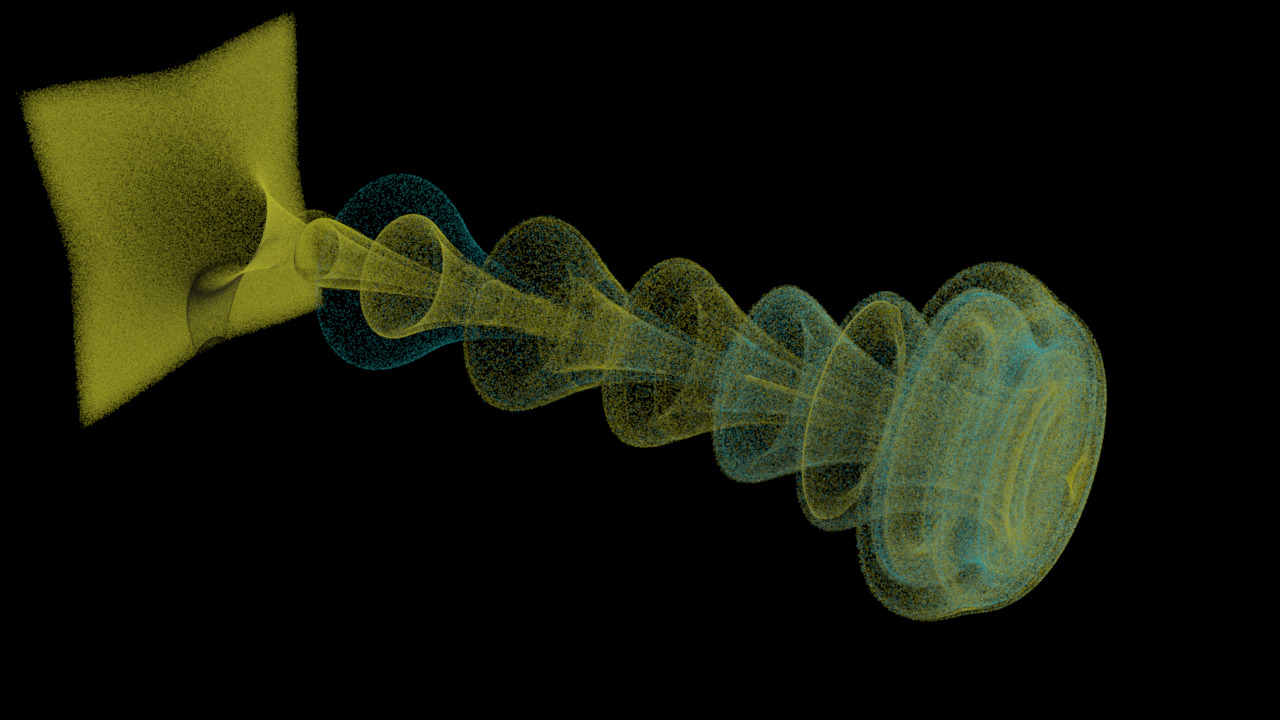}
       \caption{
           %Leapfrogging filaments and marker particles. From left to right are frames 1, 70, and 170. The initial setting is two circles located on the same plane whereas one circle has the half radius as the other.
           Two filaments leapfrog through one another, dragging marker particles into the shape of a mushroom cloud. The initial filament geometry consists of two co-planar vortex rings, one with half the radius of the other.
       }
       \label{fig:leapfrog}
    }
\end{figure}

\begin{figure}[t]
    \centering{
    \begin{subfigure}[b]{0.3\linewidth}
        \centering
        \scalebox{-1}[1]{
        \includegraphics[width=\linewidth,  trim={14cm 0cm 12cm 0cm},clip]
        {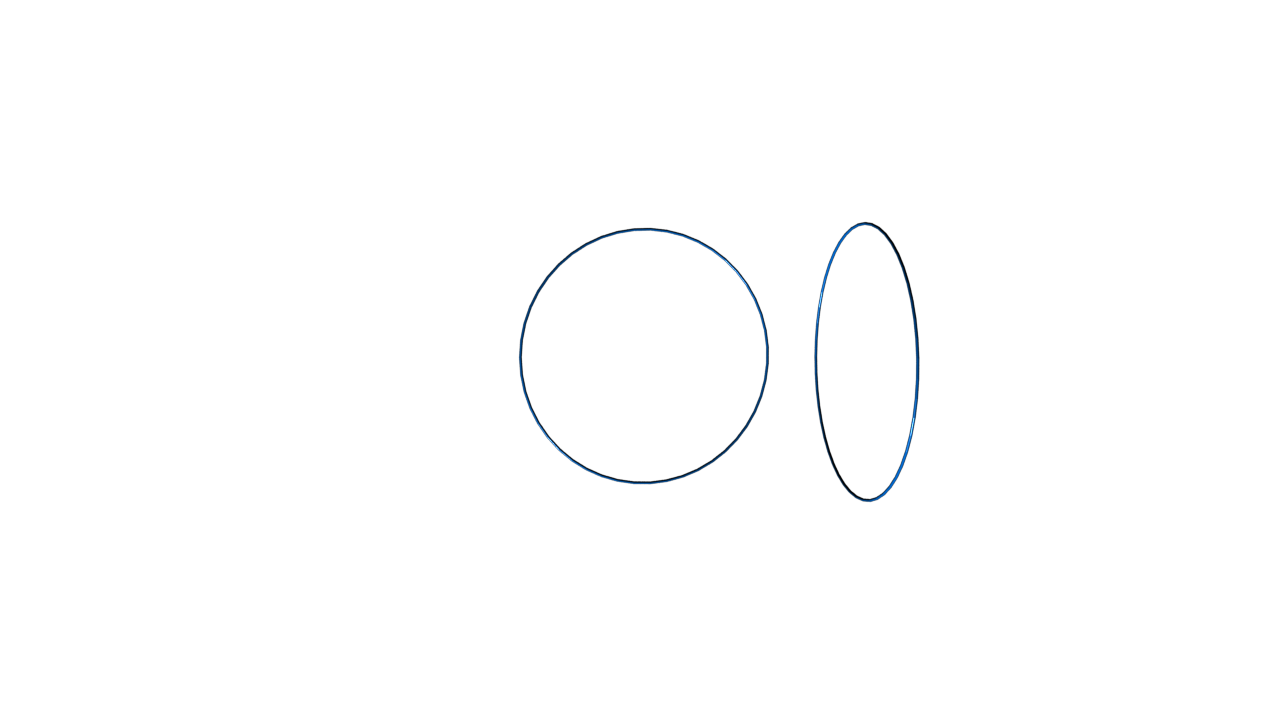}
        % {{images/obelique/obelique_curves.0001}.jpg}
        }
    \end{subfigure}
    \hfill
    \begin{subfigure}[b]{0.3\linewidth}
        \centering
        \scalebox{-1}[1]{
        \includegraphics[width=\linewidth, trim={14cm 0cm 13cm 0cm},clip]{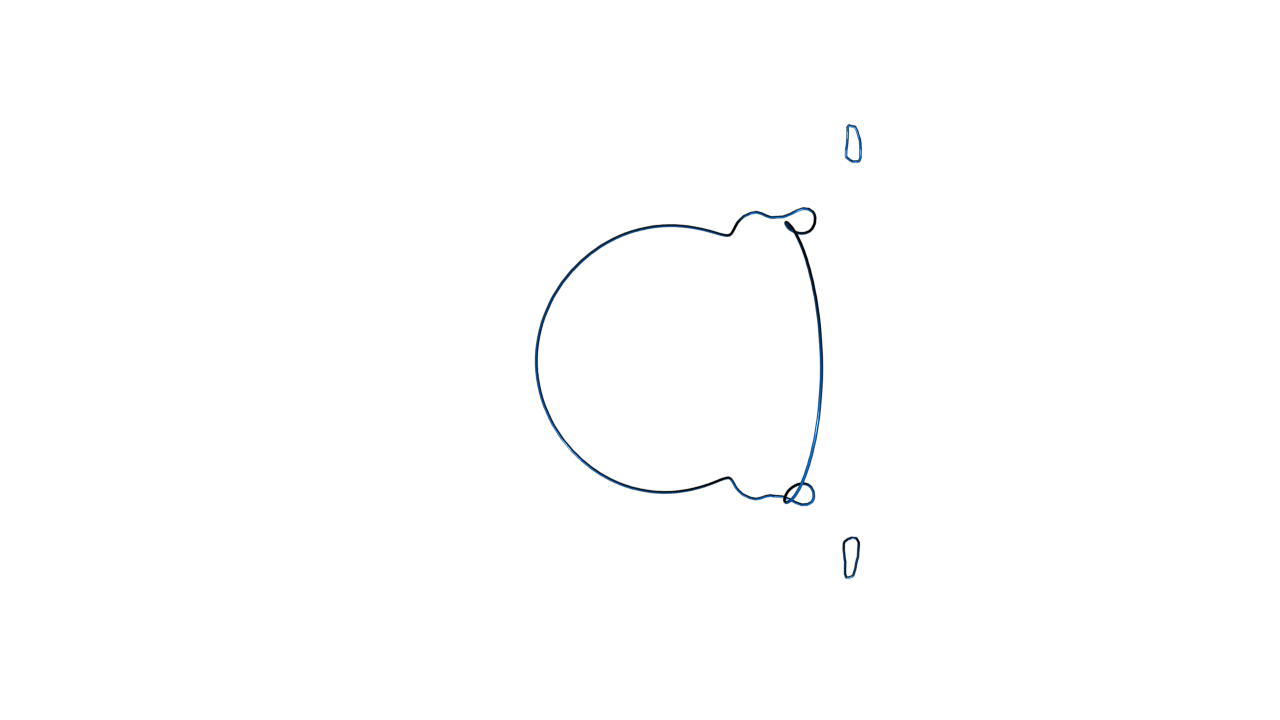}
        }
    \end{subfigure}
    \hfill
    \begin{subfigure}[b]{0.3\linewidth}
        \centering
        \scalebox{-1}[1]{
        \includegraphics[width=\linewidth, trim={14cm 0cm 14cm 0cm},clip]{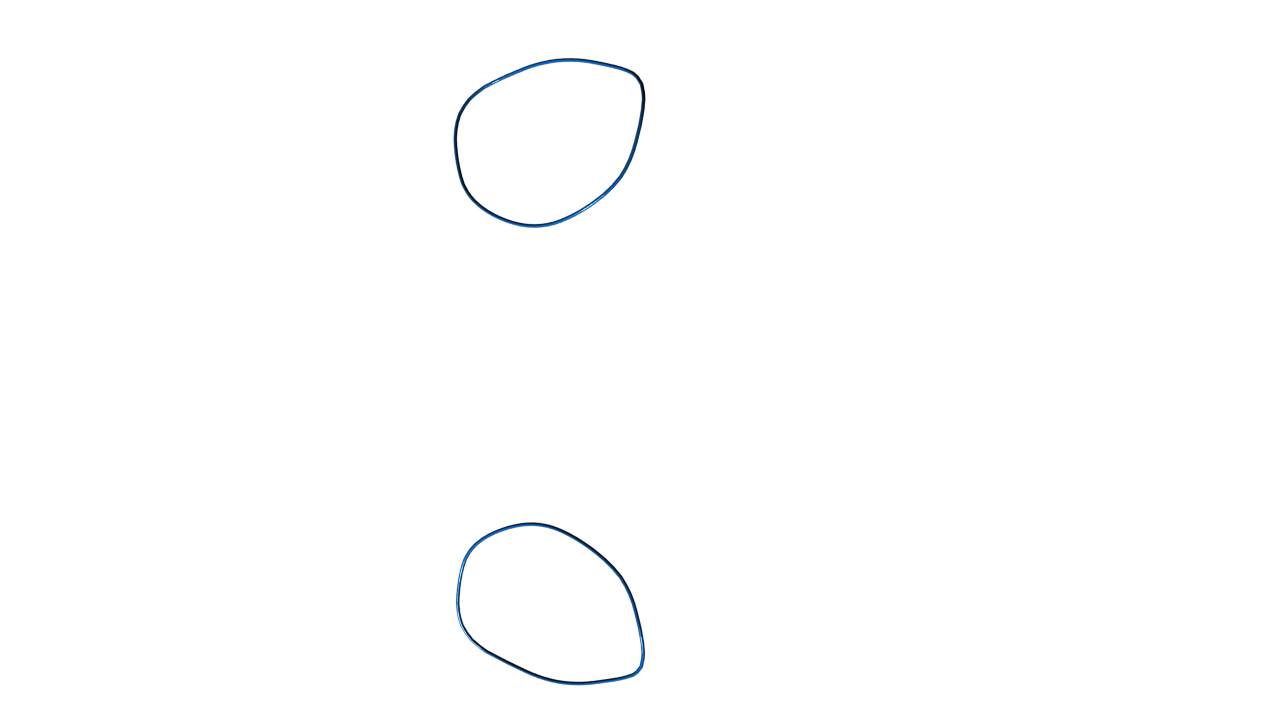}
        }
    \end{subfigure}
    }

    \centering{
    \begin{subfigure}[b]{0.3\linewidth}
        \centering
        \scalebox{-1}[1]{
        \includegraphics[width=\linewidth, trim={14cm 0cm 12cm 0cm},clip]
        % {{images/obelique/obelique_particles.0001}.png}
        {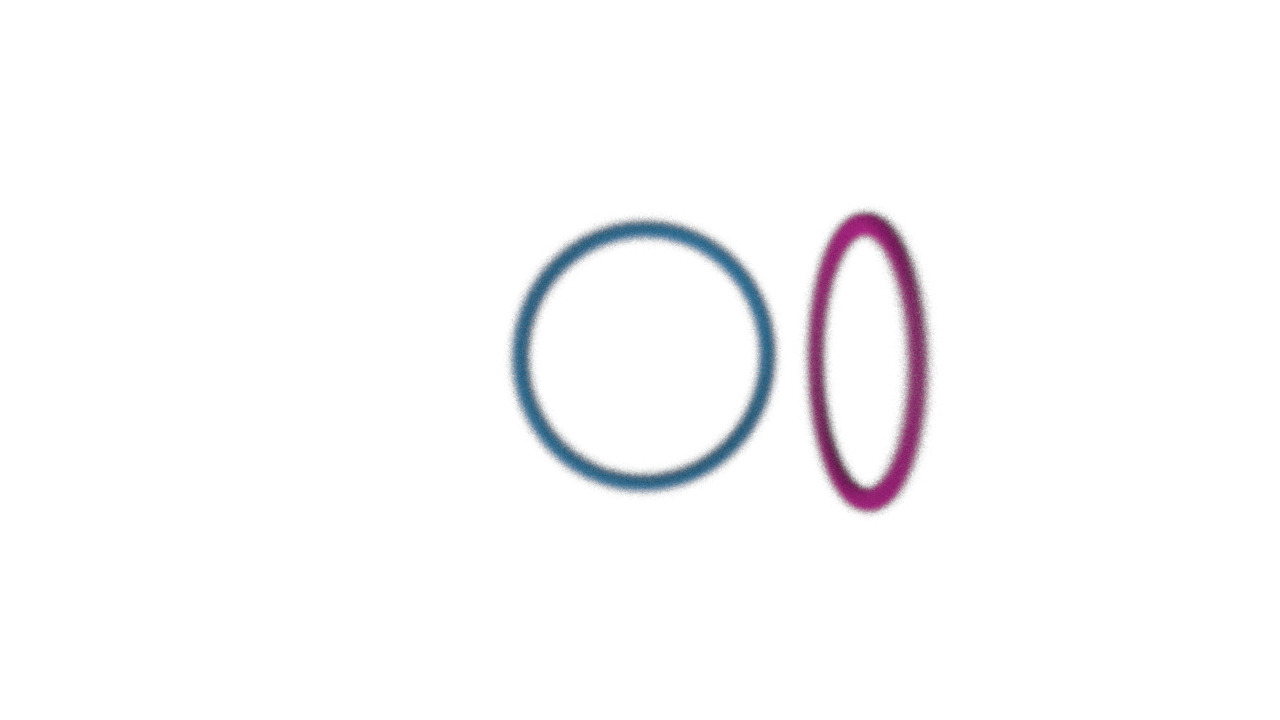}
        }
        % \caption*{Initial state}
    \end{subfigure}
    \hfill
    \begin{subfigure}[b]{0.3\linewidth}
        \centering
        \scalebox{-1}[1]{
        \includegraphics[width=\linewidth, trim={14cm 0cm 12cm 0cm},clip]{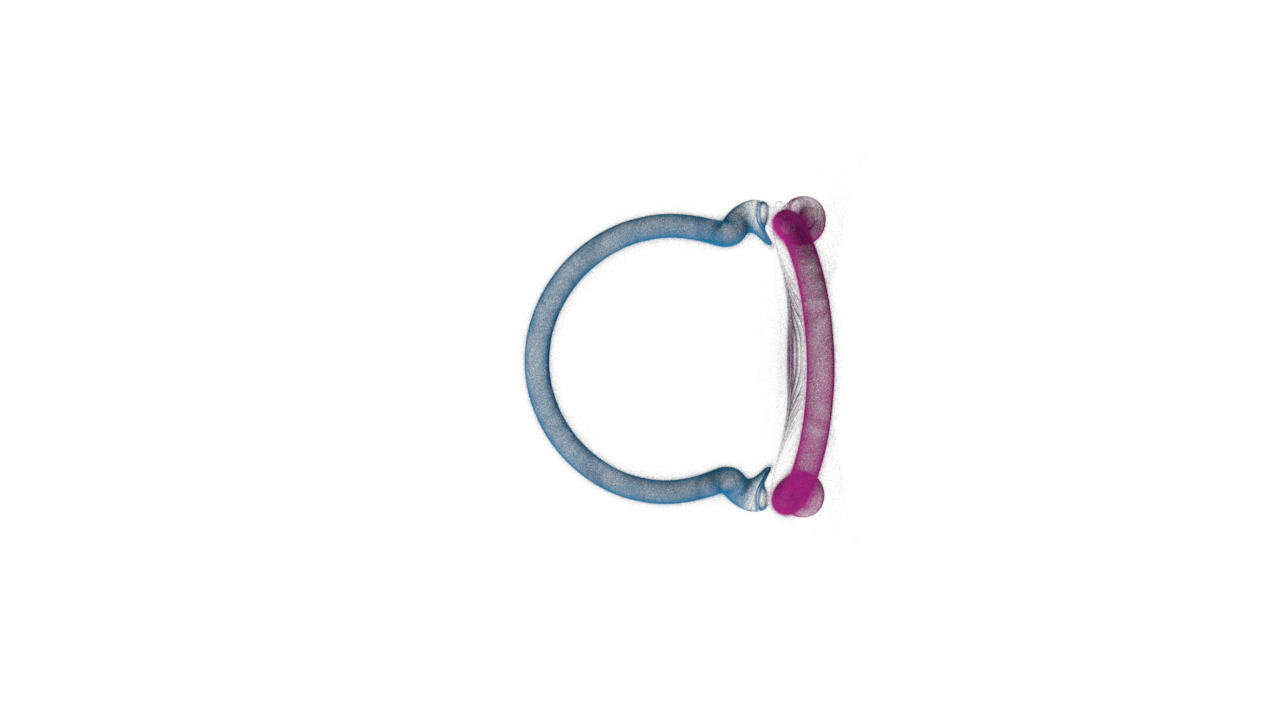}
        }
        % \caption*{Frame = 50}
    \end{subfigure}
    \hfill
    \begin{subfigure}[b]{0.3\linewidth}
        \centering
        \scalebox{-1}[1]{
        \includegraphics[width=\linewidth, trim={14cm 0cm 12cm 0cm},clip]{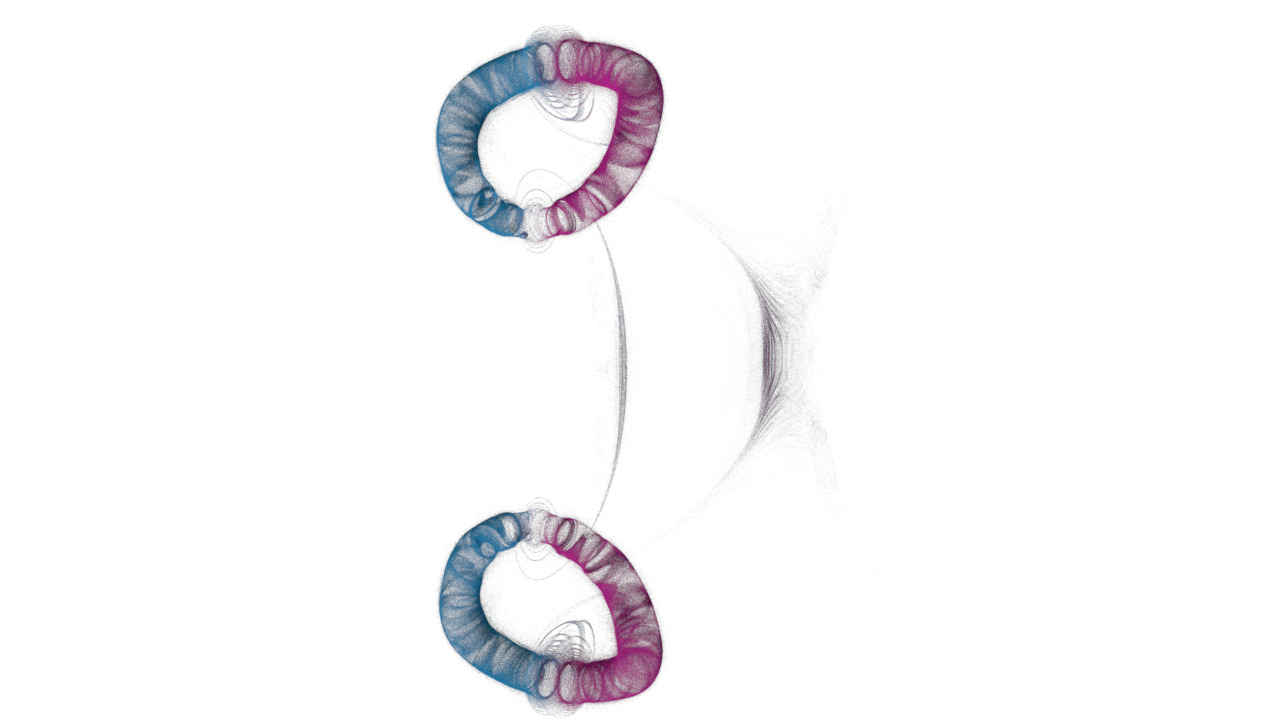}
        }
        % \caption*{Frame = 205}
    \end{subfigure}
    }

    \caption{Two smoke rings (left) colliding at orthogonal angles (middle)  and re-connecting (right). Rendered as filaments (top) and marker particles (bottom).
    % Unlike fully Eulerian methods, we can directly evaluate the velocity of each particle, so we can produce rolling patterns. 
    %The initial configuration of filaments is two circles on two vertical planes.
    \rev{The colliding rings leave swirly trails of smoke particles after their collision and reconnection.}
    }
    \label{fig:oblique_rings}
\end{figure}

We use our algorithm to animate two ``leapfrogging'' vortex rings in \figref{fig:leapfrog} and our accompanying video. We note that the system remains stable and highly symmetric even at the end of a long simulation with several high-speed ring interactions. \figref{fig:oblique_rings} visualizes two vortex rings colliding with one another at right angles, reconnecting, and detaching into two new rings. We note that the final rings retain plenty of energy after the collision event, in contrast to Eulerian simulations of this phenomenon which tend to damp out over time. The visual detail in our simulations is also practically independent of grid resolution, as the motion of marker particles are described analytically by \stt{a }Biot-Savart-style\stt{analytic } velocities \rev{(\autoref{eq:RMAll})}, instead of a vector field stored on a coarse grid.
%\rev{This produces a thin and long trail of smoke particles in these examples.}
% \cCW{I thought the revised line about smoke trails here disrupted the flow, so I moved it specifically to the figure caption where the reviewer had the question.}\cSI{Good idea.}

\begin{figure}[h]
    \begin{minipage}[b]{1.0\linewidth}
        \centering
        \includegraphics[width=0.99\linewidth, trim={0cm 2cm 0cm 2cm},clip]{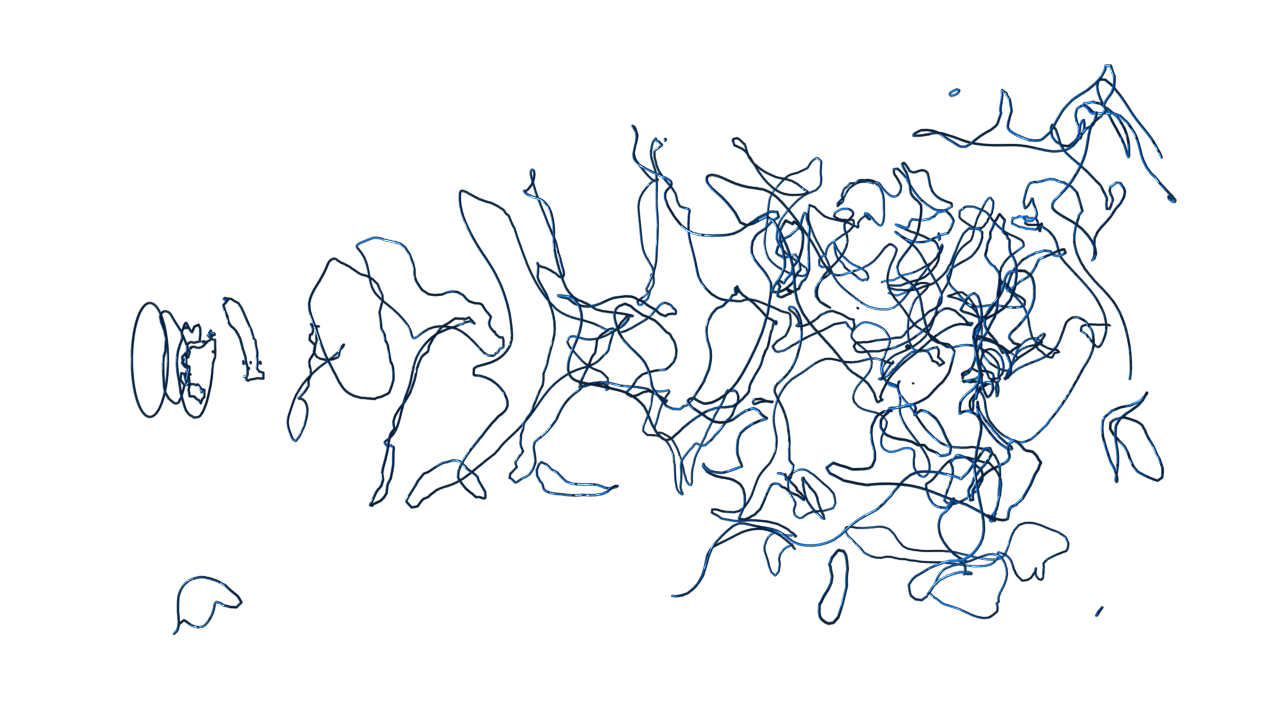}
        % \includegraphics[width=0.99\linewidth, trim={0cm 2cm 0cm 2cm},clip]{{images/jet/jet_curves.0189}.jpg}
    %   \caption{jet curves}
    \end{minipage}
    \begin{minipage}[b]{1.0\linewidth}
        \centering
        \includegraphics[width=0.99\linewidth, trim={0cm 2cm 0cm 2cm},clip]{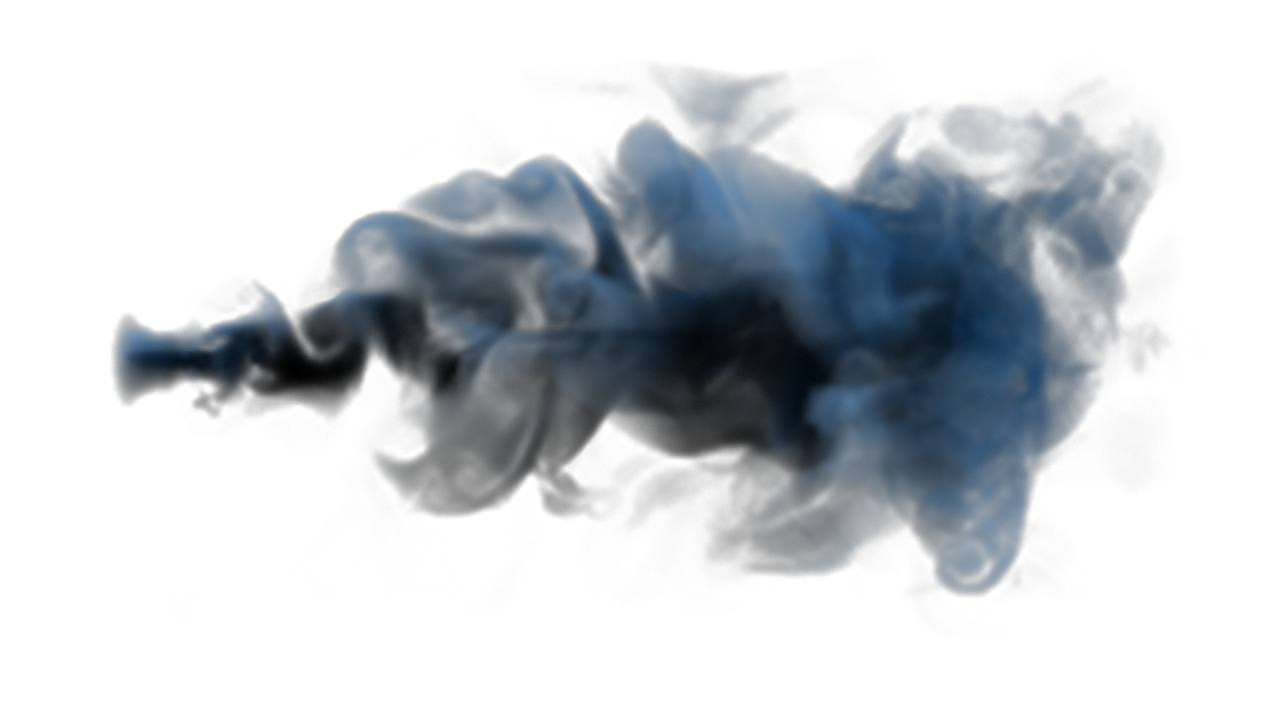}
        % \includegraphics[width=0.99\linewidth, trim={0cm 2cm 0cm 2cm},clip]{{images/jet/jet_smoke_intensity5.0189}.jpg}
    %   \caption{jet smoke}
    \end{minipage}
    \caption{A jet of smoke, rendered as raw filament geometry (top) and an advected smoke density function (bottom).
    }
    \label{fig:jet}
\end{figure}

\figref{fig:jet} illustrates a jet of smoke created by generating a new smoke ring at the left side of the domain every three time steps. 
\rev{For transporting smoke as a scalar field stored on grid points, we again used \autoref{eq:RMAll}.}
This simulation shows the robustness of our topology changes: each re-connection event is the result of a curve extraction from a level set function, so there is no possibility of any unexpected edge cases, and no need for any geometric intersection code. The simple set-up creates a large variety of chaotic motions resulting from fast leap-frogging rings squeezing in between others and reconnecting filaments causing sudden changes in direction. When rings shrink smaller than the grid resolution, our algorithm deletes them (similar to codimension-1 level set methods). 

Lastly, \figref{fig:torus} illustrates how our method can evolve intricate filament geometry, specifically the (5,8)-torus knot defined by
\begin{align}
    &\gamma(s)=\left(
        \left(\cos(q s)+2\right)\cos(ps),
        \left(\cos(q s)+2\right) \sin(ps),-sin(qs)
        \right) \notag
\end{align}
%${\gamma(s)=\left( \left(\cos(q s)+2\right)\cos(ps), \left(\cos(q s)+2\right) \sin(ps),-sin(qs) \right)}$,
with \((p,q)=(5,8)\) and \(s\in [0,2\pi)\).

\subsection{Influence of Numerical Parameters}

\begin{figure}[t!] 
    \begin{overpic}[width=0.48\linewidth,tics=0,trim={10cm 3cm 10cm 2cm},clip]
        {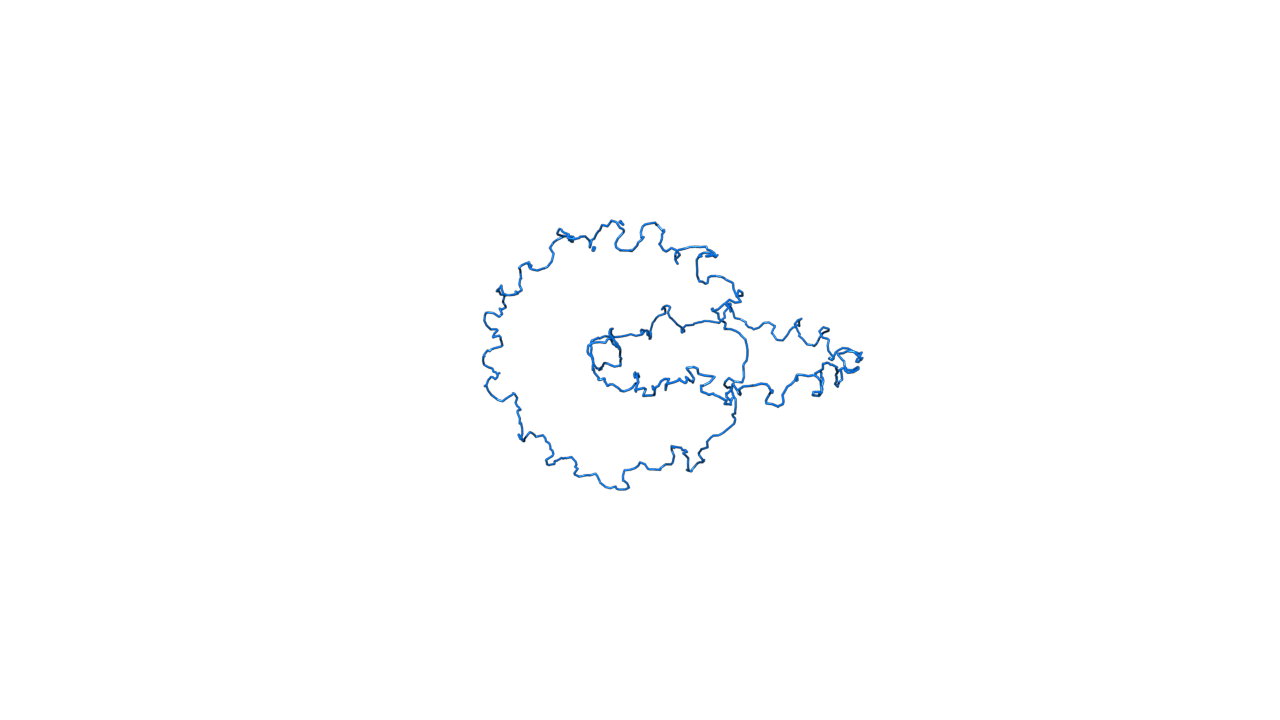}
        % {{images/twoRings/2rings_BS.0020}.jpg}
        % \put(20,0){ \small BS Frame = 20}
        % \put(40,0){\small Advection along the Biot-Savart velocity field}
    \end{overpic}
    % \hspace*{\fill}
    \begin{overpic}[width=0.48\linewidth,tics=0,trim={10cm 3cm 10cm 2cm},clip]
        {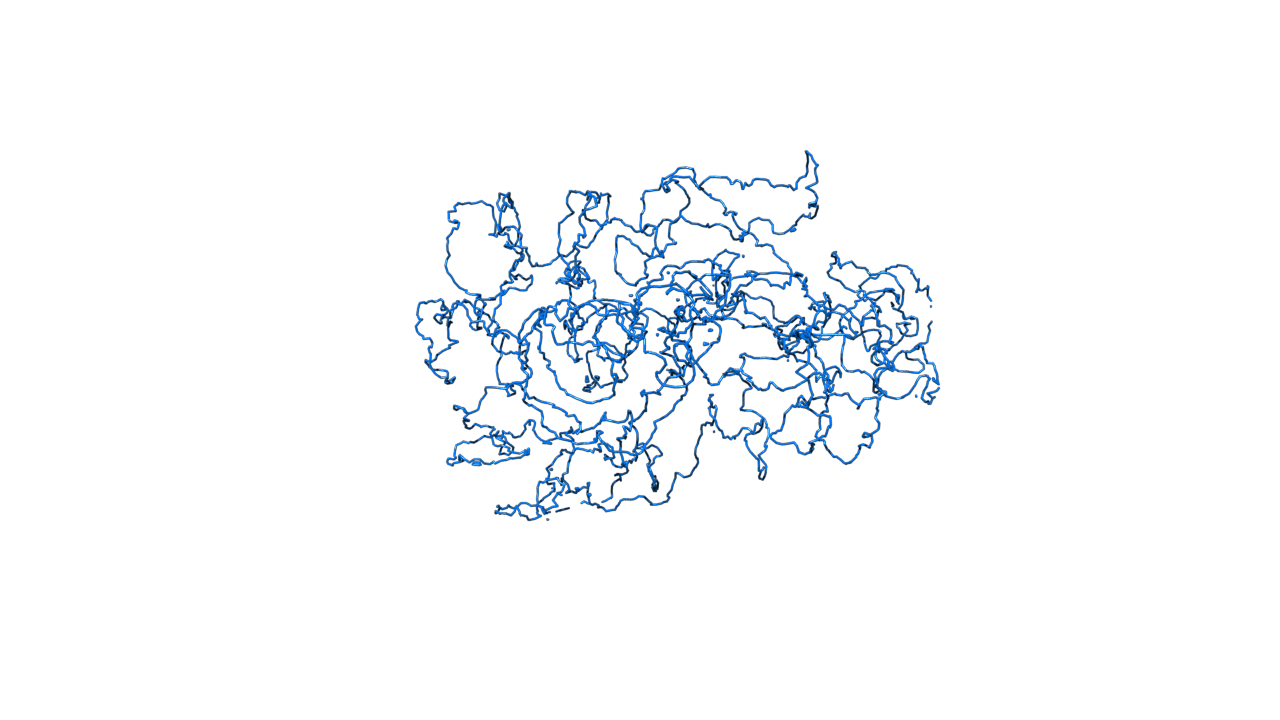}
        % {{images/twoRings/2rings_BS.0045}.jpg}
        % \put(20,0){ \small BS Frame = 45}
    \end{overpic}
    \vspace{-20pt}
    \caption*{\small Advection along the Biot-Savart velocity field}
    % \hspace*{\fill}
    % \vspace{-20pt}

    % \medskip
    \begin{overpic}[width=0.48\linewidth,tics=0,trim={10cm 3cm 10cm 2cm},clip]
        {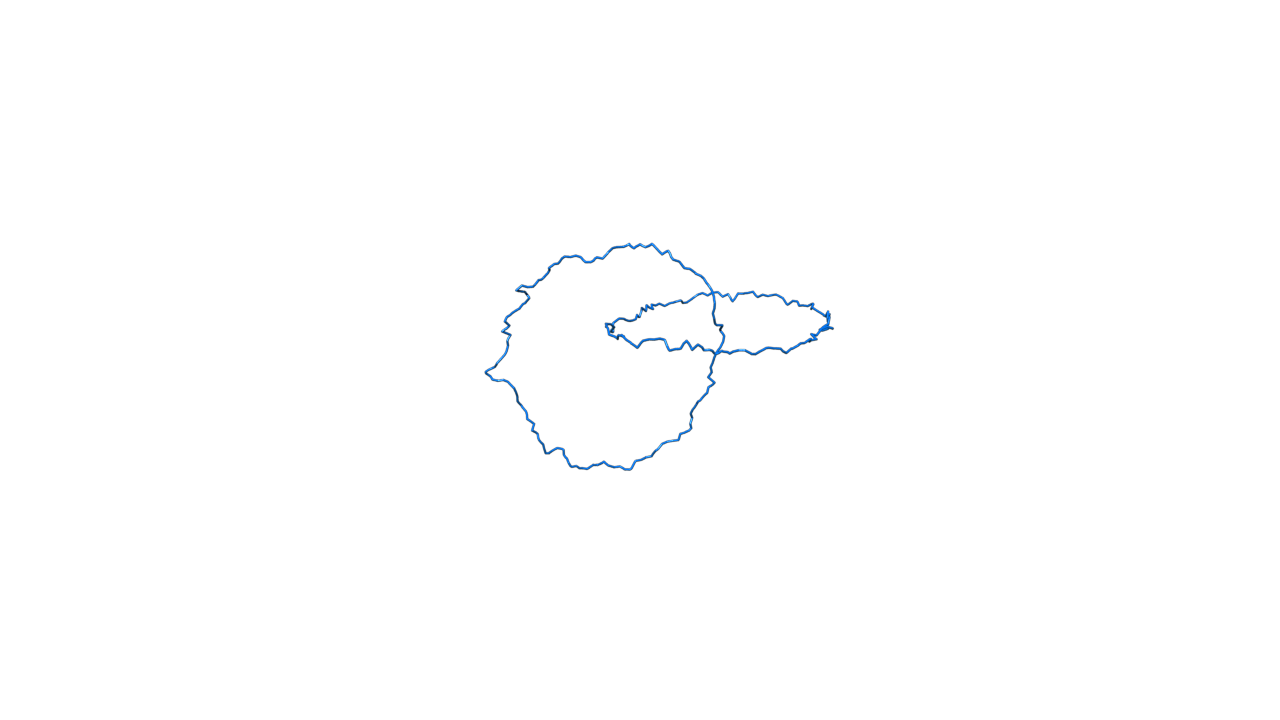}
        % {{images/twoRings/2rings_normal_extension.0020}.jpg}
        % \put(20,0){ \small NV Frame = 20}
        % \put(40,0){\small Advection along the nearest point velocity field}
    \end{overpic}
    % \hspace*{\fill}
    \begin{overpic}[width=0.48\linewidth,tics=0,trim={10cm 3cm 10cm 2cm},clip]
        {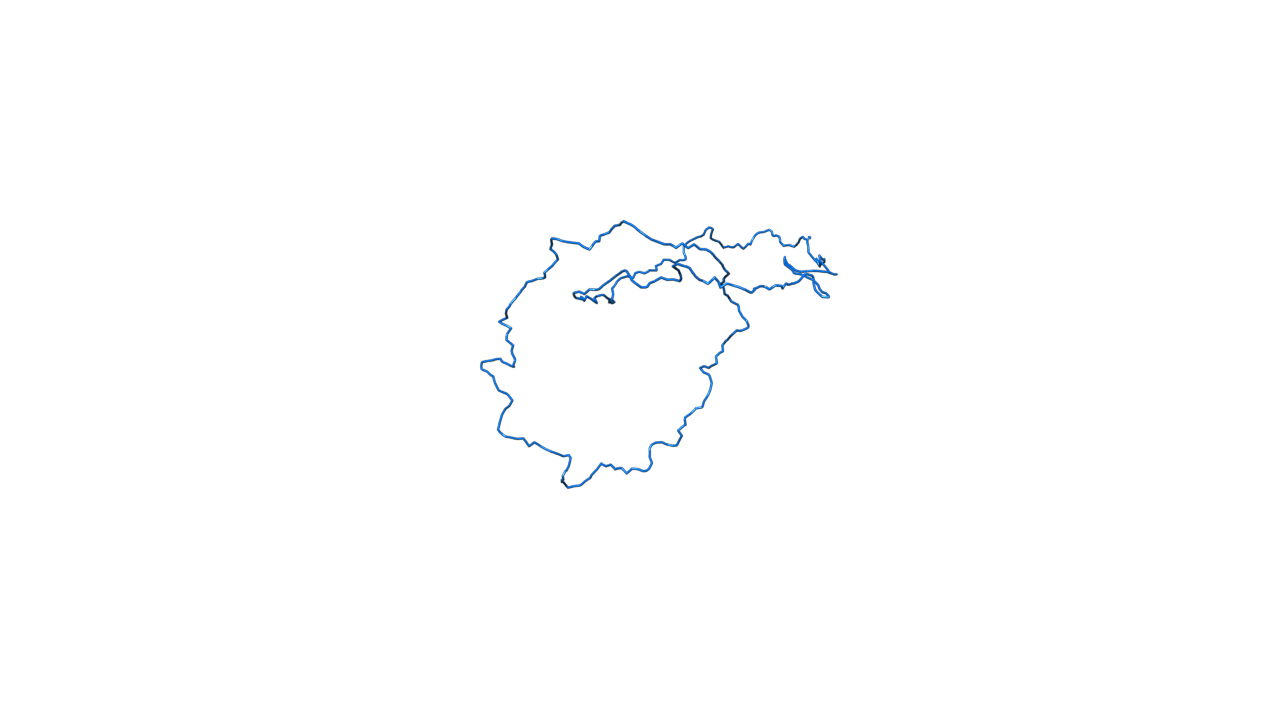}
        % {{images/twoRings/2rings_normal_extension.0045}.jpg}
        % \put(20,0){ \small NV Frame = 45}
    \end{overpic}
    % \hspace*{\fill}
    \vspace{-30pt}
    \caption*{\small Advection along the nearest point velocity field}
    % \vspace{-20pt}

    \medskip
    \begin{overpic}[width=0.48\linewidth,tics=0,trim={10cm 3cm 10cm 2cm},clip]
        {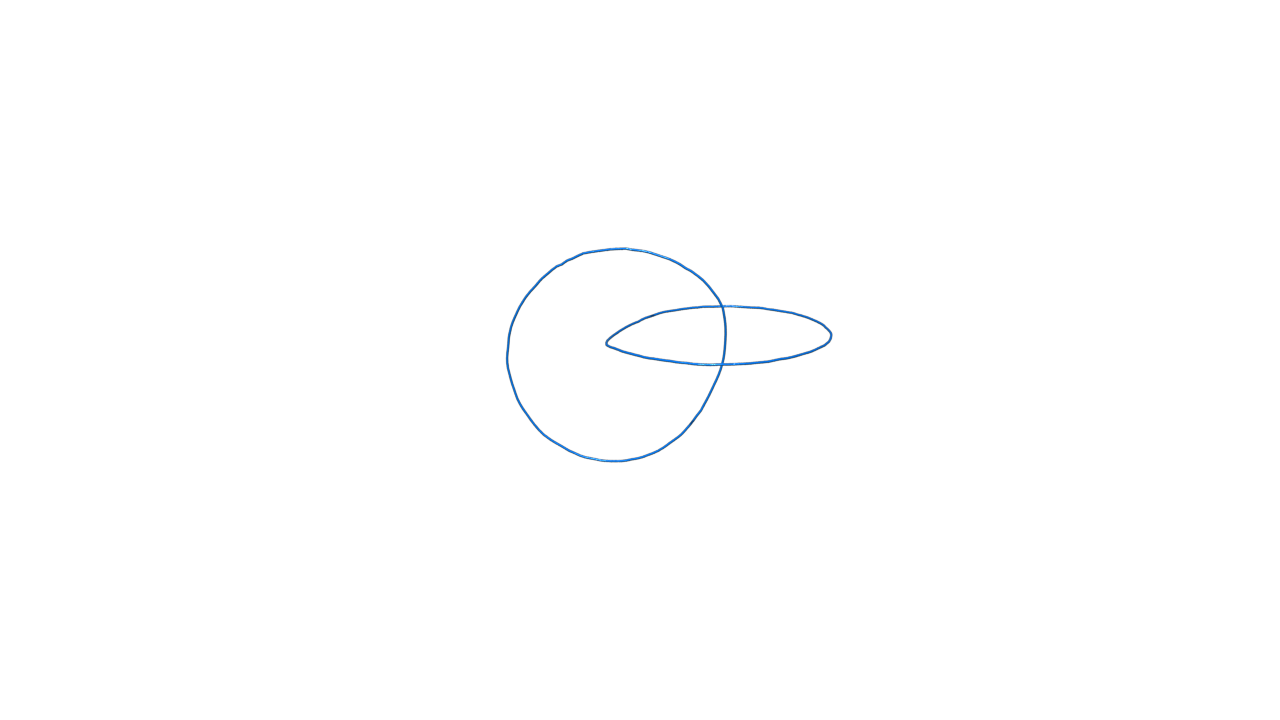}
        % {{images/twoRings/2rings_wo_redistance.0020}.jpg}
        % \put(20,0){ \small NoRedist Frame = 20}
    \end{overpic}
    % \hspace*{\fill}
    \begin{overpic}[width=0.48\linewidth,tics=0,trim={10cm 3cm 10cm 2cm},clip]
        {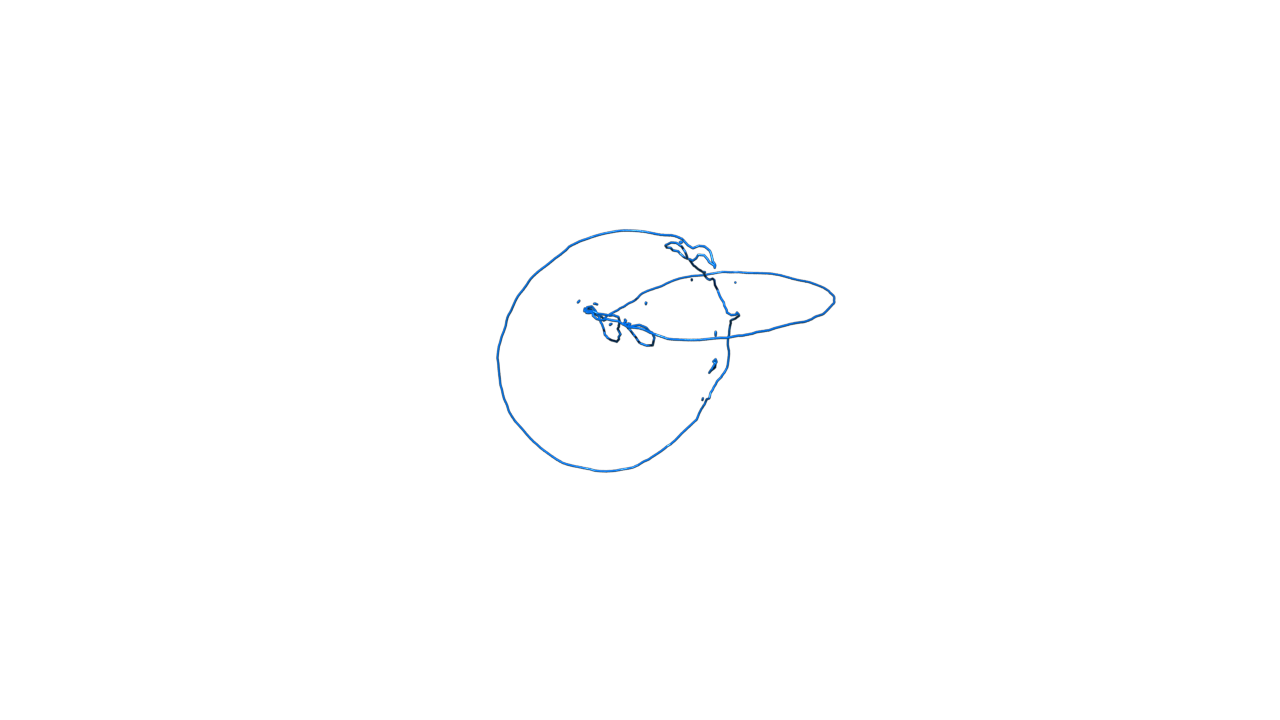}
        % {{images/twoRings/2rings_wo_redistance.0045}.jpg}
        % \put(20,0){ \small NoRedist Frame = 45}
    \end{overpic}
    % \hspace*{\fill}
    \vspace{-30pt}
    \caption*{\small Without redistancing \(\psi\)}
    % \vspace{-20pt}

    \medskip
    \begin{overpic}[width=0.48\linewidth,tics=0,trim={10cm 3cm 10cm 2cm},clip]
        {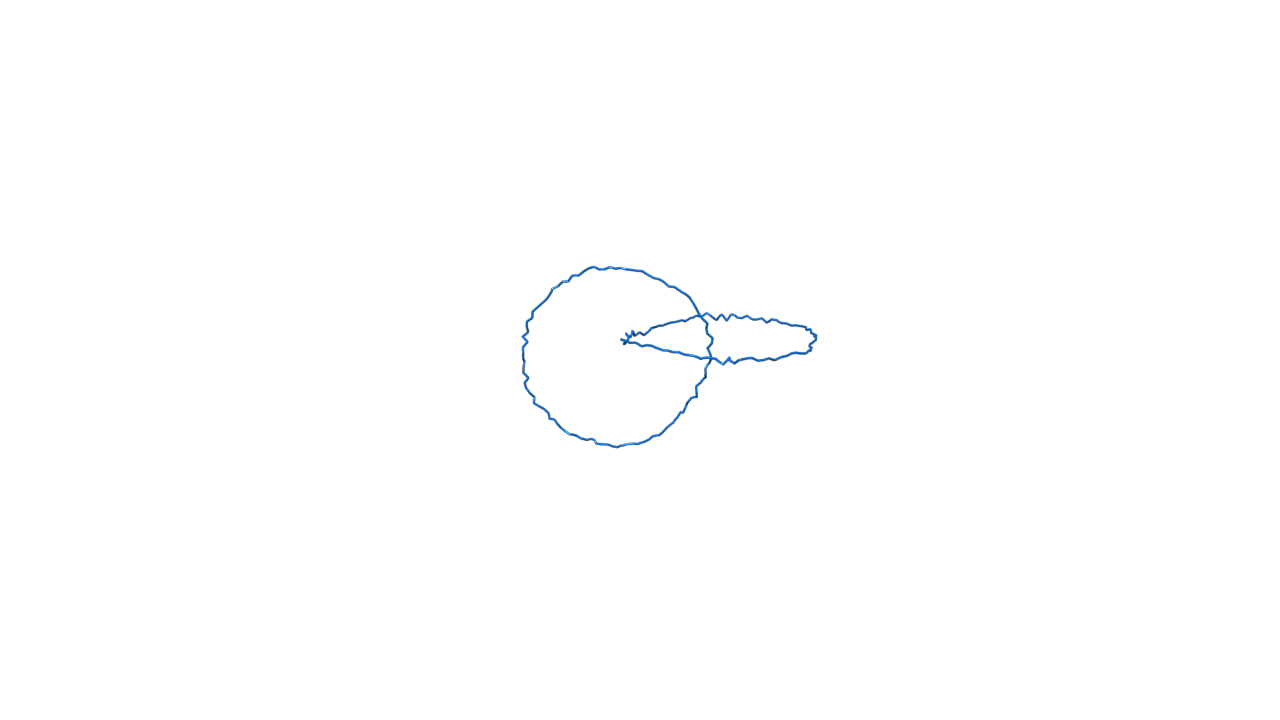}
        % {{images/twoRings/2rings_twisted.0020}.jpg}
        % \put(20,0){ \small Twisted Frame = 20}
    \end{overpic}
    % \hspace*{\fill}
    \begin{overpic}[width=0.48\linewidth,tics=0,trim={10cm 3cm 10cm 2cm},clip]
        {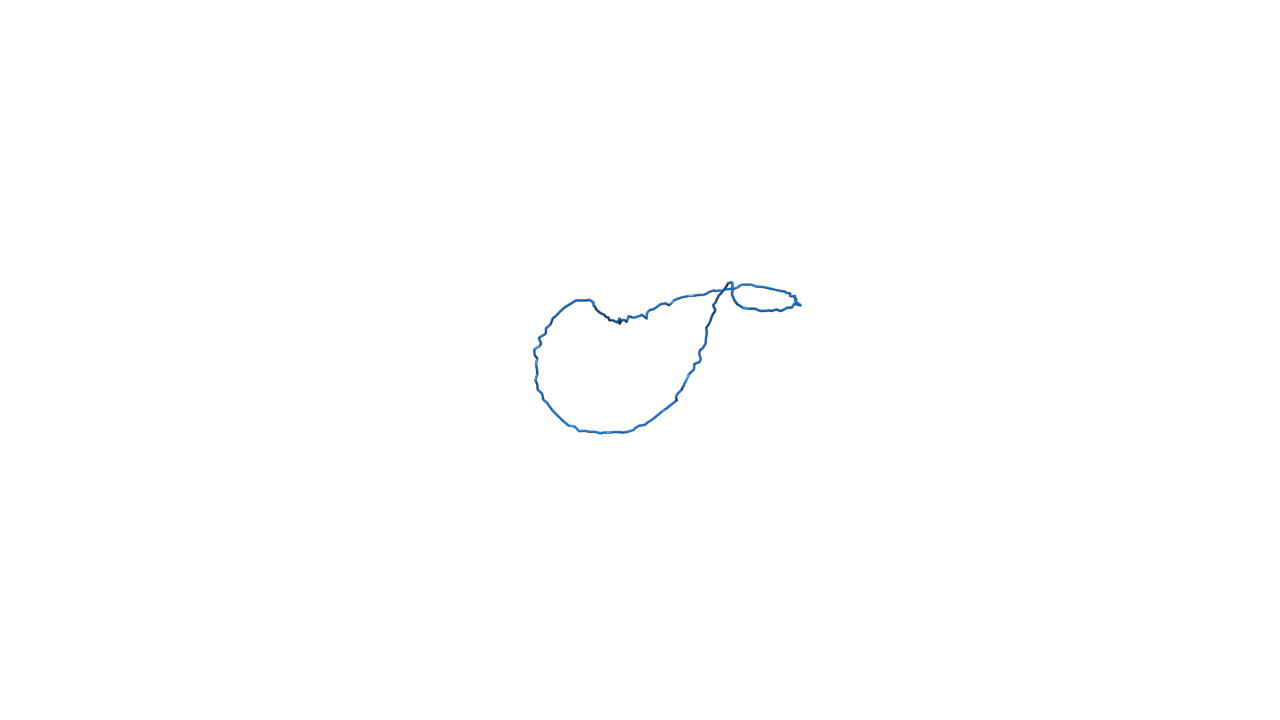}
        % {{images/twoRings/2rings_twisted.0045}.jpg}
        % \put(20,0){ \small Twisted Frame = 45}
    \end{overpic}
    % \hspace*{\fill}
    \vspace{-30pt}
    \caption*{\small Artificially twisted \(\psi\)}
    % \vspace{-20pt}

    \medskip
    \begin{overpic}[width=0.48\linewidth,tics=0,trim={10cm 3cm 10cm 2cm},clip]
        {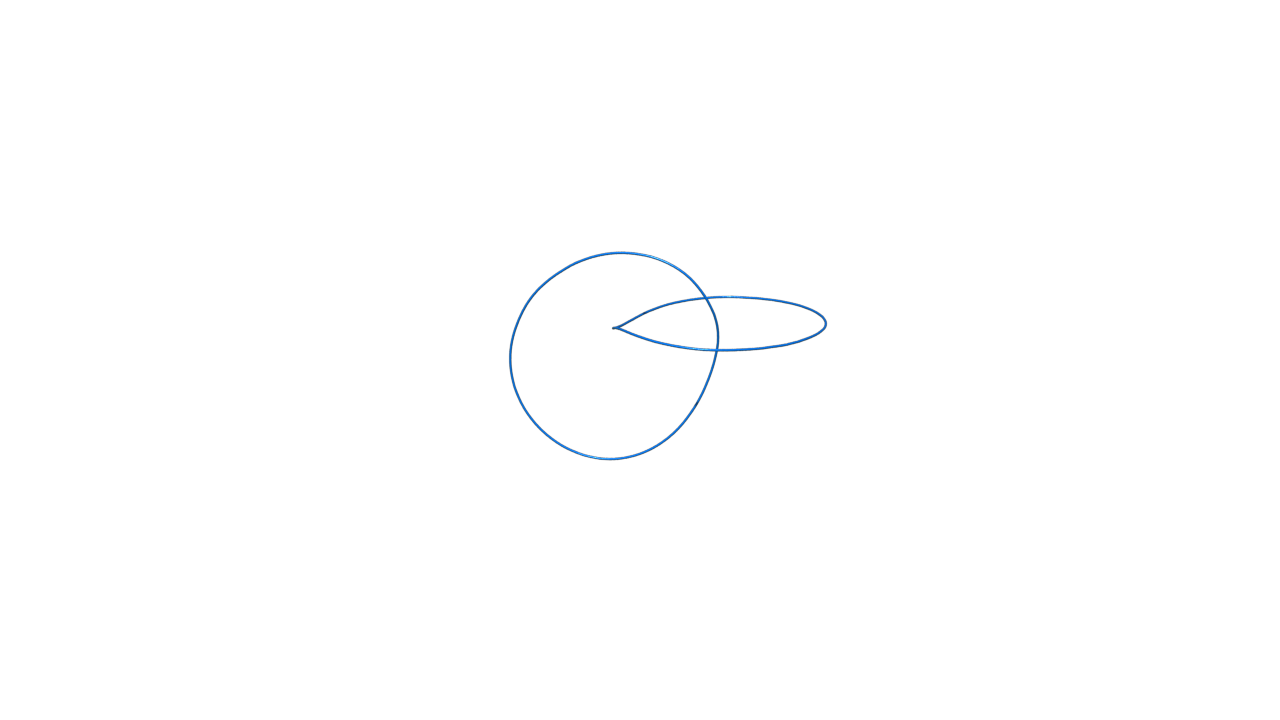}
        % {{images/twoRings/2rings_ours.0020}.jpg}
        % \put(20,0){ \small Ours Frame = 20}
        
    \end{overpic}
    % \hspace*{\fill}
    \begin{overpic}[width=0.48\linewidth,tics=0,trim={10cm 3cm 10cm 2cm},clip]
        {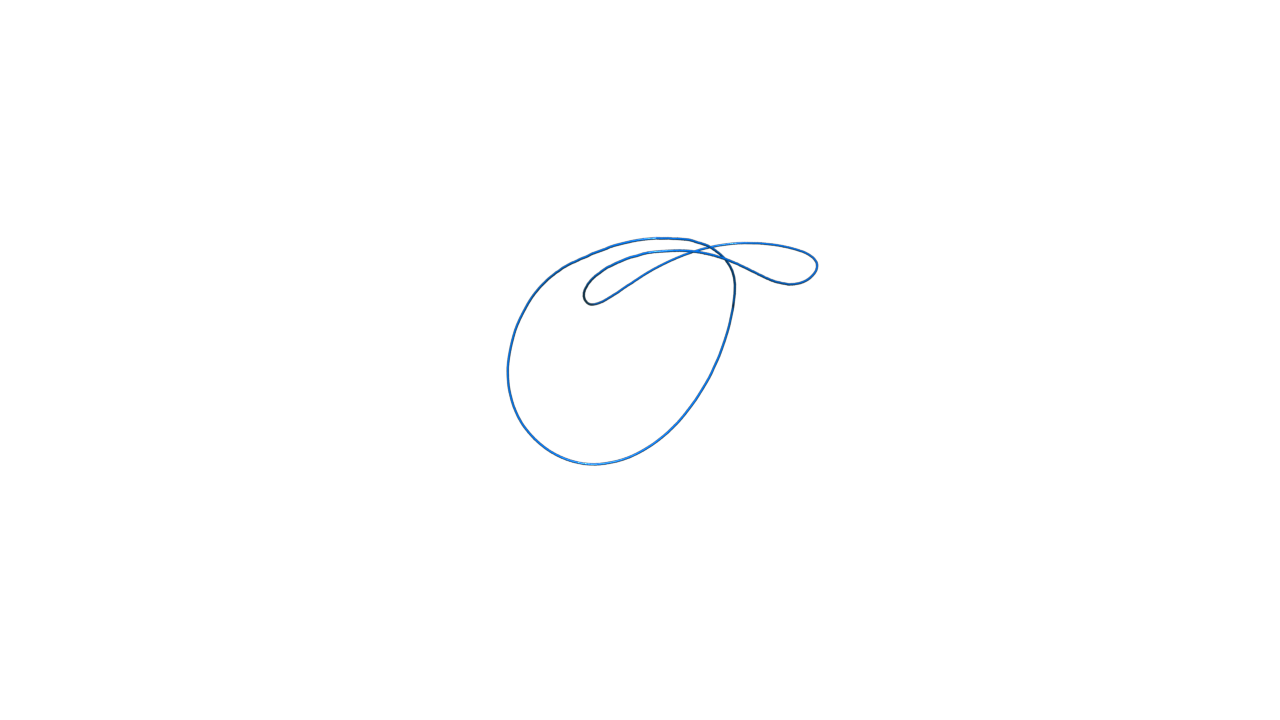}
        % {{images/twoRings/2rings_ours.0045}.jpg}
        % \put(20,0){ \small Ours Frame = 45}
   
    \end{overpic}
    % \hspace*{\fill}
        \vspace{-30pt}
    \caption*{\small Our proposed setting}
    \vspace{-10pt}
    % \caption*{\small Frame = 20 \hspace{80pt} Frame = 45}
    \caption{Comparisons with different settings. Initial state is the 1st row of \autoref{fig:two_rings}. The left and right columns show states at frames 20 and 45.  
    %\cCW{This figure uses space inefficiently. Any thoughts on improving it? We could break up variations of $\bv$ and $\psi$ into different figures, for example.}
    % In each of the following rows, the left and the right images are the configurations at frame 20 and 45 respectively. 
    % 1st row: Advection along BS velocity. Too swirling so simulation explodes. 
    % 2nd row: Advection along the velocity field taken from the nearest point of the filaments. The overall simulation works but quite jaggy. 
    % 3rd row: Advection along our velocity field, but without redistancing. Works in the beginning. After a while, gets broken and yields many topologically ill fragments. 
    % 4th row: twisted level function. Jaggy and shrinking.
    % 5th row: Advection along our velocity field with redistancing. Good.
    } \label{fig:different_settings}
\end{figure}

\figref{fig:different_settings} demonstrates the importance of each step in our approach by selectively removing \rev{or modifying} different algorithmic components and illustrating the consequences. First we illustrate what happens when we vary the free degrees of freedom in the velocity field $\bv$ used to advect the filaments. In agreement with the discussion in \secref{sec:RepresentationsForFilamentDynamics}, we see that setting $\bv$ to the fluid velocity (based on Biot-Savart kernels) field causes the level set function $\psi$ to rapidly twist up and become unstable. Setting $\bv$ to equal the velocity at the nearest point on the filament creates similar noise, presumably due to spatial discontinuities in the field. Compare these results to the smooth geometry generated by our velocity field at the bottom of \figref{fig:different_settings}.

Lower down in the same figure, we illustrate the effect of varying the free degrees of freedom in our level set function $\psi$. Starting with an initially smooth $\psi$ and advecting it without any re-distancing or regularization works well at the beginning, but it eventually accumulates topological noise. To illustrate the impact of $\psi$'s free parameters on numerical stability and accuracy, the fourth row in \figref{fig:different_settings} replaces our smooth choice of $\psi$ with one that is intentionally twisted by a phase shift of \(\Delta \theta(x):=0.05\  \text{dist}(x,\gamma)\); the twisted $\psi$ causes high-frequency geometric noise and artificially shrinks the filaments. Again, we can compare these results to the smooth geometry generated by our un-twisted $\psi$ at the bottom of \figref{fig:different_settings}.

\begin{figure}
    \centering{
        \begin{subfigure}[b]{0.48\linewidth}
            \centering
            \includegraphics[width=\linewidth,trim={6cm 2cm 6cm 2cm},clip]{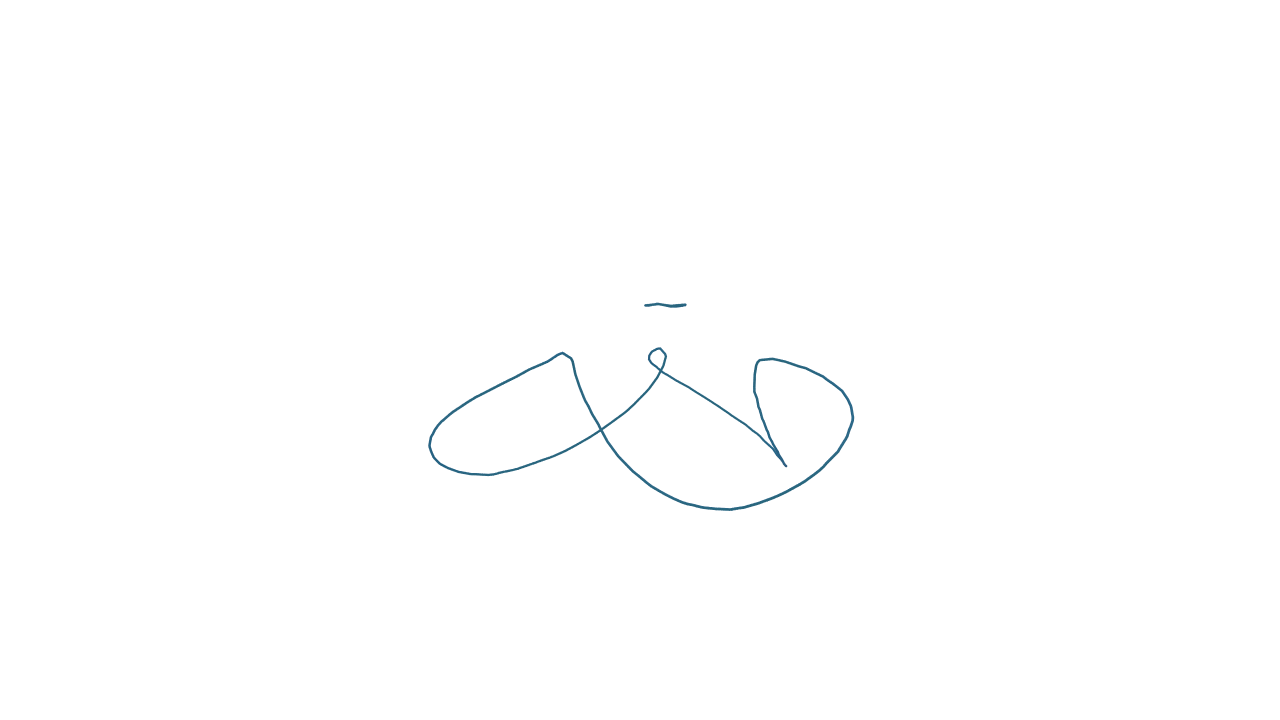}
        \end{subfigure}
        \hfill
        \begin{subfigure}[b]{0.48\linewidth}
            \centering
            \includegraphics[width=\linewidth,trim={6cm 2cm 6cm 2cm},clip]{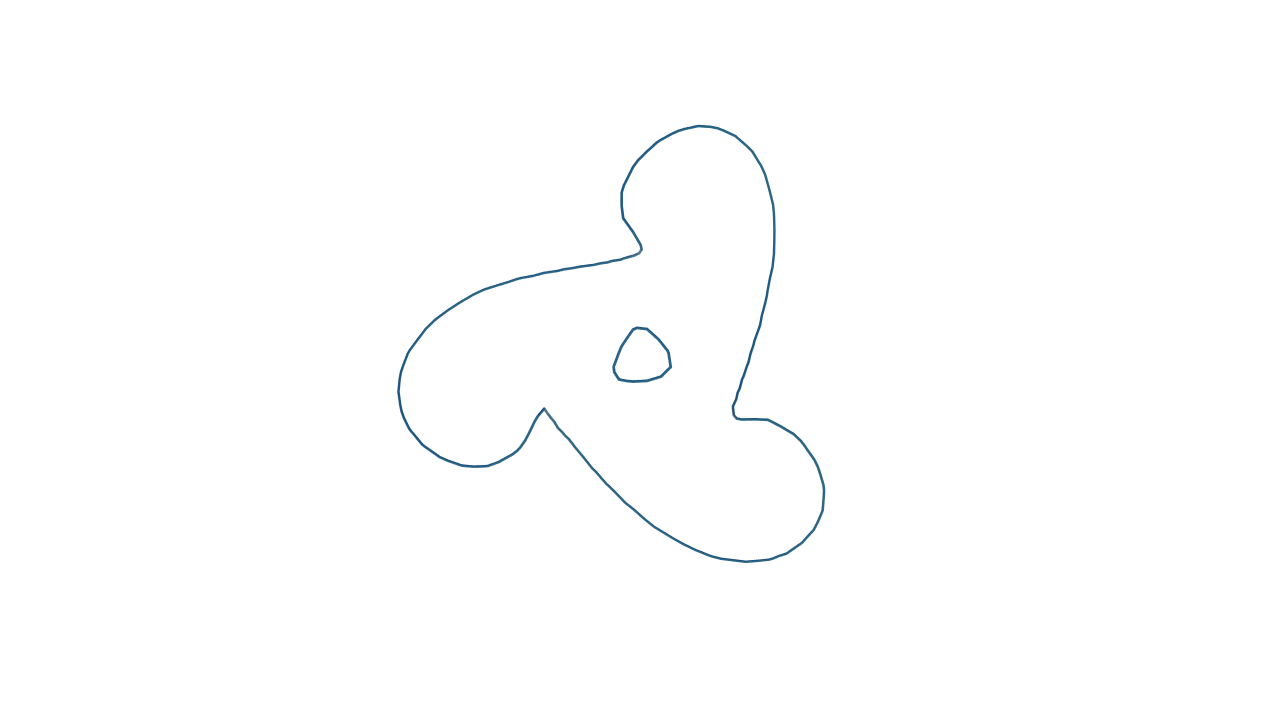}
        \end{subfigure}
        \vspace{-10pt}
        \caption*{Resolution \(50\times50\times60\)}
    }

    \centering{
    \begin{subfigure}[b]{0.48\linewidth}
        \centering
        \includegraphics[width=\linewidth,trim={6cm 2cm 6cm 2cm},clip]{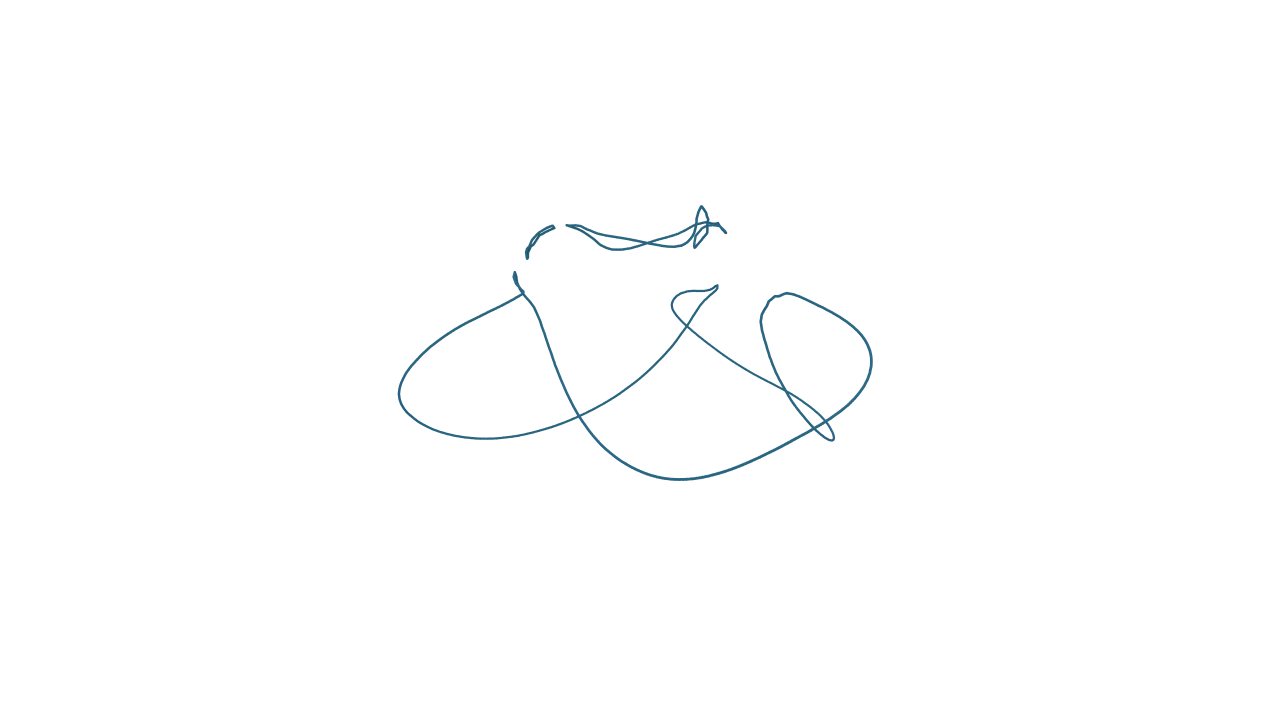}
    \end{subfigure}
    \hfill
    \begin{subfigure}[b]{0.48\linewidth}
        \centering
        \includegraphics[width=\linewidth,trim={6cm 2cm 6cm 2cm},clip]{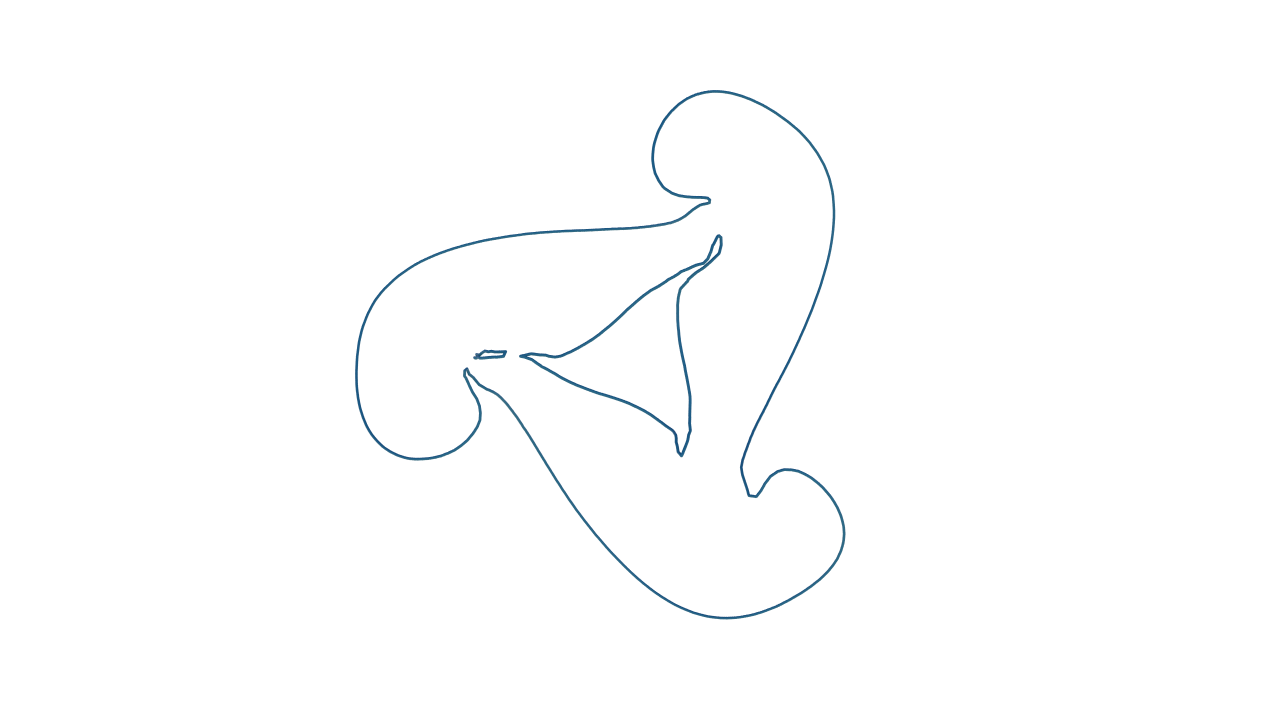}
    \end{subfigure}
    \vspace{-10pt}
    \caption*{Resolution \(100\times100\times120\)}
    }

    \caption{
        %Intermediate states of the trefoil knot in different grid resolutions. The left and right columns show a side and a front view. Filaments are under-resolved in the half resolution setting and tend to shrink and lose details.
        Simulation of a trefoil knot with different grid resolutions, as viewed from the side (left column) and front (right column).
    }
    \label{fig:different_resolutions}
\end{figure}

Another important numerical parameter is the spatial resolution of the filament. Lagrangian methods constrain the curve resolution by subdividing and collapsing edges when they become too long or short. In contrast, our method controls the curve details via the resolution of the grid used for the level set $\psi$. \stt{Figure} \figref{fig:different_resolutions} illustrates a simulation of an evolving trefoil knot on a \(50\times50\times60\) grid and one twice as detailed at \(100\times100\times120\). As expected, higher resolution grids create filaments with sharper details. Note, however, that the common way to visualize fluids is with marker particles or smoke densities, not by visualizing the filaments themselves. Thus, even very low resolution vortex filaments can still produce high resolution visual details. 
%\cCW{would be nice to have a figure or sim that showed this} 
More detailed filaments make themselves evident via there more detailed velocity fields and complex smoke dynamics.

\begin{figure}[htbp!] 
 
     \begin{overpic}[width=0.32\linewidth,tics=0,trim={12cm -2cm 12cm 8cm},clip]
        {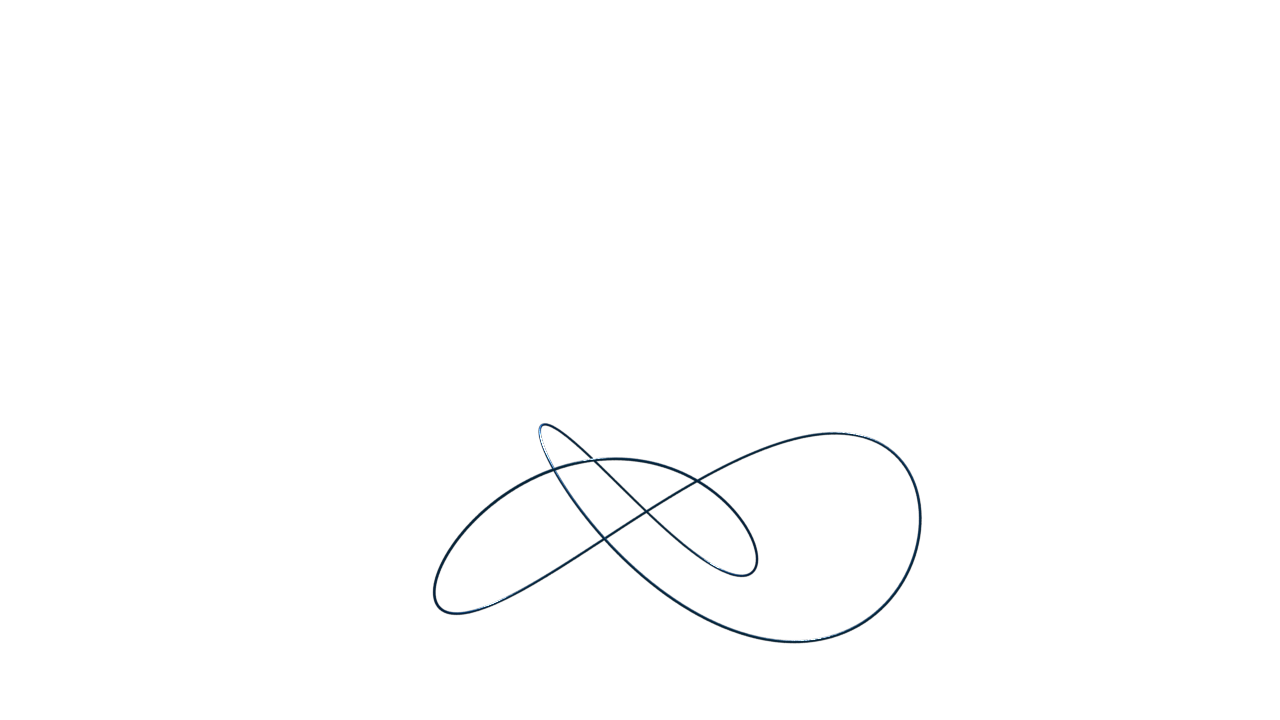}
        % {{images/trefoil/trefoil.0001}.jpg}
    \end{overpic}
    \begin{overpic}[width=0.32\linewidth,tics=0,trim={12cm 4cm 12cm 2cm},clip]
        {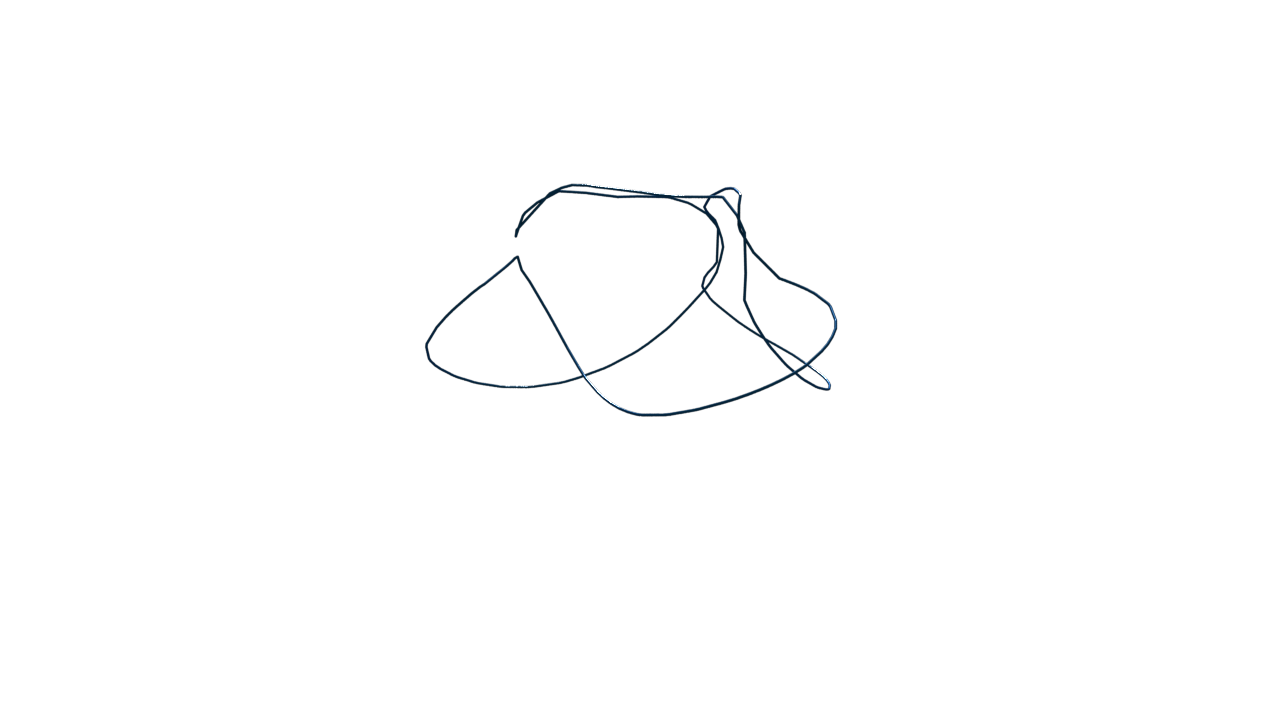}
        % {{images/trefoil/trefoil_WP.0125}.jpg}
    \end{overpic}
    \begin{overpic}[width=0.32\linewidth,tics=0,trim={12cm 4cm 12cm 2cm},clip]
        {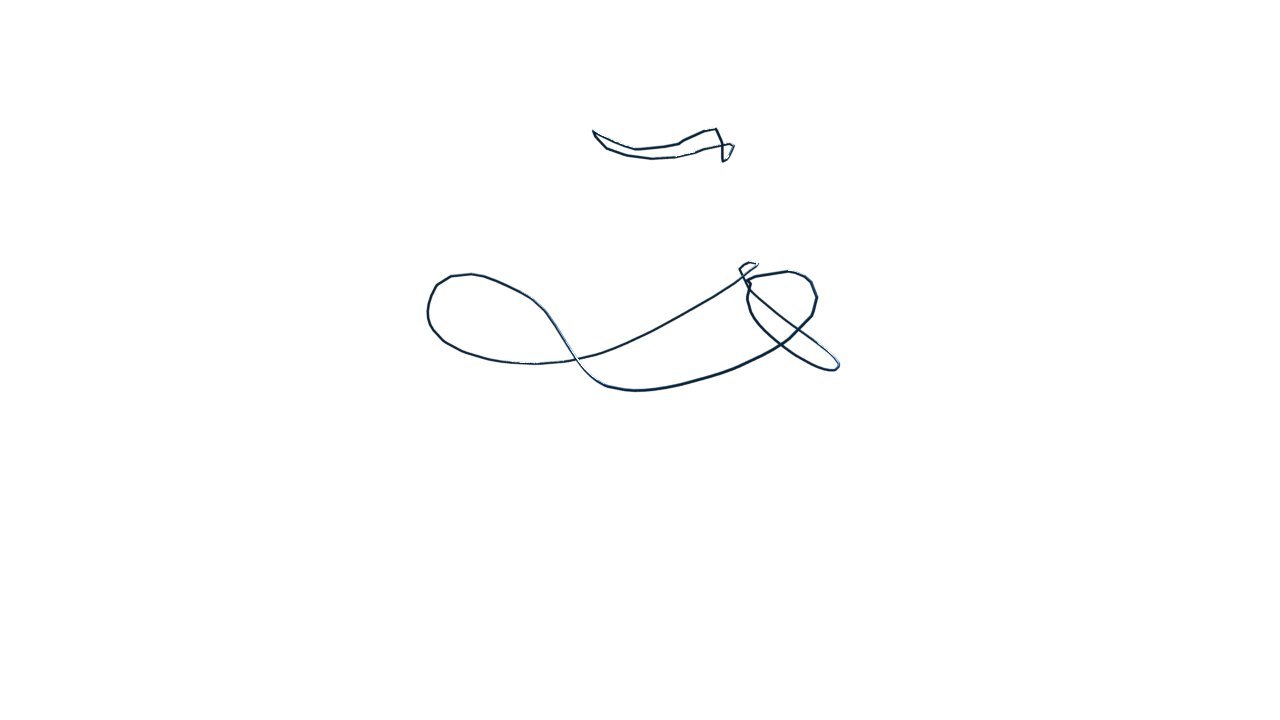}
        % {{images/trefoil/trefoil_WP.0150}.jpg}
    \end{overpic}
    \vspace{-20pt}
    \caption*{Wei\ss{}mann and Pinkall~\shortcite{Weissmann:2010:filaments}}

    \begin{overpic}[width=0.32\linewidth,tics=0,trim={12cm -2cm 12cm 8cm},clip]
        {{images/trefoil/trefoil.0001}.png}
        % {{images/trefoil/trefoil.0001}.jpg}
    \end{overpic}
    \begin{overpic}[width=0.32\linewidth,tics=0,trim={12cm 4cm 12cm 2cm},clip]
        {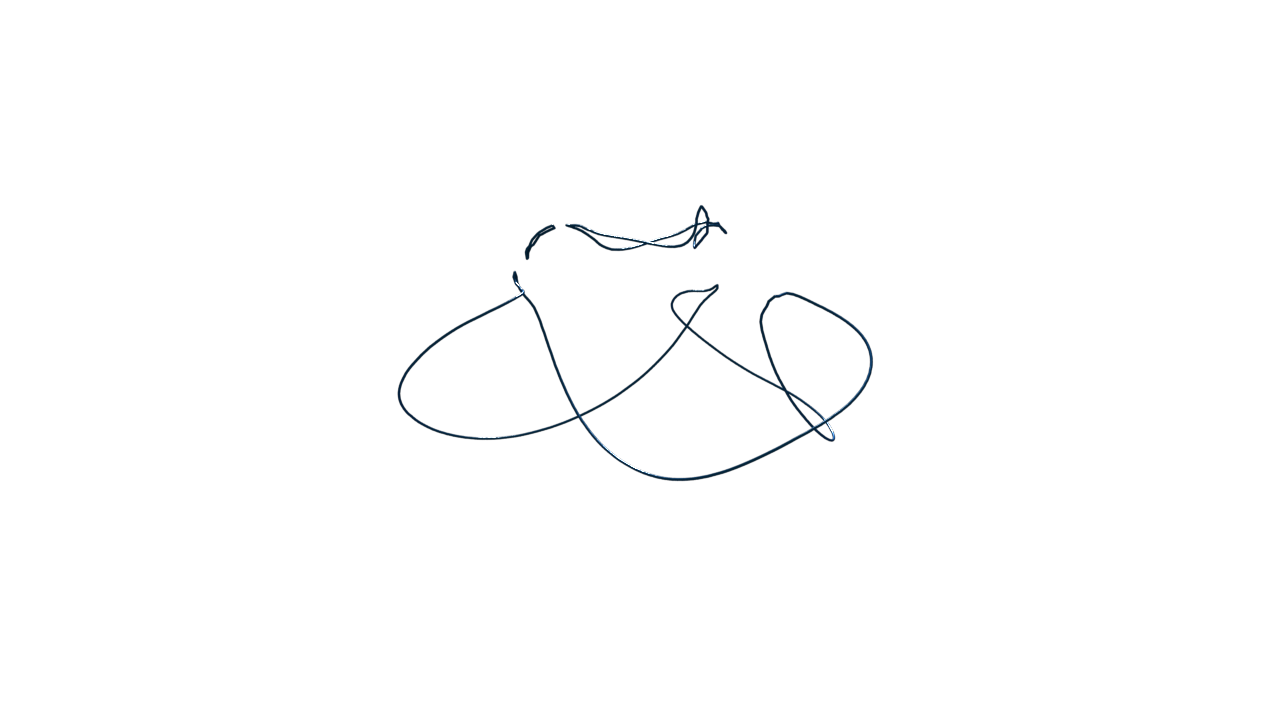}
        % {{images/trefoil/trefoil_ours.0125}.jpg}
    \end{overpic}
    \begin{overpic}[width=0.32\linewidth,tics=0,trim={12cm 4cm 12cm 2cm},clip]
        {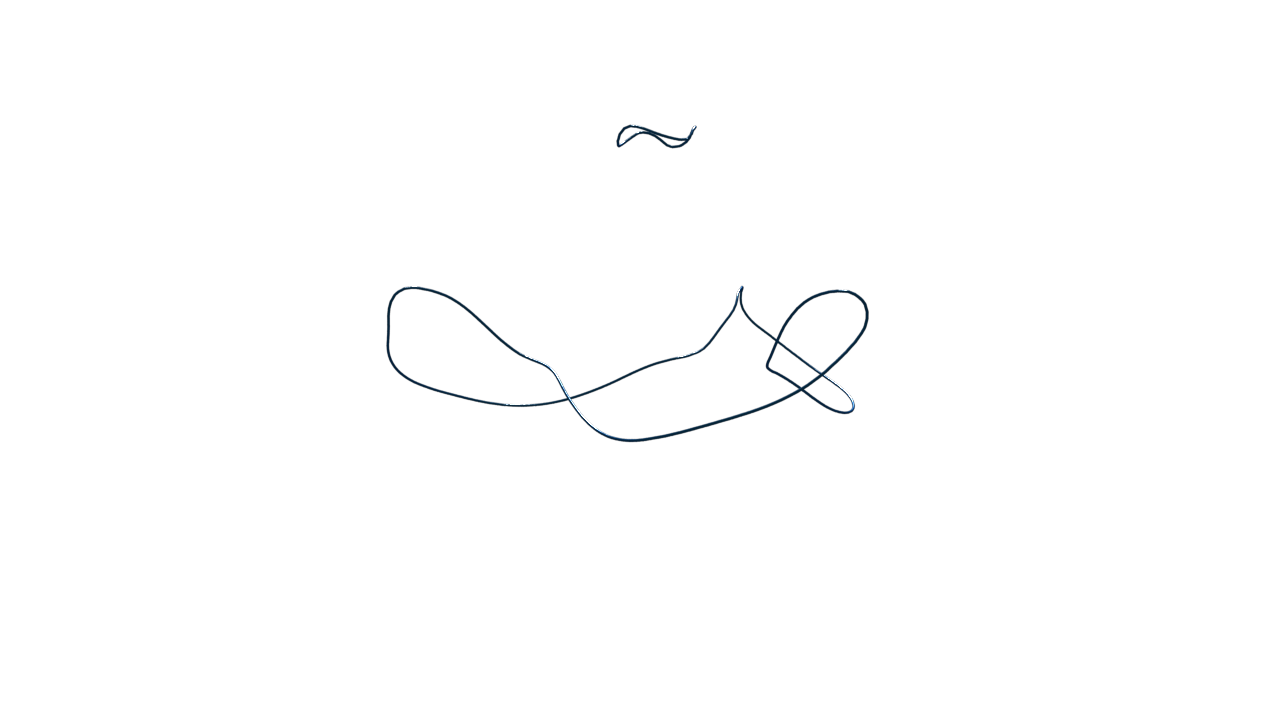}
        % {{images/trefoil/trefoil_ours.0150}.jpg}
    \end{overpic}
    \vspace{-20pt}
    \caption*{Our method}

    \caption{Comparing a buoyant trefoil knot simulation by Wei\ss{}mann and Pinkall~\shortcite{Weissmann:2010:filaments} to ours. The simulations evolve from left to right. %The left and right columns show frames  at 125 and 150.
    } 
    \label{fig:trefoil}
\end{figure}

\begin{figure}[htbp!] 
    \begin{overpic}[width=0.48\linewidth,tics=0,trim={10cm 3cm 10cm 2cm},clip]
            {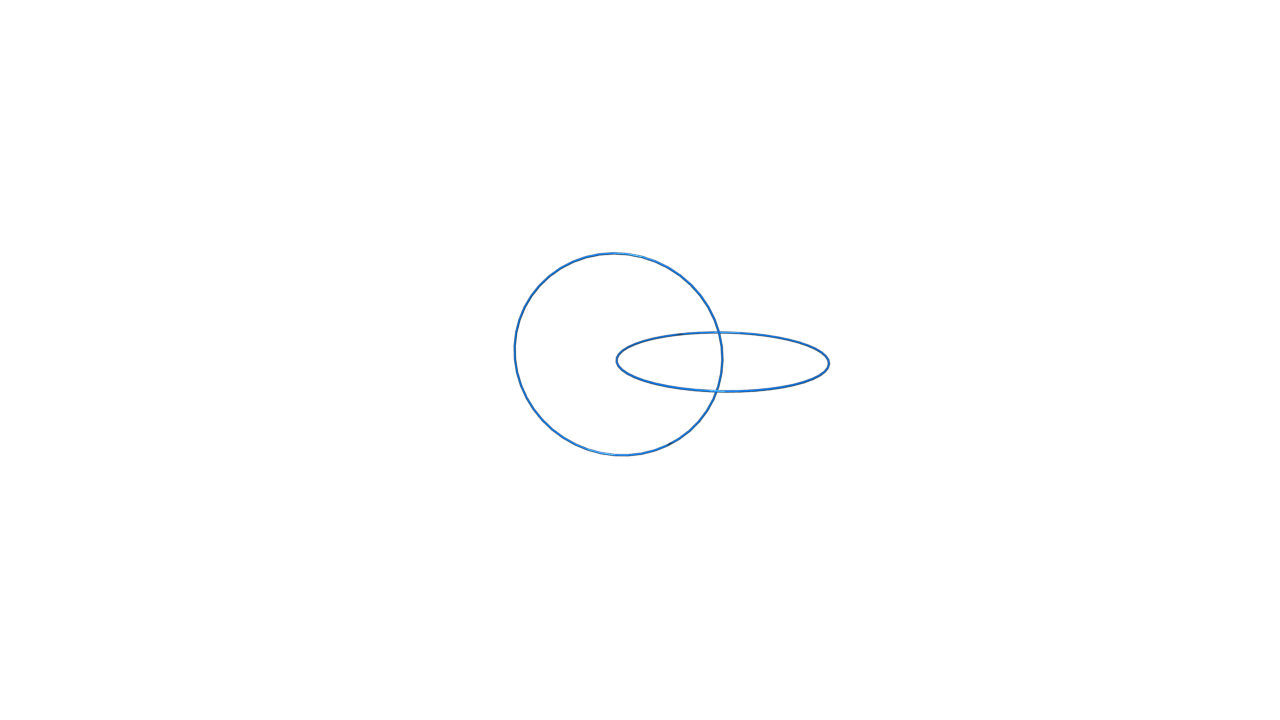}
            % {{images/twoRings/2rings_0001}.jpg}
            % \put(30,0){ \small Initial state}
    \end{overpic}
    \vspace{-20pt}
    \caption*{Initial state}
    % \medskip
    
        % \medskip
    \begin{overpic}[width=0.48\linewidth,tics=0,trim={10cm 3cm 10cm 2cm},clip]
        {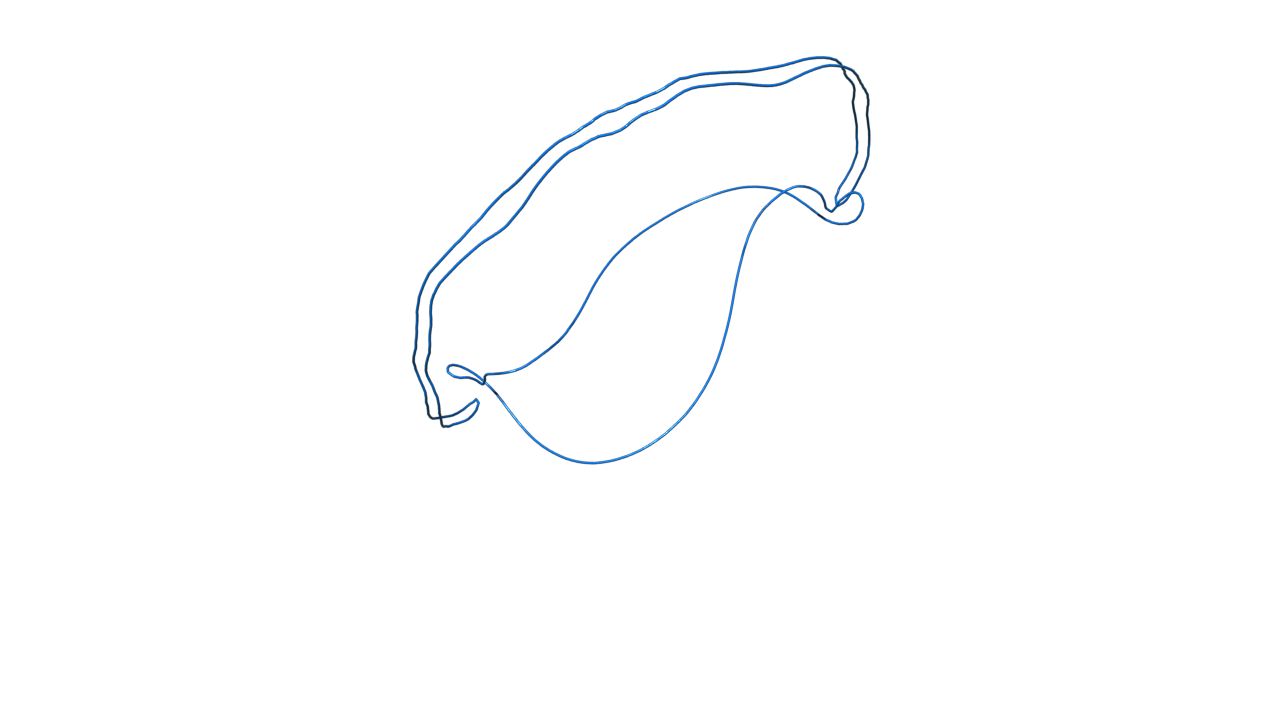}
        % {{images/twoRings/2rings_ours.0070}.jpg}
        % \put(20,0){ \small Ours Frame = 70}
    \end{overpic}
    % \hspace*{\fill}
    \begin{overpic}[width=0.48\linewidth,tics=0,trim={10cm 3cm 10cm 2cm},clip]
        {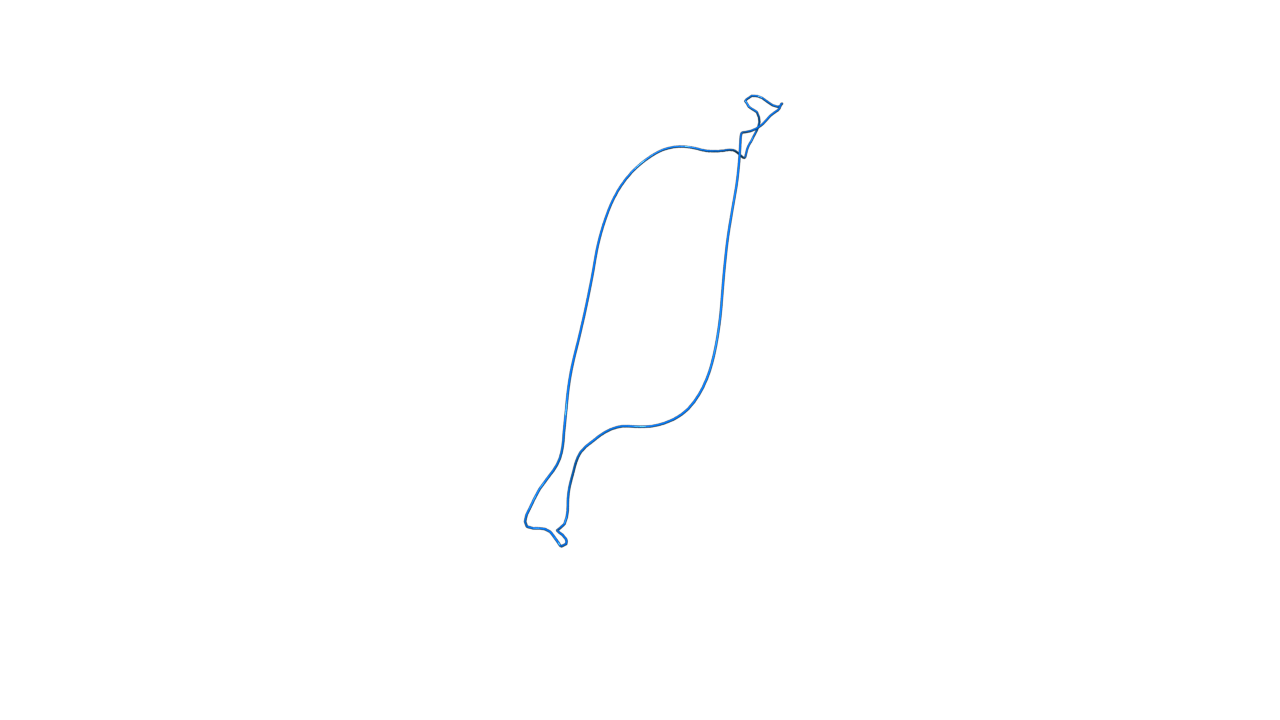}
        % {{images/twoRings/2rings_ours.0100}.jpg}
        % \put(20,0){ \small Ours Frame = 100}
    \end{overpic}
    % \hspace*{\fill}
    \vspace{-20pt}
    \caption*{Our method}

    \begin{overpic}[width=0.48\linewidth,tics=0,trim={10cm 3cm 10cm 2cm},clip]
        {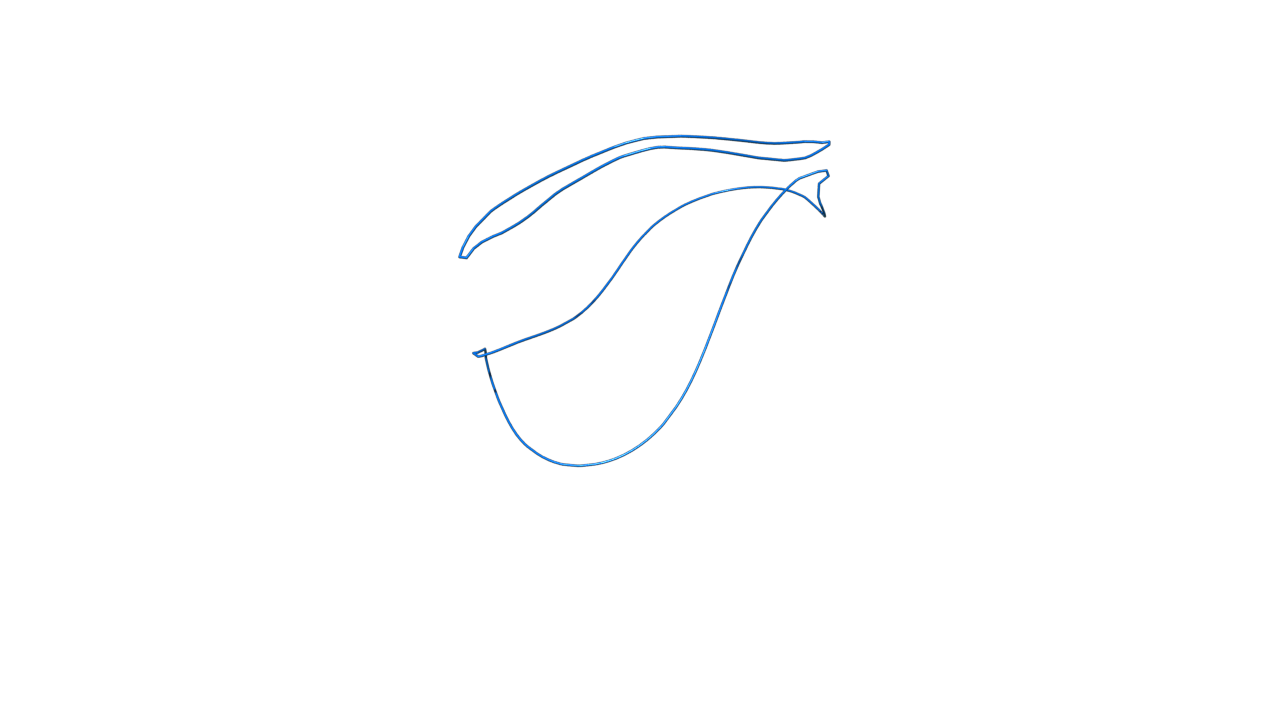}
        % {{images/twoRings/2rings_WP.0070}.jpg}
        % \put(20,0){ \small W\&P Frame = 70}
    \end{overpic}
    % \hspace*{\fill}
    \begin{overpic}[width=0.48\linewidth,tics=0,trim={10cm 3cm 10cm 2cm},clip]
        {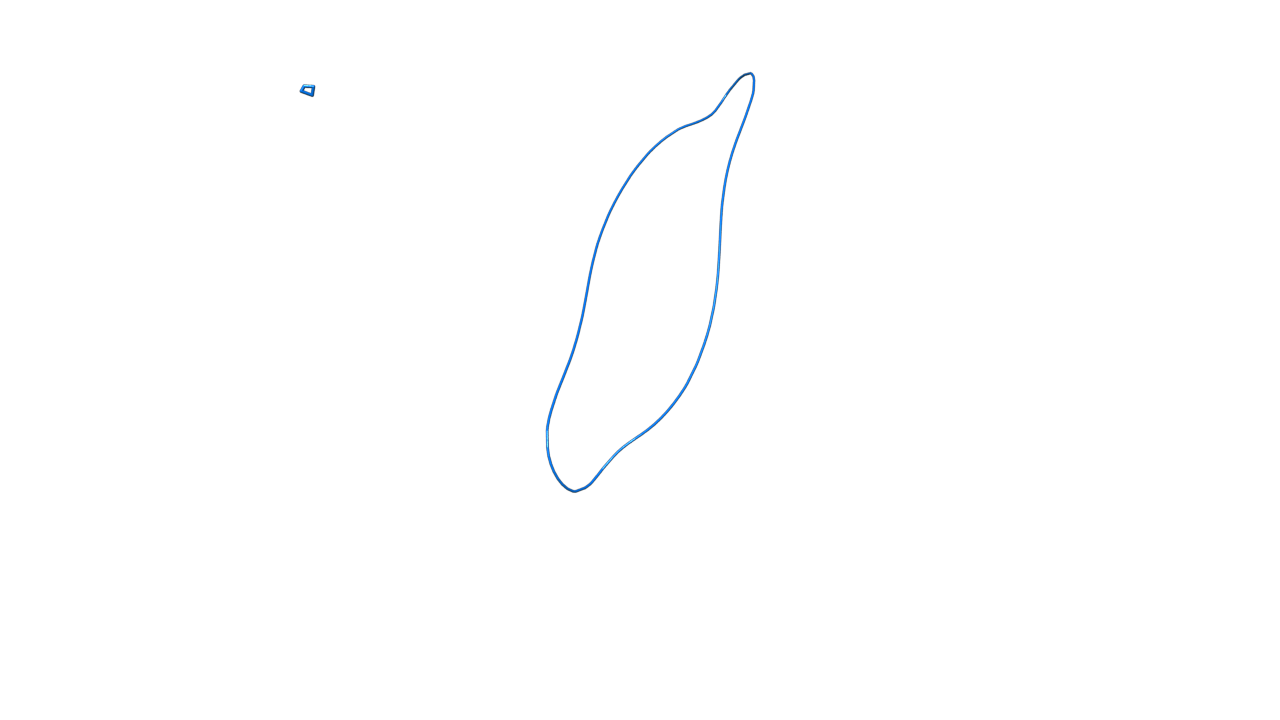}
        % {{images/twoRings/2rings_WP.0100}.jpg}
        % \put(20,0){ \small W\&P Frame = 100}
    \end{overpic}
    % \hspace*{\fill}
    \vspace{-20pt}
    \caption*{Wei\ss{}mann and Pinkall~\shortcite{Weissmann:2010:filaments}, {\em with aggressive re-connection} }

 % \medskip
    \begin{overpic}[width=0.48\linewidth,tics=0,trim={10cm 3cm 10cm 2cm},clip]
        {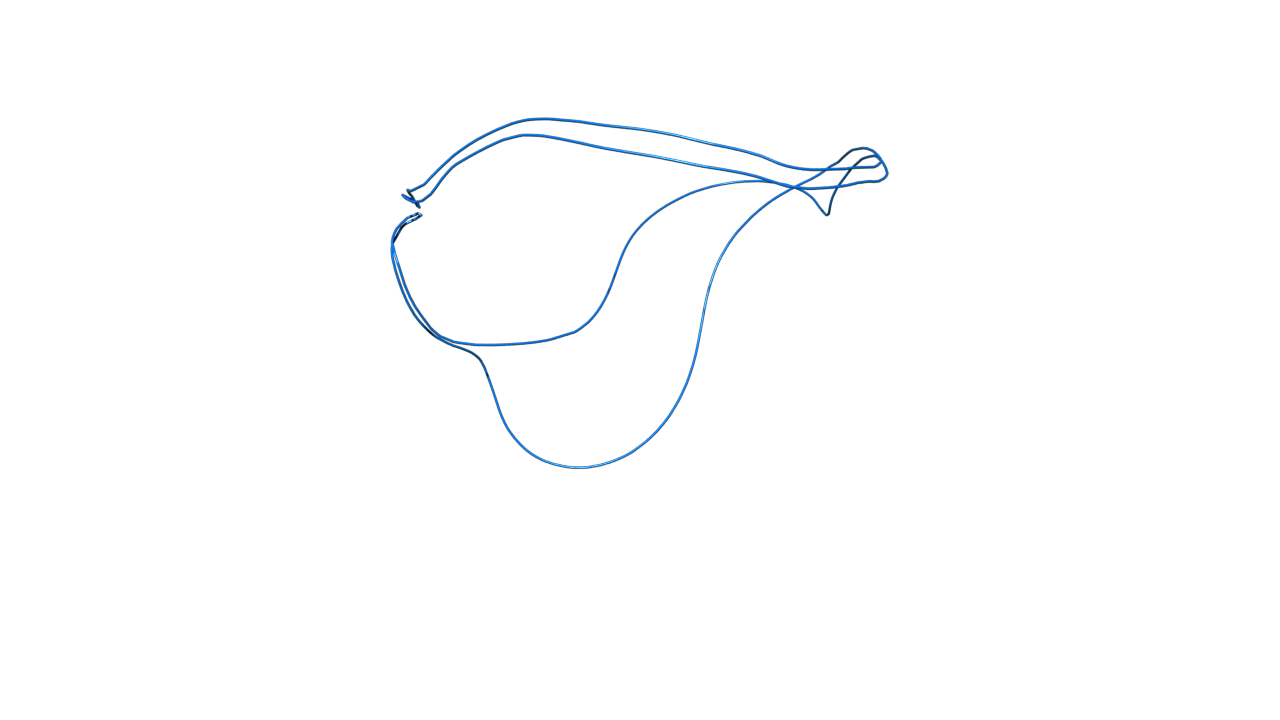}
        % {{images/twoRings/2rings.WP_sameDOF.0078}.jpg}
        % \put(20,0){ \small Ours Frame = 70}
    \end{overpic}
    % \hspace*{\fill}
    \begin{overpic}[width=0.48\linewidth,tics=0,trim={10cm 3cm 10cm 2cm},clip]
        {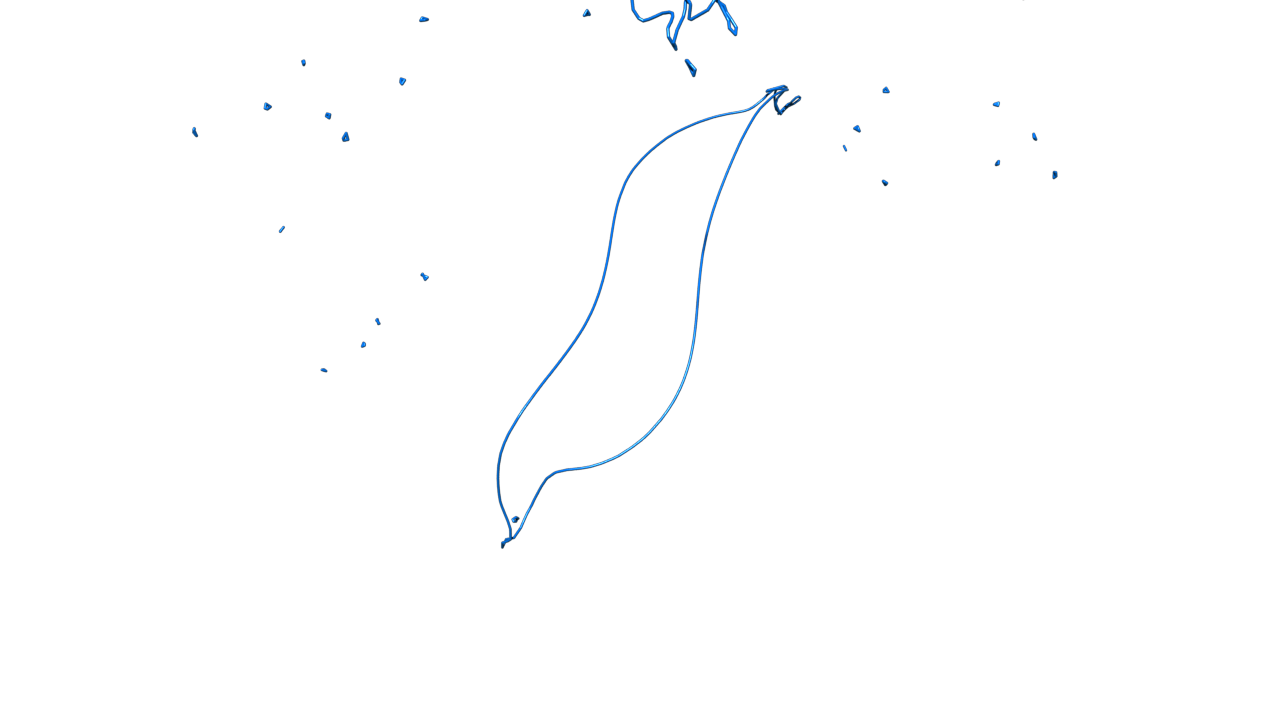}
        % {{images/twoRings/2rings.WP_sameDOF.0112}.jpg}
        % \put(20,0){ \small Ours Frame = 100}
    \end{overpic}
    % \hspace*{\fill}
    \vspace{-20pt}
    \caption*{Wei\ss{}mann and Pinkall~\shortcite{Weissmann:2010:filaments}, {\em with conservative re-connection} }
    % \caption*{}
    
    % \vspace{-10pt}
    % \caption*{\small Frame = 70 \hspace{80pt} Frame = 100}

    %\caption{Comparison with Wei\ss{}mann and Pinkall~\shortcite{Weissmann:2010:filaments} by the linked rings. The left and right columns show frames  at 70 and 100. } \label{fig:two_rings}
    \caption{Simulating two linked vortex rings (top) with our method ($2^{nd}$ panel), and with the explicit filament approach of Wei\ss{}mann and Pinkall~\shortcite{Weissmann:2010:filaments} (bottom two panels). Lagrangian methods can be sensitive to numerical parameters for topological changes.} \label{fig:two_rings}
\end{figure}
    
%\subsection{Comparisons with \cite{Weissmann:2010:filaments}}
\subsection{Comparisons with Lagrangian filaments}
Next, we \rev{qualitatively} compare \rev{implicit and explicit representaitons of curve dynamics using our algorithm and }\stt{our method for evolving implicit curves with}the Lagrangian vortex filament technique of Wei\ss{}mann and Pinkall~\shortcite{Weissmann:2010:filaments}, as implemented in Houdini software by SideFX.
 \figref{fig:trefoil} shows the evolution of a knotted vortex filament with both methods. The filament is initialized as
 \begin{align}
    \gamma(s)=\left(\sin(s)+2\sin(2s),\cos(t)-2\cos(2s),-\sin(3s)\right) \notag
 \end{align}
%${\gamma(s)=\left(\sin(s)+2\sin(2s),\cos(t)-2\cos(2s),-\sin(3s)\right)}$
with \(s\in [0,2\pi)\).
 Aside from some small differences arising from the particulars of how filaments break apart and reconnect, the two methods produce roughly the same dynamics. 
 
%On the other hand, the Lagrangian technique can create undesirably complex thin features when the curves are not cleanly aligned with each other at the start. 
On the other hand, the two methods have significantly different mechanisms for handling topological changes, which can produce divergent results. The Lagrangian method depends on user parameters like the thresholds for distances and angles between curves; the only relevant user parameter for changing topology in our method is the grid spacing, which prescribes the resolution of the level set $\psi$. \figref{fig:two_rings} shows a simulation of two linked rings: our approach both preserves long thin tendrils and filters out topological noise. The Lagrangian simulation is sensitive to re-connection parameters: setting these parameters too aggressively leads to smooth geometry but loses thin features, while setting parameters too conservatively preserves thin features but creates noisy, persistent, high-speed ``ringlets'' that dominate the fluid velocity field. 
% \cCW{I think the discussion about whether these particular artifacts are physical or not is a bit too subtle and accusatory without another test to really draw a strong conclusion. I personally think it is wrong to claim that this noise is unphysical without more evidence, but I also think it is wrong to make a benevolent and completely neutral statement like ``these are not necessarily unphysical'' because I think we can reasonably conclude that they are indeed unphysical since they depend so strongly on numerical resolution. Another test to determine that they are indeed the result of numerical parameters would solve this issue, but I don't think we should get this detailed about bashing previous work here. I'd rather just leave it alone... we did not claim they are unphysical, so we don't need to defend anything we did. So I'd like to remove the proposed statement ``although these parameter sensitivities do not necessarily mean unphysical''. The following sentence nicely sums up the point anyway. What do you think?}
% \cSI{I agree.}
%\rev{We again note that this is a qualitative comparison between methods, and that Note that although these parameter sensitivities do not necessarily mean unphysical}. 
\rev{We stress here that the purpose of this comparison is to show how explicit and implicit descriptions handle topological changes of curves result in qualitatively different ways; the accuracy of these methods are not easily comparable as they have different mechanisms, and the accuracy depends on the type of curve dynamics. }

\subsection{Obstacles}
%\cSI{It might be better to move this subsection to the Results section.}
%\cAC{I think so too.  There is no mentioning of fluids in the algorithm.  If there is a section where we can spell out the evaluation of Biot--Savart or Rosenhead--Moore, then we can insert a short subsubsection or paragraph. In that short subsubsection we just need to note that 
%\begin{itemize}
%    \item Note that the Biot--Savart and Rosenhead--Moore are explicit solution in the free space (no obstacle).
%    \item General solution when there is an obstacle is \(V_\gamma = V_\gamma^{\rm RS} + \nabla\phi\) s.t. \(\Delta\phi = 0\), \(U_\gamma^{\rm RS} + \nabla\phi\) is tangential to the boundary, and \(\phi\) is asymptotically zero at infinity.
%    \item Previous approaches are based on boundary element method (Weissmann and Pinkall). We adopt Kelvin transform to solve for this \(\phi\) using finite element on an inverted volumetric domain.
%\end{itemize}
%}

Like many methods for vortex dynamics, our method can also make filaments circumvent obstacles. 
A typical approach is to find a smooth harmonic potential \(\phi\) such that for a given obstacle \(B\subset M\), it solves 
%\cAC{\(\phi\) is harmonic}
\begin{align}\label{eq:pressure_projection}
     \langle \bv_\gamma \rev{ -\bv_{\partial B}}-\nabla \phi, \bn\rangle_{\RR^3}=0 \text{ on } \partial B \\
     \phi\rightarrow 0 \text{ at infinity} 
\end{align}
\rev{where \(\bv_{\partial B}\) is the boundary velocity at each point of \(\partial B\)}.
%There are several existing solvers for this problem. 
Wei\ss{}mann and Pinkall \shortcite{Weissmann:2012:UnderWaterRigidBody} construct such a potential by regarding points inside \(B\) as sources of localized potentials so their weighted sum solves \autoref{eq:pressure_projection}, and Brochu et al. \shortcite{brochu2012linear} and Zhang et al. \shortcite{zhang2014pppm} solve a similar system using boundary element techniques.
 Nabizadeh et al. \shortcite{nabizadeh:2021:kelvin} address an equivalent problem by solving linear PDEs on infinite domains using the Kelvin transform.%\cAC{It's not that general...} 

Another approach by Wei\ss{}mann and Pinkall \shortcite{Weissmann:2010:filaments} is more specialized to vortex filaments \rev{and phenomena like vortex shedding}; it regards the obstacle as a collection of artificial filaments \(\gamma_M\) such that the normal component of the velocity is zero.
%\begin{align}
 %    \langle \bv_\gamma+\bv_{\gamma_M}, \bn\rangle_{\RR^3}=0 \text{ on } \partial B.
%\end{align}
Other approaches, like that of Park and Kim \shortcite{park2005vortex} and Da et al. \shortcite{da2015double} add point constraints to the boundary which zero out both the normal and tangential velocity components.
 
\begin{figure}[t]
    \begin{minipage}[b]{1.0\linewidth}
        \centering
        \includegraphics[width=0.9\linewidth]{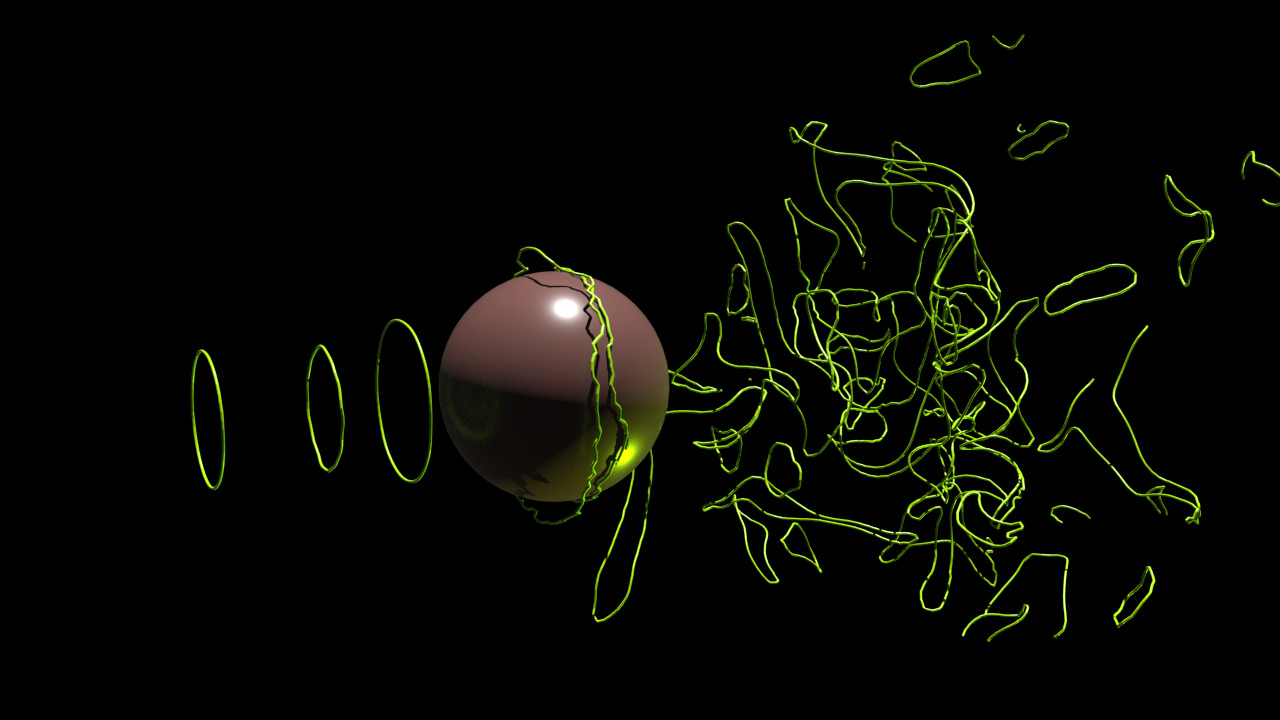}
    %   \caption{obstacle curves}
    \end{minipage}
    \begin{minipage}[b]{1.0\linewidth}
        \centering
        \includegraphics[width=0.9\linewidth]{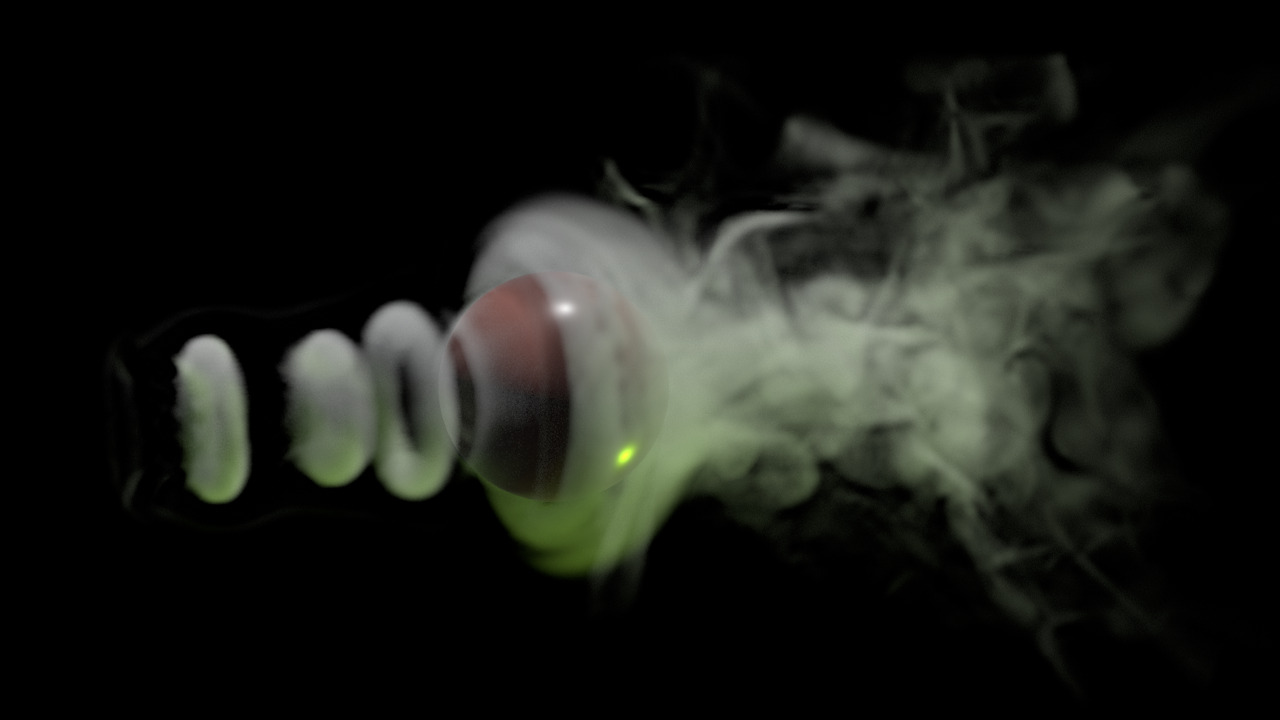}
    %   \caption{obstacle}
    \end{minipage}
    \caption{Turbulence caused by a spherical obstacle. We generate a new vortex filament ring every 10 frames.}
    \label{fig:obstacle}
\end{figure}

Our vortex filament algorithm is compatible with any of these obstacle-handling methods; our particular implementation uses \cite{nabizadeh:2021:kelvin}, as seen in \figref{fig:obstacle}. \rev{We can observe that filaments near the obstacle are accelerated due to the induced mirrored image of themselves in the obstacle (or vortex sheet on the surface). 
}

\rev{Note that for the evolution of curves, only the evaluation of the velocity field needs the treatment of the boundary. Other components of the algorithm including the construction of levelset function ignore the boundary and do not require any additional treatments. }

\begin{figure}
    \centering{
    \begin{subfigure}[b]{0.24\linewidth}
        \centering
        \includegraphics[width=\linewidth,trim={16cm 8cm 14cm 8cm},clip]{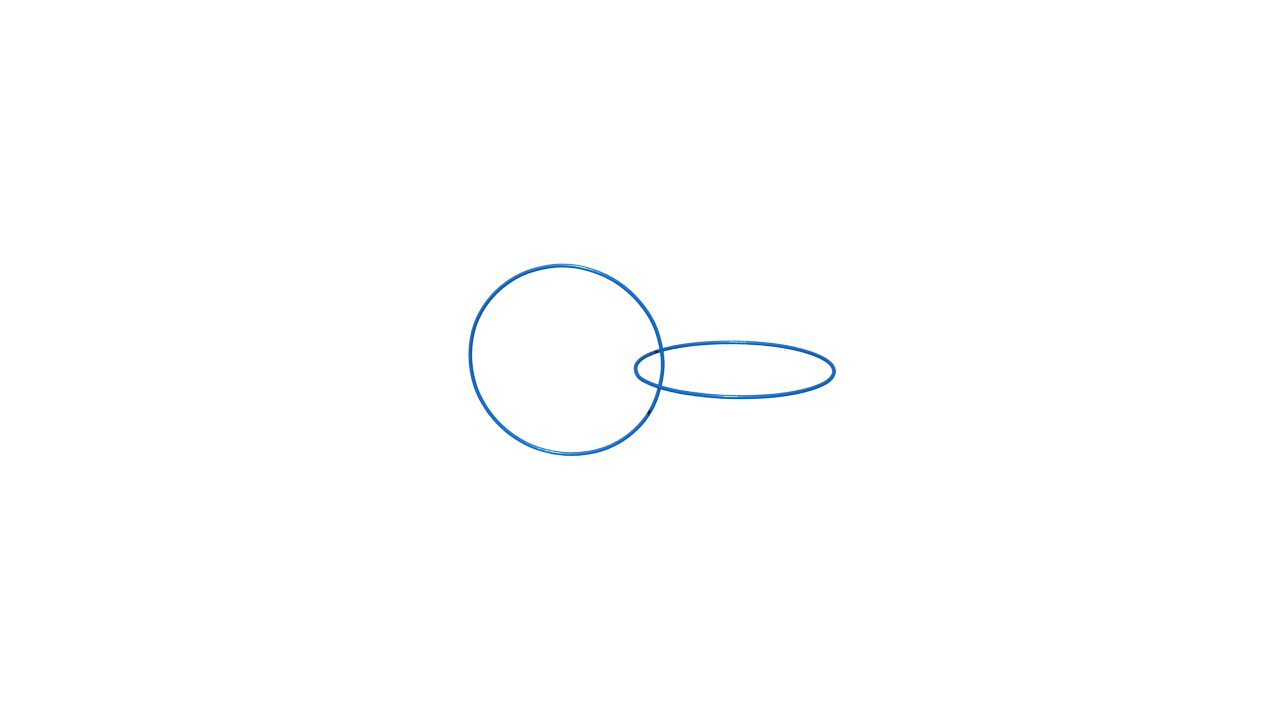}
        % \includegraphics[width=\linewidth,trim={16cm 8cm 14cm 8cm},clip]{{images/twoRings/2rings_curvature_flow_big.0090}.jpg}
        % \caption*{Frame = 90}
    \end{subfigure}
    \hfill
    \begin{subfigure}[b]{0.24\linewidth}
        \centering
        \includegraphics[width=\linewidth,trim={16cm 8cm 14cm 8cm},clip]{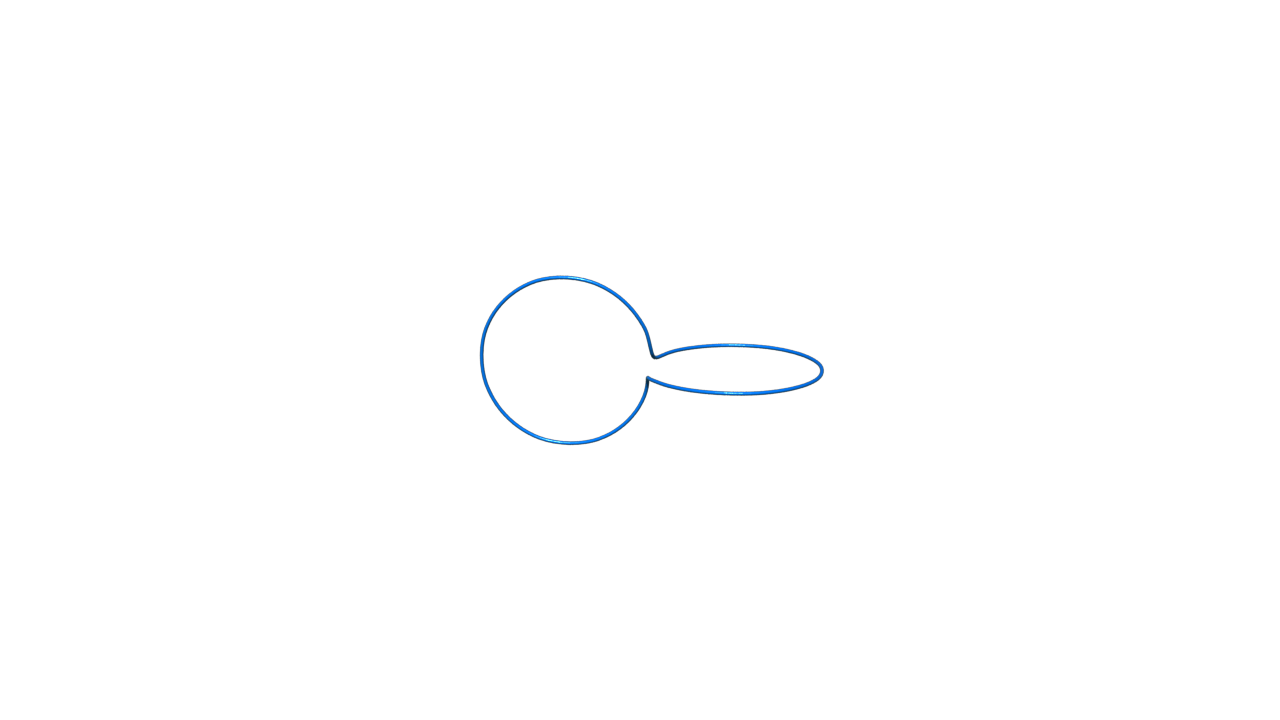}
        % \includegraphics[width=\linewidth,trim={16cm 8cm 14cm 8cm},clip]{{images/twoRings/2rings_curvature_flow_big.0100}.jpg}
        % \caption*{Frame = 100}
    \end{subfigure}
    \hfill
    \begin{subfigure}[b]{0.24\linewidth}
        \centering
        \includegraphics[width=\linewidth,trim={16cm 8cm 14cm 8cm},clip]{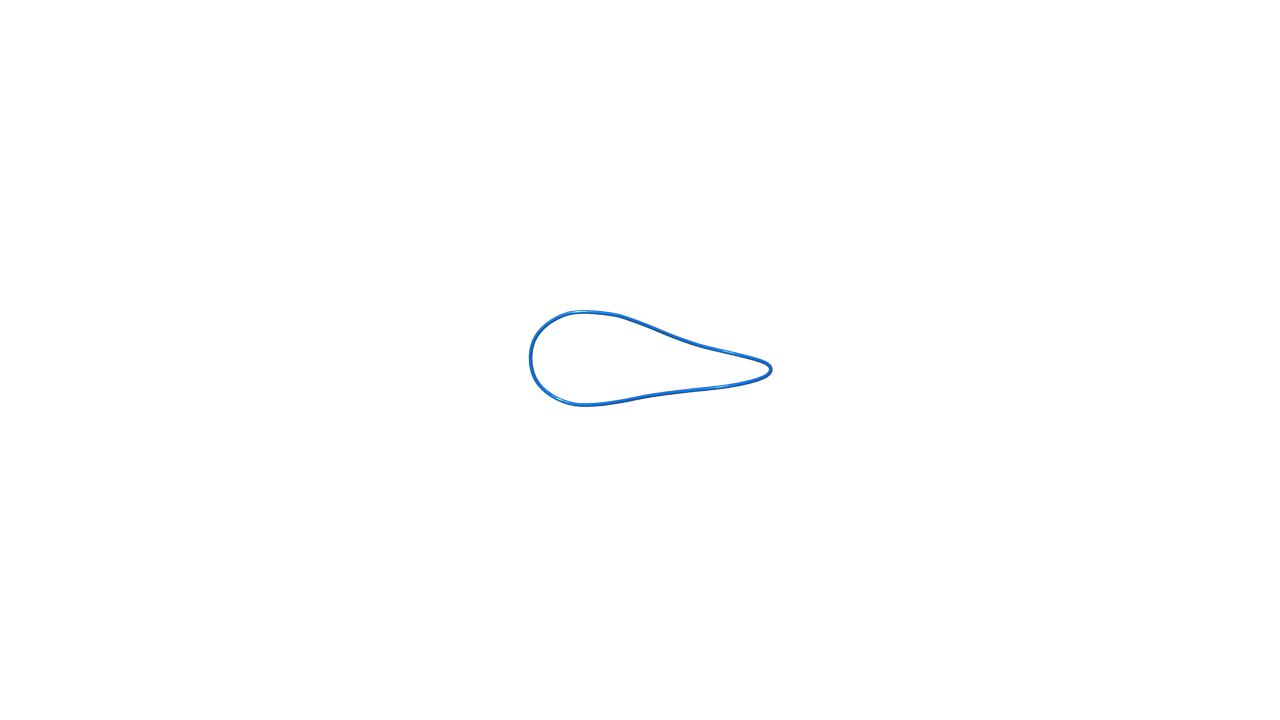}
        % \includegraphics[width=\linewidth,trim={16cm 8cm 14cm 8cm},clip]{{images/twoRings/2rings_curvature_flow_big.0130}.jpg}
        % \caption*{Frame = 130}
    \end{subfigure}
    \hfill
    \begin{subfigure}[b]{0.24\linewidth}
        \centering
        \includegraphics[width=\linewidth,trim={16cm 8cm 14cm 8cm},clip]{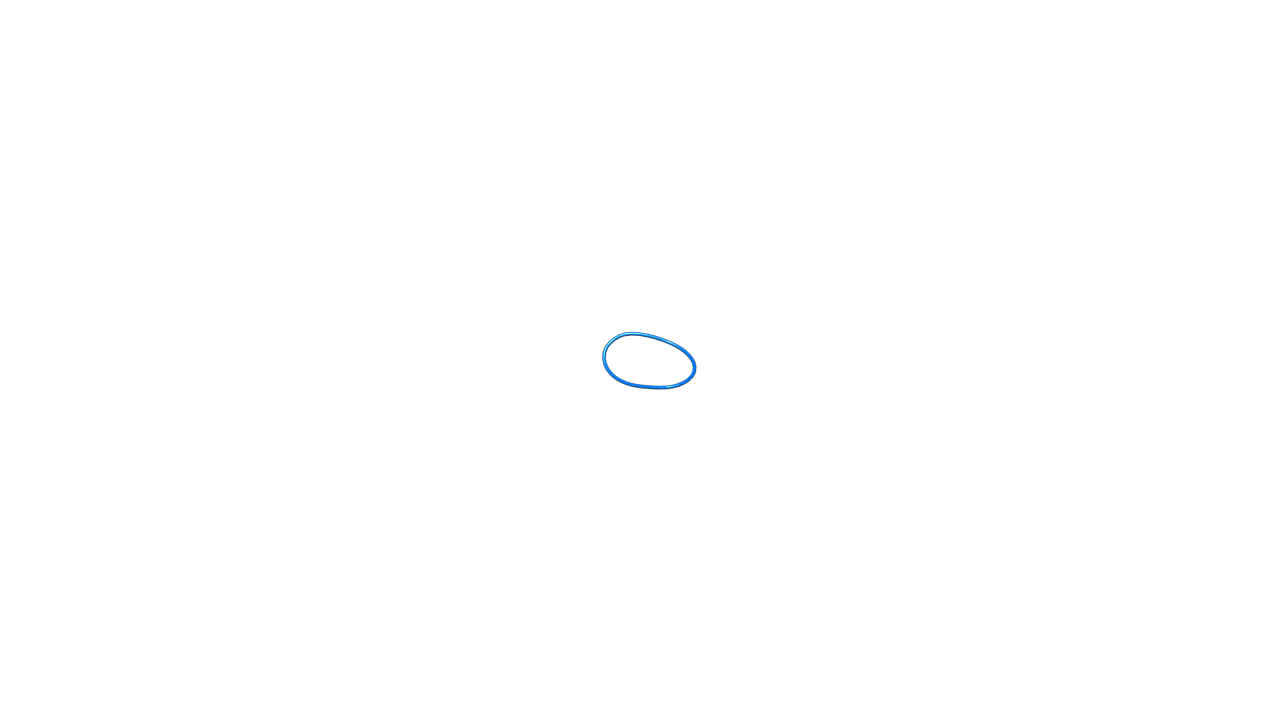}
        % \includegraphics[width=\linewidth,trim={16cm 8cm 14cm 8cm},clip]{{images/twoRings/2rings_curvature_flow_big.0155}.jpg}
        % \caption*{Frame = 130}
    \end{subfigure}
   
    }

    \caption{
        Curves evolving under the curve shortening flow, from left to right.
        %From left to right are frames 90, 100, 130, and 155. We used a finer time step \(1/120 \approx 0.0083\) than vortex filaments as points in different curves can have very different velocities even if they are very close.
        % \cCW{I think it will illustrate the flow better if we add one more figure to the right (even if it is a boring little circle).
    }
    \label{fig:curve_shortening_flow}
\end{figure}

\subsection{Other types of filament dynamics}
Our idea of implicit representation of filament dynamics is not limited to vortex filaments. In theory, this should be applicable to any first-order time evolution of curves. \autoref{fig:curve_shortening_flow} shows an example with the curve-shortening flow,
\begin{align}
    \frac{\partial\gamma}{\partial t}(s,t)=-\gamma^{\prime\prime}(s,t).
\end{align}
As expected, this velocity causes the filament to shrink over time, similar to mean-curvature flow for surfaces~\cite{osher1988fronts}.

\arrayrulecolor{black}
\begin{table*}[htbp!]
    \caption{Parameters and timing breakdown per frame for all simulations in this paper and the accompanying video. Symbols \(\Gamma\) and \(a\) are the intensity and the thickness of filaments. All simulations are 24fps and and the time step size \(1.0/24\)s except for \figref{fig:curve_shortening_flow}, which we used 120fps. Average timings are taken over the entire simulation. The ``Other'' column includes the remaining operations, including obstacle handling and generation of new curves at the sources in \figref{fig:jet} and \figref{fig:obstacle}. }

    \centering
    \begin{tabular}{|c|| c | c |c ||c|| c|c|c|c|c|c|} 
     \hline
     Scene & Resolution & \(\Gamma\) & \(a\)  & Total time & Evaluate \(\bv\) &  Advect \(\psi\) & Construct \(\gamma\) & Construct \(\psi\) & Other\\ 
    %  [0.5ex] 
     \hline
     \hline
     \figref{fig:two_rings}  & \(100 \times 100 \times 100\) & 1.0 &0.05 & 0.159s & 0.041s  & 0.017s &0.041s& 0.056s & 0.002s \\ 
     \hline
     \figref{fig:trefoil}   & \(100 \times 100 \times 120\) & 1.0 &0.05 & 0.274s  & 0.056s  & 0.026s &0.049s& 0.102s & 0.001s\\ 
     \hline
     \figref{fig:leapfrog}   & \(60 \times 60 \times 200\) & 2.0 &0.08 & 0.090s  & 0.013s &0.021s & 0.023s & 0.031s & 0.001s\\ 
     \hline
     \figref{fig:oblique_rings}   & \(80 \times 120 \times 140\) & 1.0 &0.05&  0.152s  & 0.031s  & 0.030s & 0.046s & 0.044s & 0.001s\\ 
     \hline
     \figref{fig:torus}   & \(100 \times 100 \times 140\) & 1.0 &0.05 & 0.655s & 0.144s  & 0.085s & 0.091s & 0.332s & 0.002s \\ 
     \hline
     \figref{fig:jet}   & \(100 \times 100 \times 180\) & 2.0 &0.08 & 1.423s  & 0.177s  & 0.099s & 0.812s & 0.324s & 0.011s\\ 
     \hline
     \figref{fig:obstacle}   & \(100 \times 100 \times 160\) & 2.0 &0.08 & 0.969s  & 0.047s  & 0.087s & 0.473s & 0.226s & 0.135s\\ 
     \hline
     \figref{fig:curve_shortening_flow}  & \(100 \times 100 \times 100\) & N/A &N/A & 0.152s  & 0.028s  & 0.012s & 0.068s & 0.043s & 0.001s\\ 
     \hline
    \end{tabular}
    \label{table:parameters}
\end{table*}

\begin{table}[hbtp!]
    \caption{Computational timings compared with Houdini's  built-in implementation of Weissmann \& Pinkall \shortcite{Weissmann:2010:filaments}. ``Same DOF'' refers to simulations with approximately the same number of computational degrees of freedom as our method: we set the relevant parameters (re-connection distance, minimum and maximum edge lengths ) so that the number of explicit curve vertices are similar to ours.
    % For the same DOF comparisons, we set #Vertices = 320 for the linked rings and #Vertices = 640 for the trefoil.
    }
    \centering
    \begin{tabular}{|c|| c |c|c|} 
     \hline
     Scene description &  W\&P &  W\&P (Same DOF) &  Ours  \\ 
    %  [0.5ex] 
     \hline
     \hline
     Linked rings (\figref{fig:two_rings}) & 0.052s &  0.065s& 0.159s  \\ 
     \hline
     Trefoil knot (\figref{fig:trefoil} & 0.045s & 0.101s&0.274s  \\ 
     \hline
    \end{tabular}

    \label{table:comparison_with_WP}
\end{table}

\subsection{Discussion}

Our method represents a fundamental new way to animate fluids and vortex filament dynamics. Its unique features give it some strengths and weaknesses relative to existing approaches. 

First of all, we represent our filaments with implicit functions $\psi$. While there are many ways to encode an implicit function, our implementation uses a regular grid, which implies a finite bounding box. While common for Eulerian fluid simulations, bounded domains are a constraint not shared by Lagrangian methods. This constraint could be mitigated if we use sparse grids or trees~\cite{museth2019openvdb}.

The main parameter in our method is the grid resolution. As discussed earlier, this parameter affects the geometric detail and topology of our filaments, influencing the velocity field directly, but influencing the final visual results only indirectly. 
%Effect of grid resolution on results: Curvature of curve limited by grid resolution: higher resolution grid = higher frequency levelset information (just like level set surfaces). But in our case, we don't directly visualize the curve; we visualize smoke instead. Low resolution filaments still produce high resolution visual detail (would be nice to have a figure or sim that showed this). \cCW{resolution discussed earlier}
The grid resolution also directly influences topological changes; the only mechanism for topological changes in our algorithm is to merge curves when they intersect the same grid cell. We believe this automatic and robust method for handling topological changes is a strength of our method; it minimizes the need to fine-tune parameters (especially the relationship between minimum/maximum edge length and topological change interaction lengths) and seems immune to the types of numerical blow-ups that we have seen in Lagrangian methods when filaments get close or exhibit near-degenerate geometry. 

On the other hand, this grid-based method for re-connecting curves will also delete small features when they shrink below the grid cell size (similar to level set methods for surfaces). This behavior is most evident in our jet example (\figref{fig:jet}). We believe that the deletion of small features can be reduced in the future in a number of ways. More accurate advection schemes will probably preserve higher frequency features of $\psi$ better without deleting them. Also, although we took care to introduce a $\bv$ and $\psi$ which alleviates egregious numerical stability problems, we have not quite optimized for accuracy or geometric durability for the specific dynamics of vortex filaments. Finding a pair of $\bv$ and $\psi$ that is more suitable for each curve dynamics in the degrees of freedom discussed in \secref{sec:representation} would lead to even better numerical performance.
% we have not quite optimized for accuracy or geometric durability. The free parameters discussed in \secref{sec:representation} could likely be optimized for even better performance.

Although \stt{or}\rev{our} method relies on Eulerian advection to evolve the geometry, it {\em does  not} suffer from the artificial viscosity typically associated with Eulerian fluid simulations. Our fluid velocity is reconstructed from filament dynamics, so the velocity field is not recursively re-sampled and does not accumulate damping errors. 
Consequently, our method produces swirly and energetic fluid flows even at low grid resolutions. 
% Our level set approach for handling topology changes, however, will delete small filament loops if they become smaller than a grid cell.

\tabref{table:parameters} lists the simulation parameters and timing breakdown for each of the simulations in our paper and accompanying video. We stress that our prototype implementation is meant as a proof of concept, and it has plenty of room for optimization. Our current implementation employs regular grids for ease of implementation; future implementations can make great use of sparse grids, since our filaments only use a 1-dimensional path through the 3D grid. Our implementation of ``Evaluate \(\sttEqn{\gamma}\rev{\bv}\)'' and ``Construct \(\psi\)'' iterates over the entire curve geometry for each point in the narrow band where \(\gamma\) and \(\psi\) are required; future implementations could use \rev{a sparse data structure} and \rev{approximated} fast summation \stt{or}\rev{via} tree-codes like the fast multipole method to reduce total evaluation time. Additionally, our current implementation of ``Construct \(\gamma\)'' redundantly doubles the work per curve vertex, so that step can be sped up by at least a factor of 2. 
% \rev{Such implementation efforts would allow for large-scale reconnection events like dense space-filling filaments in turbulent flows.}
% \cCW{I think we should remove this statement. It addresses a reviewer concern, but it is pure speculation. We could be wrong and will look silly later. I think it is obvious that speeding it up will let us do more detailed flows, but we shouldn't make a detailed claim like this since we don't know.}
% \cSI{Agreed. I removed the sentence.}

With these inefficiencies in mind, our implementation appears to run modestly slower than Houdini's optimized implementation of Weissmann \& Pinkall \shortcite{Weissmann:2010:filaments} for the scenarios we tested. The two methods have completely different numerical degrees of freedom, so we find it difficult to compare them directly.
\tabref{table:comparison_with_WP} compares the computational cost of the two methods for figures in this paper. 
We aimed to keep the number of curve vertices roughly the same as our method's in the ``Same DOF'' simulations, so we believe these are the most relevant for comparing timings. 

\rev{In our examples of vortex filaments,
we assumed inviscid filaments, but we can handle viscous motions too once the time-derivative of viscous curves is speficied.}
\rev{We also assumed a uniform vortex strength. Just as in the explicit description, our implicit description requires a well-defined equation of motion for curves, and it outputs only new curve configurations without processing additional quantities like vortex strengths. Simulating vortex filaments with different strengths is an independent challenge. 
%To tackle a question such as how to define the strengths of merged filaments with different circulations or how to implicitly represent them is an interesting direction. 
Merging filaments with different strengths would create a graph of filaments rather than just disjoint closed curves, which would require a new implicit representation for filament graphs. }

%% file: conclusion.tex
\section{Conclusion and future work}
\label{sec:conclusion}
%\cSI{Great writing, thanks!}
We have shown that implicit representations of geometric curves exhibit large degrees of freedom in both their mathematical representation, as well as their dynamics. We then took advantage of these redundant degrees of freedom to improve the stability of vortex filament simulations. 

We see a number of avenues that can be explored in future work. 
%Our current strategy exploits redundancy in the normal and bi-normal components of the filament velocity field, but we can also consider the 
Our current strategy exploits redundancy in the 3D velocity field and level set outside the filaments, but we can also incorporate free degrees of freedom in the tangential components of the velocity on the filament itself.
%\cSI{The normal and bi-normal components on the filaments are constrained. The redundancy we are exploiting in our specific choice of the velocity fields is the full degree (3 dim) of freedom outside the filaments. But we can even modify the velocity on the filaments in the tangent direction (1 dim for each point, thus gives us infinite degrees of freedom). This can impose optimization problems like "can we minimize some norm of the velocity field?" This is probably harder than the problem constraining the velocity on the filaments but the solution may be a nicer / more regular velocity field.}
More generally, one can attempt to formalize the regularization of $\bv$ and $\psi$ for various purposes. A possible instance is an optimization problem for certain energies aiming provable guarantees on numerical accuracy and stability. Besides, the use of redundancy in dynamics should also be possible for other codimensional cases such as level surfaces in 3D or level sets of an arbitrary codimension in a higher dimensional space.

Our numerical scheme can be made more sophisticated as well: higher order advection schemes and geometric curve representations, as well as sparse and adaptive grids can make our method both more efficient and more numerically accurate. 

Finally, this paper explores vortex filament\rev{s} and curve shortening flows, but our ideas are not limited to these specific dynamical systems. We expect that the idea of exploiting hidden degrees of freedom in implicitly represented curve dynamics will generalize to many more types of dynamics appearing in both scientific fields and engineering applications.

\begin{acks}
We thank the visual computing group at IST Austria for their valuable discussions and feedback. Houdini Education licenses were provided by SideFX software. This project was funded in part by the European Research Council (ERC Consolidator Grant 101045083 {\em CoDiNA}).
\end{acks}

%% file: appendix.tex
\appendix

\iffinalversion
\else
\color{red} 
\fi
\section{Properties of the untwisted Clebsch variables} \label{appendix:property_untwisted_Clebsch}
\iffinalversion
\else
\color{black}
\fi
\rev{
As explained in Section \ref{sec:UntwistedClebschVariables}, our untwisted Clebsch variable is composed of the distance function and the solid angle of projected curves. It has following desirable properties for curve dynamics:
\begin{enumerate}
    \item it has an explicit formula and is uniquely computed from given curves;
    \item it is locally little twisted and \(d\psi\vert_{\gamma^\bot}\) is locally nearly-isometry;
    \item it is harmonic;
    % i.e. minimizes the Dirichlet energy \(\int_{M\setminus \gamma} |\nabla \theta|^2 dx\) among functions in the equivalence class;
    \item the zeros of the real and the imaginary parts intersect orthogonally, which ensures the zeros to be always codimension-2 and makes numerically extracting zeros robust. 
\end{enumerate}
}
% \cSI{I tried to migrate this part to Section 4.1, but couldn't do that successfully without distracting flow. So I think we can keep the proerties of untisted Clebsch variables here.}

\rev{\paragraph{Open curves}
We also note that our construction of \(\psi\) is not limited to closed curves but is also valid for open curves 
% whose two end points are located at the same topologically-connected component of a boundary such as an obstacle or the end of a domain of interest.
with end points located on the boundary or obstacles subject to the following integrability condition.  The orientation of each curve assigns a positive or negative signature for its two end points.  The collection of curves is said to be integrable if all ends of the curves lie on the boundary and each connected component of the boundary contains an equal number of positive and negative ends.  Note that vortex filaments must be integrable since this integrability condition is precisely the integrability condition for curl: The vorticity 2-form is the exterior derivative of a velocity 1-form (i.e. exact) if and only if it is closed (divergence-free) and its restriction to every boundary component has zero total flux. In this case, the integrability of curl ensures that the endpoints of open filaments landing on boundaries must give equal positive and negative ends per boundary component. Therefore, it is possible to pair the endpoints along the boundary and complete the filaments as closed curves, from which we know how to construct \(\psi\). 
 } 
% \cSI{Can we make the explanation more step-by-step for readers with less background? If it becomes long, we can make a seperate paragraph "Open curves".}

%% file: VortexUntwisted_paper.bbl
%%% -*-BibTeX-*-
%%% Do NOT edit. File created by BibTeX with style
%%% ACM-Reference-Format-Journals [18-Jan-2012].

\begin{thebibliography}{50}

%%% ====================================================================
%%% NOTE TO THE USER: you can override these defaults by providing
%%% customized versions of any of these macros before the \bibliography
%%% command.  Each of them MUST provide its own final punctuation,
%%% except for \shownote{}, \showDOI{}, and \showURL{}.  The latter two
%%% do not use final punctuation, in order to avoid confusing it with
%%% the Web address.
%%%
%%% To suppress output of a particular field, define its macro to expand
%%% to an empty string, or better, \unskip, like this:
%%%
%%% \newcommand{\showDOI}[1]{\unskip}   % LaTeX syntax
%%%
%%% \def \showDOI #1{\unskip}           % plain TeX syntax
%%%
%%% ====================================================================

\ifx \showCODEN    \undefined \def \showCODEN     #1{\unskip}     \fi
\ifx \showDOI      \undefined \def \showDOI       #1{#1}\fi
\ifx \showISBNx    \undefined \def \showISBNx     #1{\unskip}     \fi
\ifx \showISBNxiii \undefined \def \showISBNxiii  #1{\unskip}     \fi
\ifx \showISSN     \undefined \def \showISSN      #1{\unskip}     \fi
\ifx \showLCCN     \undefined \def \showLCCN      #1{\unskip}     \fi
\ifx \shownote     \undefined \def \shownote      #1{#1}          \fi
\ifx \showarticletitle \undefined \def \showarticletitle #1{#1}   \fi
\ifx \showURL      \undefined \def \showURL       {\relax}        \fi
% The following commands are used for tagged output and should be
% invisible to TeX
\providecommand\bibfield[2]{#2}
\providecommand\bibinfo[2]{#2}
\providecommand\natexlab[1]{#1}
\providecommand\showeprint[2][]{arXiv:#2}

\bibitem[Ambrosio and Soner(1996)]%
        {Ambrosio:1996:LSA}
\bibfield{author}{\bibinfo{person}{Luigi Ambrosio} {and}
  \bibinfo{person}{Halil~Mete Soner}.} \bibinfo{year}{1996}\natexlab{}.
\newblock \showarticletitle{Level set approach to mean curvature flow in
  arbitrary codimension}.
\newblock \bibinfo{journal}{\emph{Journal of differential geometry}}
  \bibinfo{volume}{43}, \bibinfo{number}{4} (\bibinfo{year}{1996}),
  \bibinfo{pages}{693--737}.
\newblock


\bibitem[Angelidis(2017)]%
        {Angelidis:2017:MVF}
\bibfield{author}{\bibinfo{person}{Alexis Angelidis}.}
  \bibinfo{year}{2017}\natexlab{}.
\newblock \showarticletitle{Multi-scale vorticle fluids}.
\newblock \bibinfo{journal}{\emph{ACM Transactions on Graphics (TOG)}}
  \bibinfo{volume}{36}, \bibinfo{number}{4} (\bibinfo{year}{2017}),
  \bibinfo{pages}{1--12}.
\newblock


\bibitem[Angelidis and Neyret(2005)]%
        {Angelidis:2005:SSB}
\bibfield{author}{\bibinfo{person}{Alexis Angelidis} {and}
  \bibinfo{person}{Fabrice Neyret}.} \bibinfo{year}{2005}\natexlab{}.
\newblock \showarticletitle{Simulation of smoke based on vortex filament
  primitives}. In \bibinfo{booktitle}{\emph{Proceedings of the 2005 ACM
  SIGGRAPH/Eurographics symposium on Computer animation}}.
  \bibinfo{pages}{87--96}.
\newblock


\bibitem[Arnold and Khesin(1998)]%
        {Arnold:1998:TMH}
\bibfield{author}{\bibinfo{person}{Vladimir~I. Arnold} {and}
  \bibinfo{person}{Boris~A. Khesin}.} \bibinfo{year}{1998}\natexlab{}.
\newblock
  \bibinfo{booktitle}{\emph{\href{http://www.springer.com/us/book/9780387949475}{Topological
  Methods in Hydrodynamics}}}.
\newblock \bibinfo{publisher}{Springer}.
\newblock


\bibitem[Arvo(1995)]%
        {arvo1995applications}
\bibfield{author}{\bibinfo{person}{James Arvo}.}
  \bibinfo{year}{1995}\natexlab{}.
\newblock \showarticletitle{Applications of irradiance tensors to the
  simulation of non-lambertian phenomena}.
\newblock \bibinfo{journal}{\emph{Proceedings of the 22nd annual conference on
  Computer graphics and interactive techniques}}, \bibinfo{pages}{335--342}.
\newblock


\bibitem[Bergou et~al\mbox{.}(2008)]%
        {bergou2008elastic}
\bibfield{author}{\bibinfo{person}{Mikl\'{o}s Bergou}, \bibinfo{person}{Max
  Wardetzky}, \bibinfo{person}{Stephen Robinson}, \bibinfo{person}{Basile
  Audoly}, {and} \bibinfo{person}{Eitan Grinspun}.}
  \bibinfo{year}{2008}\natexlab{}.
\newblock \showarticletitle{Discrete Elastic Rods}.
\newblock , Article \bibinfo{articleno}{63} (\bibinfo{year}{2008}),
  \bibinfo{numpages}{12}~pages.
\newblock
\showISBNx{9781450301121}
\urldef\tempurl%
\url{https://doi.org/10.1145/1399504.1360662}
\showDOI{\tempurl}


\bibitem[Bethuel et~al\mbox{.}(1994)]%
        {Bethuel:1994:GLV}
\bibfield{author}{\bibinfo{person}{Fabrice Bethuel}, \bibinfo{person}{Ha{\"\i}m
  Brezis}, \bibinfo{person}{Fr{\'e}d{\'e}ric H{\'e}lein}, {et~al\mbox{.}}}
  \bibinfo{year}{1994}\natexlab{}.
\newblock \bibinfo{booktitle}{\emph{Ginzburg-landau vortices}}.
  Vol.~\bibinfo{volume}{13}.
\newblock \bibinfo{publisher}{Springer}.
\newblock


\bibitem[Bevis and Cambareri(1987)]%
        {Bevis1987ComputingTA}
\bibfield{author}{\bibinfo{person}{Michael~G. Bevis} {and}
  \bibinfo{person}{Greg Cambareri}.} \bibinfo{year}{1987}\natexlab{}.
\newblock \showarticletitle{Computing the area of a spherical polygon of
  arbitrary shape}.
\newblock \bibinfo{journal}{\emph{Mathematical Geology}}  \bibinfo{volume}{19}
  (\bibinfo{year}{1987}), \bibinfo{pages}{335--346}.
\newblock


\bibitem[Binysh and Alexander(2018)]%
        {Binysh_2018solidangle}
\bibfield{author}{\bibinfo{person}{Jack Binysh} {and} \bibinfo{person}{Gareth~P
  Alexander}.} \bibinfo{year}{2018}\natexlab{}.
\newblock \showarticletitle{Maxwell's theory of solid angle and the
  construction of knotted fields}.
\newblock \bibinfo{journal}{\emph{Journal of Physics A: Mathematical and
  Theoretical}} \bibinfo{volume}{51}, \bibinfo{number}{38} (\bibinfo{date}{aug}
  \bibinfo{year}{2018}), \bibinfo{pages}{385202}.
\newblock
\urldef\tempurl%
\url{https://doi.org/10.1088/1751-8121/aad8c6}
\showDOI{\tempurl}


\bibitem[Brochu et~al\mbox{.}(2012)]%
        {brochu2012linear}
\bibfield{author}{\bibinfo{person}{Tyson Brochu}, \bibinfo{person}{Todd
  Keeler}, {and} \bibinfo{person}{Robert Bridson}.}
  \bibinfo{year}{2012}\natexlab{}.
\newblock \showarticletitle{Linear-time smoke animation with vortex sheet
  meshes}. In \bibinfo{booktitle}{\emph{Proceedings of the ACM
  SIGGRAPH/Eurographics Symposium on Computer Animation}}. Citeseer,
  \bibinfo{pages}{87--95}.
\newblock


\bibitem[Burchard et~al\mbox{.}(2001)]%
        {Burchard:2001:MOC}
\bibfield{author}{\bibinfo{person}{Paul Burchard}, \bibinfo{person}{Li-Tien
  Cheng}, \bibinfo{person}{Barry Merriman}, {and} \bibinfo{person}{Stanley
  Osher}.} \bibinfo{year}{2001}\natexlab{}.
\newblock \showarticletitle{Motion of curves in three spatial dimensions using
  a level set approach}.
\newblock \bibinfo{journal}{\emph{J. Comput. Phys.}} \bibinfo{volume}{170},
  \bibinfo{number}{2} (\bibinfo{year}{2001}), \bibinfo{pages}{720--741}.
\newblock


\bibitem[Chern(2017)]%
        {Chern:2017:FD}
\bibfield{author}{\bibinfo{person}{Albert Chern}.}
  \bibinfo{year}{2017}\natexlab{}.
\newblock \emph{\bibinfo{title}{Fluid dynamics with incompressible
  Schr{\"o}dinger flow}}.
\newblock \bibinfo{thesistype}{Ph.\,D. Dissertation}.
  \bibinfo{school}{California Institute of Technology}.
\newblock


\bibitem[Chern et~al\mbox{.}(2017)]%
        {Chern:2017:IF}
\bibfield{author}{\bibinfo{person}{Albert Chern}, \bibinfo{person}{Felix
  Kn{\"o}ppel}, \bibinfo{person}{Ulrich Pinkall}, {and} \bibinfo{person}{Peter
  Schr{\"o}der}.} \bibinfo{year}{2017}\natexlab{}.
\newblock \showarticletitle{Inside fluids: Clebsch maps for visualization and
  processing}.
\newblock \bibinfo{journal}{\emph{ACM Transactions on Graphics (TOG)}}
  \bibinfo{volume}{36}, \bibinfo{number}{4} (\bibinfo{year}{2017}),
  \bibinfo{pages}{1--11}.
\newblock


\bibitem[Chern et~al\mbox{.}(2016)]%
        {Chern:2016:SS}
\bibfield{author}{\bibinfo{person}{Albert Chern}, \bibinfo{person}{Felix
  Kn{\"o}ppel}, \bibinfo{person}{Ulrich Pinkall}, \bibinfo{person}{Peter
  Schr{\"o}der}, {and} \bibinfo{person}{Steffen Wei{\ss}mann}.}
  \bibinfo{year}{2016}\natexlab{}.
\newblock \showarticletitle{Schr{\"o}dinger's smoke}.
\newblock \bibinfo{journal}{\emph{ACM Transactions on Graphics (TOG)}}
  \bibinfo{volume}{35}, \bibinfo{number}{4} (\bibinfo{year}{2016}),
  \bibinfo{pages}{1--13}.
\newblock


\bibitem[Chorin(1990)]%
        {Chorin:1990:HRV}
\bibfield{author}{\bibinfo{person}{Alexandre~Joel Chorin}.}
  \bibinfo{year}{1990}\natexlab{}.
\newblock \showarticletitle{Hairpin removal in vortex interactions}.
\newblock \bibinfo{journal}{\emph{J. Comput. Phys.}} \bibinfo{volume}{91},
  \bibinfo{number}{1} (\bibinfo{year}{1990}), \bibinfo{pages}{1--21}.
\newblock


\bibitem[Clebsch(1859)]%
        {Clebsch:1859:IHG}
\bibfield{author}{\bibinfo{person}{A. Clebsch}.}
  \bibinfo{year}{1859}\natexlab{}.
\newblock \showarticletitle{Ueber die Integration der hydrodynamischen
  Gleichungen}.
\newblock \bibinfo{journal}{\emph{Journal f\"ur die reine und angewandte
  Mathematik}}  \bibinfo{volume}{56} (\bibinfo{year}{1859}),
  \bibinfo{pages}{1--10}.
\newblock
\newblock
\shownote{English translation by {D.\ H.\ Delphenich},
  \url{http://www.neo-classical-physics.info/uploads/3/4/3/6/34363841/clebsch_-_clebsch_variables.pdf}}.


\bibitem[Cottet et~al\mbox{.}(2000)]%
        {Cottet:2000:VM}
\bibfield{author}{\bibinfo{person}{Georges-Henri Cottet},
  \bibinfo{person}{Petros~D Koumoutsakos}, {et~al\mbox{.}}}
  \bibinfo{year}{2000}\natexlab{}.
\newblock \bibinfo{booktitle}{\emph{Vortex methods: theory and practice}}.
  Vol.~\bibinfo{volume}{8}.
\newblock \bibinfo{publisher}{Cambridge university press Cambridge}.
\newblock


\bibitem[Da et~al\mbox{.}(2015)]%
        {da2015double}
\bibfield{author}{\bibinfo{person}{Fang Da}, \bibinfo{person}{Christopher
  Batty}, \bibinfo{person}{Chris Wojtan}, {and} \bibinfo{person}{Eitan
  Grinspun}.} \bibinfo{year}{2015}\natexlab{}.
\newblock \showarticletitle{Double bubbles sans toil and trouble: Discrete
  circulation-preserving vortex sheets for soap films and foams}.
\newblock \bibinfo{journal}{\emph{ACM Transactions on Graphics (TOG)}}
  \bibinfo{volume}{34}, \bibinfo{number}{4} (\bibinfo{year}{2015}),
  \bibinfo{pages}{1--9}.
\newblock


\bibitem[Elcott et~al\mbox{.}(2007)]%
        {Elcott:2007:SCP}
\bibfield{author}{\bibinfo{person}{Sharif Elcott}, \bibinfo{person}{Yiying
  Tong}, \bibinfo{person}{Eva Kanso}, \bibinfo{person}{Peter Schr{\"o}der},
  {and} \bibinfo{person}{Mathieu Desbrun}.} \bibinfo{year}{2007}\natexlab{}.
\newblock \showarticletitle{Stable, circulation-preserving, simplicial fluids}.
\newblock \bibinfo{journal}{\emph{ACM Transactions on Graphics (TOG)}}
  \bibinfo{volume}{26}, \bibinfo{number}{1} (\bibinfo{year}{2007}),
  \bibinfo{pages}{4--es}.
\newblock


\bibitem[Gamito et~al\mbox{.}(1995)]%
        {gamito1995two}
\bibfield{author}{\bibinfo{person}{Manuel~Noronha Gamito},
  \bibinfo{person}{Pedro~Faria Lopes}, {and} \bibinfo{person}{M{\'a}rio~Rui
  Gomes}.} \bibinfo{year}{1995}\natexlab{}.
\newblock \showarticletitle{Two-dimensional simulation of gaseous phenomena
  using vortex particles}.
\newblock In \bibinfo{booktitle}{\emph{Computer Animation and Simulation’95}}.
  \bibinfo{publisher}{Springer}, \bibinfo{pages}{3--15}.
\newblock


\bibitem[Kuznetsov and Mikhailov(1980)]%
        {Kuznetsov:1980:TMC}
\bibfield{author}{\bibinfo{person}{Evgenii~A Kuznetsov} {and}
  \bibinfo{person}{Aleksandr~V Mikhailov}.} \bibinfo{year}{1980}\natexlab{}.
\newblock \showarticletitle{On the topological meaning of canonical Clebsch
  variables}.
\newblock \bibinfo{journal}{\emph{Physics Letters A}} \bibinfo{volume}{77},
  \bibinfo{number}{1} (\bibinfo{year}{1980}), \bibinfo{pages}{37--38}.
\newblock


\bibitem[Lamb(1895)]%
        {Lamb:1895:HD}
\bibfield{author}{\bibinfo{person}{Horace Lamb}.}
  \bibinfo{year}{1895}\natexlab{}.
\newblock \bibinfo{booktitle}{\emph{Hydrodynamics}}.
\newblock \bibinfo{publisher}{Cambridge University Press}.
\newblock


\bibitem[Lorensen and Cline(1987)]%
        {Lorensen1987marchingcubes}
\bibfield{author}{\bibinfo{person}{William~E. Lorensen} {and}
  \bibinfo{person}{Harvey~E. Cline}.} \bibinfo{year}{1987}\natexlab{}.
\newblock \showarticletitle{Marching Cubes: A High Resolution 3D Surface
  Construction Algorithm}.
\newblock \bibinfo{journal}{\emph{SIGGRAPH Comput. Graph.}}
  \bibinfo{volume}{21}, \bibinfo{number}{4} (\bibinfo{date}{aug}
  \bibinfo{year}{1987}), \bibinfo{pages}{163^^e2^^80^^93169}.
\newblock
\showISSN{0097-8930}
\urldef\tempurl%
\url{https://doi.org/10.1145/37402.37422}
\showDOI{\tempurl}


\bibitem[Marsden and Weinstein(1983)]%
        {Marsden:1983:COV}
\bibfield{author}{\bibinfo{person}{Jerrold Marsden} {and} \bibinfo{person}{Alan
  Weinstein}.} \bibinfo{year}{1983}\natexlab{}.
\newblock \showarticletitle{Coadjoint Orbits, Vortices, and Clebsch Variables
  for Incompressible Fluids}.
\newblock \bibinfo{journal}{\emph{Physica D: Nonlinear Phenomena}}
  \bibinfo{volume}{7}, \bibinfo{number}{1} (\bibinfo{year}{1983}),
  \bibinfo{pages}{305--323}.
\newblock


\bibitem[Min(2004)]%
        {Min:2004:LLS}
\bibfield{author}{\bibinfo{person}{Chohong Min}.}
  \bibinfo{year}{2004}\natexlab{}.
\newblock \showarticletitle{Local level set method in high dimension and
  codimension}.
\newblock \bibinfo{journal}{\emph{Journal of computational physics}}
  \bibinfo{volume}{200}, \bibinfo{number}{1} (\bibinfo{year}{2004}),
  \bibinfo{pages}{368--382}.
\newblock


\bibitem[Morrison(1998)]%
        {Morrison:1998:HDI}
\bibfield{author}{\bibinfo{person}{Philip~J Morrison}.}
  \bibinfo{year}{1998}\natexlab{}.
\newblock \showarticletitle{Hamiltonian description of the ideal fluid}.
\newblock \bibinfo{journal}{\emph{Reviews of modern physics}}
  \bibinfo{volume}{70}, \bibinfo{number}{2} (\bibinfo{year}{1998}),
  \bibinfo{pages}{467}.
\newblock


\bibitem[Murasugi(1996)]%
        {musasugi2008knottheory}
\bibfield{author}{\bibinfo{person}{Kunio Murasugi}.} \bibinfo{year}{2008 -
  1996}\natexlab{}.
\newblock \bibinfo{booktitle}{\emph{Knot theory \& its applications}}.
\newblock \bibinfo{publisher}{Birkh\"auser}, \bibinfo{address}{Boston}.
\newblock


\bibitem[Museth et~al\mbox{.}(2019)]%
        {museth2019openvdb}
\bibfield{author}{\bibinfo{person}{Ken Museth}, \bibinfo{person}{Nick
  Avramoussis}, {and} \bibinfo{person}{Dan Bailey}.}
  \bibinfo{year}{2019}\natexlab{}.
\newblock \showarticletitle{OpenVDB}.
\newblock In \bibinfo{booktitle}{\emph{ACM SIGGRAPH 2019 Courses}}.
  \bibinfo{pages}{1--56}.
\newblock


\bibitem[Nabizadeh et~al\mbox{.}(2021)]%
        {nabizadeh:2021:kelvin}
\bibfield{author}{\bibinfo{person}{Mohammad~Sina Nabizadeh},
  \bibinfo{person}{Albert Chern}, {and} \bibinfo{person}{Ravi Ramamoorthi}.}
  \bibinfo{year}{2021}\natexlab{}.
\newblock \showarticletitle{Kelvin Transformations for Simulations on Infinite
  Domains.}
\newblock \bibinfo{journal}{\emph{ACM Transactions on Graphics (TOG)}}
  \bibinfo{volume}{40}, \bibinfo{number}{4} (\bibinfo{year}{2021}),
  \bibinfo{pages}{97:1--97:15}.
\newblock


\bibitem[Osher and Sethian(1988)]%
        {osher1988fronts}
\bibfield{author}{\bibinfo{person}{Stanley Osher} {and}
  \bibinfo{person}{James~A Sethian}.} \bibinfo{year}{1988}\natexlab{}.
\newblock \showarticletitle{Fronts propagating with curvature-dependent speed:
  Algorithms based on Hamilton-Jacobi formulations}.
\newblock \bibinfo{journal}{\emph{Journal of computational physics}}
  \bibinfo{volume}{79}, \bibinfo{number}{1} (\bibinfo{year}{1988}),
  \bibinfo{pages}{12--49}.
\newblock


\bibitem[Padilla et~al\mbox{.}(2019)]%
        {Padilla:2019:BRIC}
\bibfield{author}{\bibinfo{person}{Marcel Padilla}, \bibinfo{person}{Albert
  Chern}, \bibinfo{person}{Felix Kn{\"o}ppel}, \bibinfo{person}{Ulrich
  Pinkall}, {and} \bibinfo{person}{Peter Schr{\"o}der}.}
  \bibinfo{year}{2019}\natexlab{}.
\newblock \showarticletitle{On bubble rings and ink chandeliers}.
\newblock \bibinfo{journal}{\emph{ACM Transactions on Graphics (TOG)}}
  \bibinfo{volume}{38}, \bibinfo{number}{4} (\bibinfo{year}{2019}),
  \bibinfo{pages}{1--14}.
\newblock


\bibitem[Palmer et~al\mbox{.}(2020)]%
        {Palmer:2020:ARV}
\bibfield{author}{\bibinfo{person}{David Palmer}, \bibinfo{person}{David
  Bommes}, {and} \bibinfo{person}{Justin Solomon}.}
  \bibinfo{year}{2020}\natexlab{}.
\newblock \showarticletitle{Algebraic representations for volumetric frame
  fields}.
\newblock \bibinfo{journal}{\emph{ACM Transactions on Graphics (TOG)}}
  \bibinfo{volume}{39}, \bibinfo{number}{2} (\bibinfo{year}{2020}),
  \bibinfo{pages}{1--17}.
\newblock


\bibitem[Park and Kim(2005)]%
        {park2005vortex}
\bibfield{author}{\bibinfo{person}{Sang~Il Park} {and}
  \bibinfo{person}{Myoung~Jun Kim}.} \bibinfo{year}{2005}\natexlab{}.
\newblock \showarticletitle{Vortex fluid for gaseous phenomena}. In
  \bibinfo{booktitle}{\emph{Proceedings of the 2005 ACM SIGGRAPH/Eurographics
  symposium on Computer animation}}. \bibinfo{pages}{261--270}.
\newblock


\bibitem[Pfaff et~al\mbox{.}(2012)]%
        {Pfaff:2012:LVS}
\bibfield{author}{\bibinfo{person}{Tobias Pfaff}, \bibinfo{person}{Nils
  Thuerey}, {and} \bibinfo{person}{Markus Gross}.}
  \bibinfo{year}{2012}\natexlab{}.
\newblock \showarticletitle{Lagrangian vortex sheets for animating fluids}.
\newblock \bibinfo{journal}{\emph{ACM Transactions on Graphics (TOG)}}
  \bibinfo{volume}{31}, \bibinfo{number}{4} (\bibinfo{year}{2012}),
  \bibinfo{pages}{1--8}.
\newblock


\bibitem[Pismen et~al\mbox{.}(1999)]%
        {Pismen:1999:VNF}
\bibfield{author}{\bibinfo{person}{Len~M Pismen}, \bibinfo{person}{Len~M
  Pismen}, {et~al\mbox{.}}} \bibinfo{year}{1999}\natexlab{}.
\newblock \bibinfo{booktitle}{\emph{Vortices in nonlinear fields: from liquid
  crystals to superfluids, from non-equilibrium patterns to cosmic strings}}.
  Vol.~\bibinfo{volume}{100}.
\newblock \bibinfo{publisher}{Oxford University Press}.
\newblock


\bibitem[Qu et~al\mbox{.}(2019)]%
        {Qu:2019:ECF}
\bibfield{author}{\bibinfo{person}{Ziyin Qu}, \bibinfo{person}{Xinxin Zhang},
  \bibinfo{person}{Ming Gao}, \bibinfo{person}{Chenfanfu Jiang}, {and}
  \bibinfo{person}{Baoquan Chen}.} \bibinfo{year}{2019}\natexlab{}.
\newblock \showarticletitle{Efficient and conservative fluids using
  bidirectional mapping}.
\newblock \bibinfo{journal}{\emph{ACM Transactions on Graphics (TOG)}}
  \bibinfo{volume}{38}, \bibinfo{number}{4} (\bibinfo{year}{2019}),
  \bibinfo{pages}{1--12}.
\newblock


\bibitem[Ruuth et~al\mbox{.}(2001)]%
        {Ruuth:2001:DGM}
\bibfield{author}{\bibinfo{person}{Steven~J Ruuth}, \bibinfo{person}{Barry
  Merriman}, \bibinfo{person}{Jack Xin}, {and} \bibinfo{person}{Stanley
  Osher}.} \bibinfo{year}{2001}\natexlab{}.
\newblock \showarticletitle{Diffusion-generated motion by mean curvature for
  filaments}.
\newblock \bibinfo{journal}{\emph{Journal of Nonlinear Science}}
  \bibinfo{volume}{11}, \bibinfo{number}{6} (\bibinfo{year}{2001}),
  \bibinfo{pages}{473--493}.
\newblock


\bibitem[Saffman(1992)]%
        {Saffman:1995:VD}
\bibfield{author}{\bibinfo{person}{Philip~Geoffrey Saffman}.}
  \bibinfo{year}{1992}\natexlab{}.
\newblock
  \bibinfo{booktitle}{\emph{\href{https://books.google.com/books?id=FyxqMmCPu4AC\&dq=vortex+dynamics+saffman\&source=gbs_navlinks_s}{Vortex
  Dynamics}}}.
\newblock


\bibitem[Seifert(1935)]%
        {Seifert1935seifertsurface}
\bibfield{author}{\bibinfo{person}{H. Seifert}.}
  \bibinfo{year}{1935}\natexlab{}.
\newblock \showarticletitle{\"Uber das Geschlecht von Knoten}.
\newblock \bibinfo{journal}{\emph{Math. Ann.}} \bibinfo{volume}{110},
  \bibinfo{number}{1} (\bibinfo{year}{1935}), \bibinfo{pages}{571--592}.
\newblock
\showISSN{0025-5831}


\bibitem[Selle et~al\mbox{.}(2008)]%
        {Selle:2008:MCM}
\bibfield{author}{\bibinfo{person}{Andrew Selle}, \bibinfo{person}{Ronald
  Fedkiw}, \bibinfo{person}{Byungmoon Kim}, \bibinfo{person}{Yingjie Liu},
  {and} \bibinfo{person}{Jarek Rossignac}.} \bibinfo{year}{2008}\natexlab{}.
\newblock \showarticletitle{An unconditionally stable MacCormack method}.
\newblock \bibinfo{journal}{\emph{Journal of Scientific Computing}}
  \bibinfo{volume}{35}, \bibinfo{number}{2} (\bibinfo{year}{2008}),
  \bibinfo{pages}{350--371}.
\newblock


\bibitem[Selle et~al\mbox{.}(2005)]%
        {Selle:2005:VPM}
\bibfield{author}{\bibinfo{person}{Andrew Selle}, \bibinfo{person}{Nick
  Rasmussen}, {and} \bibinfo{person}{Ronald Fedkiw}.}
  \bibinfo{year}{2005}\natexlab{}.
\newblock \showarticletitle{A vortex particle method for smoke, water and
  explosions}.
\newblock In \bibinfo{booktitle}{\emph{ACM SIGGRAPH 2005 Papers}}.
  \bibinfo{pages}{910--914}.
\newblock


\bibitem[Solomon et~al\mbox{.}(2017)]%
        {Solomon:2017:BEO}
\bibfield{author}{\bibinfo{person}{Justin Solomon}, \bibinfo{person}{Amir
  Vaxman}, {and} \bibinfo{person}{David Bommes}.}
  \bibinfo{year}{2017}\natexlab{}.
\newblock \showarticletitle{Boundary element octahedral fields in volumes}.
\newblock \bibinfo{journal}{\emph{ACM Transactions on Graphics (TOG)}}
  \bibinfo{volume}{36}, \bibinfo{number}{4} (\bibinfo{year}{2017}),
  \bibinfo{pages}{1}.
\newblock


\bibitem[Wei{\ss}mann and Pinkall(2009)]%
        {Weissmann:2009:FBS}
\bibfield{author}{\bibinfo{person}{Steffen Wei{\ss}mann} {and}
  \bibinfo{person}{Ulrich Pinkall}.} \bibinfo{year}{2009}\natexlab{}.
\newblock \showarticletitle{{Real-time Interactive Simulation of Smoke Using
  Discrete Integrable Vortex Filaments}}. In \bibinfo{booktitle}{\emph{Workshop
  in Virtual Reality Interactions and Physical Simulation "VRIPHYS" (2009)}},
  \bibfield{editor}{\bibinfo{person}{Hartmut Prautzsch},
  \bibinfo{person}{Alfred Schmitt}, \bibinfo{person}{Jan Bender}, {and}
  \bibinfo{person}{Matthias Teschner}} (Eds.). \bibinfo{publisher}{The
  Eurographics Association}.
\newblock
\showISBNx{978-3-905673-73-9}
\urldef\tempurl%
\url{https://doi.org/10.2312/PE/vriphys/vriphys09/001-010}
\showDOI{\tempurl}


\bibitem[Wei\ss{}mann and Pinkall(2010)]%
        {Weissmann:2010:filaments}
\bibfield{author}{\bibinfo{person}{Steffen Wei\ss{}mann} {and}
  \bibinfo{person}{Ulrich Pinkall}.} \bibinfo{year}{2010}\natexlab{}.
\newblock \showarticletitle{Filament-Based Smoke with Vortex Shedding and
  Variational Reconnection}.
\newblock \bibinfo{journal}{\emph{ACM Trans. Graph.}} \bibinfo{volume}{29},
  \bibinfo{number}{4}, Article \bibinfo{articleno}{115} (\bibinfo{date}{jul}
  \bibinfo{year}{2010}), \bibinfo{numpages}{12}~pages.
\newblock
\showISSN{0730-0301}
\urldef\tempurl%
\url{https://doi.org/10.1145/1778765.1778852}
\showDOI{\tempurl}


\bibitem[Wei\ss{}mann and Pinkall(2012)]%
        {Weissmann:2012:UnderWaterRigidBody}
\bibfield{author}{\bibinfo{person}{Steffen Wei\ss{}mann} {and}
  \bibinfo{person}{Ulrich Pinkall}.} \bibinfo{year}{2012}\natexlab{}.
\newblock \showarticletitle{Underwater Rigid Body Dynamics}.
\newblock \bibinfo{journal}{\emph{ACM Trans. Graph.}} \bibinfo{volume}{31},
  \bibinfo{number}{4}, Article \bibinfo{articleno}{104} (\bibinfo{date}{jul}
  \bibinfo{year}{2012}), \bibinfo{numpages}{7}~pages.
\newblock
\showISSN{0730-0301}
\urldef\tempurl%
\url{https://doi.org/10.1145/2185520.2185600}
\showDOI{\tempurl}


\bibitem[Wei\ss{}mann et~al\mbox{.}(2014)]%
        {weissmann2014smokerings}
\bibfield{author}{\bibinfo{person}{Steffen Wei\ss{}mann},
  \bibinfo{person}{Ulrich Pinkall}, {and} \bibinfo{person}{Peter
  Schr\"{o}der}.} \bibinfo{year}{2014}\natexlab{}.
\newblock \showarticletitle{Smoke Rings from Smoke}.
\newblock \bibinfo{journal}{\emph{ACM Trans. Graph.}} \bibinfo{volume}{33},
  \bibinfo{number}{4}, Article \bibinfo{articleno}{140} (\bibinfo{date}{jul}
  \bibinfo{year}{2014}), \bibinfo{numpages}{8}~pages.
\newblock
\showISSN{0730-0301}
\urldef\tempurl%
\url{https://doi.org/10.1145/2601097.2601171}
\showDOI{\tempurl}


\bibitem[Xiong et~al\mbox{.}(2021)]%
        {Xiong:2021:VSC}
\bibfield{author}{\bibinfo{person}{Shiying Xiong}, \bibinfo{person}{Rui Tao},
  \bibinfo{person}{Yaorui Zhang}, \bibinfo{person}{Fan Feng}, {and}
  \bibinfo{person}{Bo Zhu}.} \bibinfo{year}{2021}\natexlab{}.
\newblock \showarticletitle{Incompressible Flow Simulation on Vortex Segment
  Clouds}.
\newblock \bibinfo{journal}{\emph{ACM Transactions on Graphics (TOG)}}
  \bibinfo{volume}{40}, \bibinfo{number}{4} (\bibinfo{year}{2021}),
  \bibinfo{pages}{98:1--98:11}.
\newblock


\bibitem[Yang et~al\mbox{.}(2021)]%
        {Yang:2021:CGF}
\bibfield{author}{\bibinfo{person}{Shuqi Yang}, \bibinfo{person}{Shiying
  Xiong}, \bibinfo{person}{Yaorui Zhang}, \bibinfo{person}{Fan Feng},
  \bibinfo{person}{Jinyuan Liu}, {and} \bibinfo{person}{Bo Zhu}.}
  \bibinfo{year}{2021}\natexlab{}.
\newblock \showarticletitle{Clebsch gauge fluid}.
\newblock \bibinfo{journal}{\emph{ACM Transactions on Graphics (TOG)}}
  \bibinfo{volume}{40}, \bibinfo{number}{4} (\bibinfo{year}{2021}),
  \bibinfo{pages}{1--11}.
\newblock


\bibitem[Zhang and Bridson(2014)]%
        {zhang2014pppm}
\bibfield{author}{\bibinfo{person}{Xinxin Zhang} {and} \bibinfo{person}{Robert
  Bridson}.} \bibinfo{year}{2014}\natexlab{}.
\newblock \showarticletitle{A PPPM fast summation method for fluids and
  beyond}.
\newblock \bibinfo{journal}{\emph{ACM Transactions on Graphics (TOG)}}
  \bibinfo{volume}{33}, \bibinfo{number}{6} (\bibinfo{year}{2014}),
  \bibinfo{pages}{1--11}.
\newblock


\bibitem[Zhang et~al\mbox{.}(2015)]%
        {Zhang:2015:RMV}
\bibfield{author}{\bibinfo{person}{Xinxin Zhang}, \bibinfo{person}{Robert
  Bridson}, {and} \bibinfo{person}{Chen Greif}.}
  \bibinfo{year}{2015}\natexlab{}.
\newblock \showarticletitle{Restoring the missing vorticity in
  advection-projection fluid solvers}.
\newblock \bibinfo{journal}{\emph{ACM Transactions on Graphics (TOG)}}
  \bibinfo{volume}{34}, \bibinfo{number}{4} (\bibinfo{year}{2015}),
  \bibinfo{pages}{1--8}.
\newblock


\end{thebibliography}
